\newcount\Comments  
\documentclass[longauth]{aa}  

\usepackage{natbib}
\usepackage{graphicx}
\usepackage{amsmath}
\usepackage{amssymb}
\usepackage{txfonts}
\usepackage{longtable}
\usepackage[switch]{lineno}
\usepackage{color}
\definecolor{darkgreen}{rgb}{0,0.5,0}
\definecolor{darkred}{rgb}{0.5,0,0}

\usepackage{dcolumn}
\newcolumntype{.}{D{.}{.}{-1}}

\bibpunct{(}{)}{;}{a}{}{,} 
\newcommand{\kibitz}[2]{\ifnum\Comments=1\textcolor{#1}{#2}\fi}
\newcommand{\kk}[1]{#1}
\newcommand{\lr}[1]{#1}

\newcommand{\mgclsdoi}{\url{https://doi.org/10.48479/7epd-w356}}

\usepackage[]{hyperref}
\newcommand{\per}{{$^{-1}$}}

\newcommand{\kms}{{\,km\,s$^{-1}$}}

\def\arrvline{\hfil\kern\arraycolsep\vline\kern-\arraycolsep\hfilneg}

\usepackage{soul} 
\newcommand{\editout}[1]{}
\newcommand{\editnew}[1]{\textcolor{black}{{#1}}}

\begin{document} 
   \setstcolor{red}

   \title{The MeerKAT Galaxy Cluster Legacy Survey}

   \subtitle{I. Survey Overview and Highlights}

    \author{\small 
        K.~Knowles\inst{1,2,3}\fnmsep\thanks{\email{k.knowles@ru.ac.za}} \and W.~D.~Cotton\inst{4,3} \and L.~Rudnick\inst{5} \and F.~Camilo\inst{3} \and S.~Goedhart\inst{3} \and R.~Deane\inst{6,7} \and M.~Ramatsoku\inst{2,8} 
        \and M.~F.~Bietenholz\inst{9,10} \and M.~Br\"{u}ggen\inst{11} \and C.~Button\inst{7}
% ...
        }

    \institute{Astrophysics Research Centre, University of KwaZulu-Natal, Durban 4041, South Africa 
        \and 
            Centre for Radio Astronomy Techniques and Technologies, Department of Physics and Electronics, Rhodes University, P.O. Box 94, Makhanda 6140, South Africa
        \and 
            South African Radio Astronomy Observatory, 2 Fir Street, Observatory 7925, South Africa
        \and 
            National Radio Astronomy Observatory, Charlottesville, VA 22903, USA
        \and 
            Minnesota Institute for Astrophysics, University of Minnesota, 116 Church St SE, Minneapolis, MN 55455, USA 
        \and 
            Wits Centre for Astrophysics, School of Physics, University of the Witwatersrand, 1 Jan Smuts Avenue, Johannesburg 2050, South Africa
        \and 
            Department of Physics, University of Pretoria, Hatfield 0028, South Africa
        \and 
            INAF -- Osservatorio Astronomico di Cagliari, Via della Scienza 5, I-09047 Selargius (CA), Italy
        \and 
            Hartebeesthoek Radio Astronomy Observatory, SARAO, Krugersdorp 1740, South Africa
        \and 
            York University, Toronto, ON, Canada
        \and 
            Hamburger Sternwarte, Universit\"at Hamburg, Gojenbergsweg 112, 21029 Hamburg, Germany
% ...
}

   \date{Received XXX; accepted YYY}

 
  \abstract{
 MeerKAT's large number (64) of 13.5\,m-diameter antennas, spanning 8\,km with a densely packed 1\,km core, create a powerful instrument for wide-area surveys, with high sensitivity over a wide range of angular scales. The MeerKAT Galaxy Cluster Legacy Survey (MGCLS) is a programme of long-track MeerKAT L-band (900--1670\,MHz) observations of 115 galaxy clusters, observed for $\sim$\,6--10 hours each in full polarisation. 
 The first legacy product data release (DR1), made available with this paper, includes the MeerKAT visibilities, basic image cubes at $\sim$\,8\arcsec\ resolution, and enhanced spectral and polarisation image cubes at $\sim$\,8\arcsec\ and 15\arcsec\ resolutions. Typical sensitivities for the full-resolution MGCLS image products range over $\sim$\,3--5\,$\mu$Jy\,beam\per. The basic cubes are full-field and span 2$^\circ\,\times\,2^\circ$. The enhanced products consist of the inner 1.2$^\circ\,\times\,1.2^\circ$ field of view, corrected for the primary beam. The survey is fully sensitive to structures up to $\sim$\,10\arcmin\ scales and the wide bandwidth allows spectral and Faraday rotation mapping.  Relatively narrow frequency channels (209\,kHz) are also used to provide \textup{H\,\textsc{\lowercase{i}}} mapping in windows of $0 < z < 0.09$ and $0.19 < z < 0.48$. 
 In this paper, we provide an overview of the survey and the DR1 products, including caveats for usage. 
 We present some initial results from the survey, both for their intrinsic scientific value and to highlight the capabilities for further exploration with these data. 
 These include a primary beam-corrected compact source catalogue of $\sim$\,626,000 sources for the full survey, and an optical/infrared cross-matched catalogue for compact sources in the primary-beam corrected areas of Abell~209 and Abell~S295. We examine dust unbiased star-formation rates as a function of clustercentric radius in Abell~209, extending out to 3.5\,$R_{200}$. We \kk{find no dependence of the star formation rate on distance from the cluster centre}\editout{confirm earlier studies of reduced star formation near the centre}, and observe a small excess of the radio-to-100\,$\mu$m flux ratio towards the centre of Abell~209 that may reflect a ram pressure enhancement in the denser environment.  
 We detect diffuse cluster radio emission in 62 of the surveyed systems and present a catalogue of the 99 diffuse cluster emission structures, of which 56 are new. These include mini-halos, halos, relics, and other diffuse structures for which no suitable characterization currently exists.
 We highlight some of the radio galaxies which challenge current paradigms such as trident-shaped structures, jets that remain well-collimated far beyond their bending radius, and filamentary features linked to radio galaxies that likely illuminate magnetic flux tubes in the intracluster medium. 
 We also present early results from the \textup{H\,\textsc{\lowercase{i}}} analysis of four clusters, showing a wide variety of \textup{H\,\textsc{\lowercase{i}}} mass distributions reflecting both sensitivity and intrinsic cluster effects, and the serendipitous discovery of a group in the foreground of Abell~3365. 
  } 


   \maketitle
   
   {
   \begingroup
   \let\clearpage\relax
   \tableofcontents
   \endgroup
   }
%


\section{Introduction}\label{sec:intro}

\editout{MeerKAT}
\editout{is a 64-dish radio interferometer that can observe the sky below a declination of $+45^\circ$ (with an elevation limit of 15$^\circ$), operating in the UHF (580--1015\,MHz), L (900--1670\,MHz), and S bands (1.75--3.5\,GHz). Its specifications are described in detail in }
\editout{. MeerKAT's L-band system, with a primary beam full-width at half-maximum (FWHM) of 1.2$^\circ$ at 1.28\,GHz, was the first to be commissioned, and in 2018 MeerKAT began a programme of long-track observations of galaxy clusters. This programme became the MeerKAT Galaxy Cluster Legacy Survey (MGCLS), using $\sim$\,1000 hours at L-band to observe 115 galaxy clusters in full polarisation between $-85^\circ$ and $0^\circ$ declination, spread out over the full range of right ascension. }

Galaxy clusters are the largest gravitationally-bound structures in the Universe, and as such are powerful tools for a variety of research areas in both astrophysics and cosmology. Their composition is dominated by dark matter, with $\sim$13\% of their mass coming from  the ionised plasma of the intracluster medium (ICM) and only $\sim$2\% from the stars of their constituent galaxies and cold gas. Some of these galaxies radiate at radio frequencies, either through star formation processes or from nuclear activity in the galaxy's core \citep[][]{Condon1992, 2006MNRAS.372..741S, 2015AJ....150...87L, 2017ApJ...842...95M}. Some radio galaxies with active galactic nuclei (AGN) exhibit large-scale radio jets or lobes \citep{FR}, which can be disrupted by interaction with the ICM through merger-related or other processes \citep[see e.g.,][]{1972ApJ...176....1G, 1972Natur.237..269M, 1975A&A....43..337C, 2003AJ....125.1635B}.

Radio observations of clusters have also revealed steep-spectrum, diffuse radio emission \citep[see reviews by][]{Feretti2012,2019SSRv..215...16V}, which can be used to study the distributed populations of cosmic ray particles and magnetic fields in the ICM, outside of individual radio or star-forming galaxies. These diffuse structures are closely linked to cluster mergers \citep{2010ApJ...721L..82C, 2011JApA...32..505V}, so can also be used to study shock physics, merger-related turbulence, and other particle re-acceleration processes within the ICM \citep[see reviews by][and references therein]{2014IJMPD..2330007B,2019SSRv..215...16V}. 
Cluster observations carried out by wide-field instruments \editout{such as MeerKAT }also contain many field sources, both along the cluster line of sight and in the surrounding area. These provide important information on the clusters themselves, e.g., by using background sources as Faraday rotation probes, and on radio galaxy physics outside of the dense cluster environments. Wide-field imaging\editout{, such as employed here,} enables both statistical studies, such as environment-sensitive properties of galaxy populations, and serendipitous studies of individual field sources \citep[e.g.,][]{2021A&A...647A...3B}.

\kk{MeerKAT\footnote{Operated by the South African Radio Astronomy Observatory (SARAO).} is a 64-dish radio interferometer that can observe the sky below a declination of $+45^\circ$ (with an elevation limit of 15$^\circ$), operating in the UHF (580--1015\,MHz), L (900--1670\,MHz), and S bands (1.75--3.5\,GHz). Its specifications are described in detail in \citet{Jonas2016} and \citet{Camilo2018}. MeerKAT's L-band system, with a primary beam full-width at half-maximum (FWHM) of 1.2$^\circ$ at 1.28\,GHz, was the first to be commissioned, and in 2018 MeerKAT began a programme of long-track observations of galaxy clusters. This programme became the MeerKAT Galaxy Cluster Legacy Survey (MGCLS), using $\sim$\,1000 hours at L-band to observe 115 galaxy clusters in full polarisation between $-80^\circ$ and $0^\circ$ declination, spread out over the full range of right ascension. }

In addition to continuum and polarimetric studies, the deep, broadband, wide-field, sub-10{\arcsec} resolution MGCLS observations provide a rich resource to study neutral hydrogen in galaxies. Studies of \textup{H\,\textsc{\lowercase{i}}} morphologies in  dense cluster environments and in the field, distributions of \textup{H\,\textsc{\lowercase{i}}} masses in different types of clusters, and the cosmic evolution of cluster \textup{H\,\textsc{\lowercase{i}}} out to redshifts of $z = 0.48$ are all enabled with these data, with a velocity resolution of $\sim$ 44\,\kms\ at $z=0$.

Here we present an overview of the MGCLS and the various legacy products being made available to the astronomical community. These data are a rich resource for many scientific studies, both cluster-specific and involving field sources. We provide some initial science findings in the areas of cluster diffuse emission, radio galaxy physics, star-forming systems, and neutral hydrogen mapping. In addition to their intrinsic value, these examples also demonstrate the potential of the legacy products for a wide range of astrophysical investigations. 
   
The paper is organised as follows. In Section~\ref{sec:sampleobs} we describe the target sample, with the observations and initial data processing described in Section~\ref{sec:initialproc}. A discussion of the legacy data products, including caveats for use and some primary use cases, is provided in Section~\ref{sec:products}. Source catalogues are presented in Section~\ref{sec:srccats}. The next four sections present highlights of various science investigations that have been or can be carried out using the legacy products and visibilities: Section~\ref{sec:clusterde} focuses on cluster diffuse emission; Section~\ref{sec:individsrcs} highlights some interesting individual radio sources; Section~\ref{sec:galaxies} presents results based on star-forming galaxies; and Section~\ref{sec:hiscience} highlights some \textup{H\,\textsc{\lowercase{i}}} science capabilities. A summary and concluding remarks are presented in Section~\ref{sec:conclusion}. In this paper we assume a flat $\Lambda$CDM cosmology with $H_0\,=\,70$\,km\,s{\per}\,Mpc{\per}, $\Omega_{\rm m}\,=\,0.3$, and $\Omega_{\rm \Lambda}\,=\,0.7$. We define the radio spectral index $\alpha$ such that $S_\nu\,\propto\,\nu^\alpha$, where $S_\nu$ is the flux density at frequency $\nu$. $R_{200}$ denotes the radius within which the average density is 200 times the critical density of the Universe. Unless otherwise noted, we give all synthesised beams in terms of FWHM values, and redshifts are taken from NED\footnote{The NASA/IPAC Extragalactic Database (NED) is operated by the Jet Propulsion Laboratory, California Institute of Technology, under contract with the National Aeronautics and Space Administration.} \citep{NED} or VizieR \citep{vizier}.

\section{Cluster Sample}\label{sec:sampleobs}
The MGCLS sample consists of 115 galaxy clusters spanning a declination range of $-85$ to 0 degrees. The targeted clusters form a heterogeneous sample, with no mass or redshift selection criteria applied, and consist of two groups: ``radio-selected'' and ``X-ray-selected''. The full list of MGCLS clusters is given in Table \ref{tab:sample}, presented at the end of this paper, where the listed right ascension (R.A.) and declination (Dec.) are that of the MeerKAT pointing. The median redshift of the sample is 0.14, with only four clusters at $z > 0.4$.

\longtab{
\begin{longtable}{lrrcccccccl}
\caption{\label{tab:sample}Observed cluster sample. Cols: (1) Cluster name, listed alphabetically: Radio-selected targets are indicated by their common name (top panel), X-ray-selected targets are indicated by their MCXC catalogue designation (bottom panel). See Section~\ref{sec:sampleobs} for details; (2--3) MeerKAT pointing coordinates: J2000 R.A. and Dec.; (4) Cluster redshift; (5--7) Product status: Astrometry (see Section~\ref{sec:issues-time_err}) --- corrected mapping (Fix) and positional offsets (Posn); Polarisation mapped (Pol.); (8) Image sigma-clipped standard deviation; (9) Data quality flag: 0 --- Good dynamic range, 1 --- Moderate dynamic range with some artefacts around bright sources, 2 --- Poor dynamic range with high contamination by bright source artefacts, 3 --- Poor dynamic range with ripples across image; (10) Presence of diffuse cluster emission, see Table \ref{tab:diffuse} for more details; (11) Alternate cluster name. }  \\           
\hline\hline  
(1) & \multicolumn{1}{c}{(2)} & \multicolumn{1}{c}{(3)} & (4) & (5) & (6) & (7) & (8) & (9) & (10) & (11) \\
Cluster Name & \multicolumn{1}{c}{R.A.$_{\rm J2000}$} & \multicolumn{1}{c}{Dec.$_{\rm J2000}$} & $z$ & \multicolumn{2}{c}{Astrometry} & Pol. & RMS & DQF & D.E. & Alternate name\\ 
     & \multicolumn{1}{c}{(deg)} & \multicolumn{1}{c}{(deg)} & & Fix & Posn & & \multicolumn{1}{c}{($\mu$Jy\,beam\per)}  \\
\hline 
\endfirsthead
\endfoot
\caption{continued.}\\
\hline\hline
(1) & \multicolumn{1}{c}{(2)} & \multicolumn{1}{c}{(3)} & (4) & (5) & (6) & (7) & (8) & (9) & (10) & (11) \\
Cluster Name & \multicolumn{1}{c}{R.A.$_{\rm J2000}$} & \multicolumn{1}{c}{Dec.$_{\rm J2000}$} & $z$ & \multicolumn{2}{c}{Astrometry} & Pol. & RMS & DQF & D.E. & Alternate name\\ 
     & \multicolumn{1}{c}{(deg)} & \multicolumn{1}{c}{(deg)} & & Fix & Posn & & \multicolumn{1}{c}{($\mu$Jy\,beam\per)}  \\
\hline     
\endhead
\hline
\endlastfoot
\multicolumn{2}{l}{\textit{\qquad Radio-selected sample}} &\\
Abell 13            &   3.3842 & $-$19.5010 & 0.094 & -            & $\checkmark$ & -            & 3.5 & 1 & $\checkmark$ & MCXC J0013.6$-$1930 \\
Abell 22            &   5.1608 & $-$25.7220 & 0.142 & $\checkmark$ & $\checkmark$ & $\checkmark$ & 2.9 & 0 & $\checkmark$ & MCXC J0020.7$-$2542 \\
Abell 33            &   6.7792 & $-$19.5067 & 0.280 & -            & $\checkmark$ & -            & 5.7 & 1 & -           &     \\
Abell 85            &  10.4529 &  $-$9.3180 & 0.055 & $\checkmark$ & $\checkmark$ & $\checkmark$ & 3.3 & 1 & $\checkmark$ & MCXC J0041.8$-$0918\\
Abell 133           &  15.6879 & $-$21.8800 & 0.057 & $\checkmark$ & $\checkmark$ & $\checkmark$ & 6.7 & 1 & -            & MCXC J0102.7$-$2152  \\
Abell 168           &  18.7908 &     0.2475 & 0.045 & $\checkmark$ & $\checkmark$ & $\checkmark$ & 3.6 & 2 & $\checkmark$ & MCXC J0115.2$+$0019 \\
Abell 194           &  21.4458 &  $-$1.3731 & 0.018 & $\checkmark$ & $\checkmark$ & $\checkmark$ & 5.7 & 1 & -            & MCXC J0125.6$-$0124 \\
Abell 209           &  22.9896 & $-$13.5764 & 0.206 & $\checkmark$ & $\checkmark$ & $\checkmark$ & 3.6 & 1 & $\checkmark$  & MCXC J0131.8$-$1336 \\
Abell 370           &  39.9604 &  $-$1.5856 & 0.375 & $\checkmark$ & $\checkmark$ & -            & 6.9 & 2 & $\checkmark$  & ZwCl 0237.2$-$0146 \\
Abell 521           &  73.5358 & $-$10.2442 & 0.253 & -            & $\checkmark$ & -            & 3.4 & 0 & $\checkmark$ & MCXC J0454.1$-$1014 \\
Abell 545           &  83.1017 & $-$11.5431 & 0.154 & $\checkmark$ & $\checkmark$ & -            & 3.1 & 1 & $\checkmark$ & MCXC J0532.3$-$1131 \\
Abell 548           &  86.7571 & $-$25.6164 & 0.042 & $\checkmark$ & $\checkmark$ & $\checkmark$ & 2.8 & 1 & -             & \\
Abell 2485          & 342.1371 & $-$16.1064 & 0.247 & -            & $\checkmark$ & -            & 2.8 & 0 & -            & MCXC J2248.5$-$1606 \\
Abell 2597          & 351.3321 & $-$12.1244 & 0.085 & -            & $\checkmark$ & -            & 6.0 & 2 & -            & MCXC J2325.3$-$1207 \\
Abell 2645          & 355.3200 &  $-$9.0275 & 0.251 & -            & $\checkmark$ & -            & 4.3 & 2 & $\checkmark$ & MCXC J2341.2$-$0901 \\
Abell 2667          & 357.9196 & $-$26.0836 & 0.230 & -            & $\checkmark$ & -            & 2.7 & 0 & $\checkmark$ & MCXC J2351.6$-$2605 \\
Abell 2744          &   3.5671 & $-$30.3830 & 0.308 & $\checkmark$ & $\checkmark$ & $\checkmark$ & 2.9 & 0 & $\checkmark$  & MCXC J0014.3$-$3023 \\
Abell 2751          &   4.0580 & $-$31.3885 & 0.107 & $\checkmark$ & $\checkmark$ & $\checkmark$ & 2.6 & 0 & $\checkmark$  & MCXC J0016.3$-$3121 \\
Abell 2811          &  10.5368 & $-$28.5358 & 0.108 & -            & $\checkmark$ & -            & 2.6 & 0 & $\checkmark$ & MCXC J0042.1$-$2832 \\
Abell 2813          &  10.8517 & $-$20.6214 & 0.292 & -            & $\checkmark$ & -            & 3.4 & 2 & $\checkmark$ & MCXC J0043.4$-$2037 \\
Abell 2895          &  19.5463 & $-$26.9731 & 0.227 & -            & $\checkmark$ & -            & 3.0 & 1 & $\checkmark$ & MCXC J0118.1$-$2658 \\
Abell 3365          &  87.0500 & $-$21.9350 & 0.093 & -            & $\checkmark$ & -            & 2.8 & 0 & $\checkmark$ & \\
Abell 3376          &  90.4256 & $-$39.9851 & 0.046 & $\checkmark$ & $\checkmark$ & $\checkmark$ & 3.1 & 1 & $\checkmark$  & MCXC J0601.7$-$3959 \\
Abell 3558          & 201.9783 & $-$31.4922 & 0.048 & $\checkmark$ & $\checkmark$ & -            & 2.9 & 1 & $\checkmark$ & MCXC J1327.9$-$3130 \\
Abell 3562          & 202.7833 & $-$31.6731 & 0.049 & $\checkmark$ & $\checkmark$ & $\checkmark$ & 3.3 & 0 & $\checkmark$  & MCXC J1333.6$-$3139 \\
Abell 3667          & 303.1403 & $-$56.8406 & 0.056 & $\checkmark$ & $\checkmark$ & $\checkmark$ & 4.2 & 1 & $\checkmark$  & MCXC J2012.5$-$5649 \\
Abell 4038          & 356.8796 & $-$28.2028 & 0.028 & $\checkmark$ & $\checkmark$ & $\checkmark$ & 3.0 & 0 & $\checkmark$  & MCXC J2347.7$-$2808 \\
Abell S295          &  41.4000 & $-$53.0380 & 0.300 & $\checkmark$ & $\checkmark$ & $\checkmark$ & 2.3 & 0 & $\checkmark$  & PSZ1 G271.48$-$56.57 \\
Abell S1063         & 342.1813 & $-$44.5289 & 0.348 & -            & $\checkmark$ & -            & 2.6 & 0 & $\checkmark$ & MCXC J2248.7$-$4431 \\
Abell S1121         & 351.2844 & $-$41.2118 & 0.190 & -            & $\checkmark$ & -            & 5.4 & 2 & $\checkmark$ & PSZ2 G348.90$-$67.37 \\
Bullet$^\dagger$    & 104.6579 & $-$55.9500 & 0.296 & $\checkmark$ & $\checkmark$ & -            & 2.8 & 0 & $\checkmark$ & MCXC J0658.5$-$5556\\
El Gordo            &  15.7188 & $-$49.2495 & 0.870 & $\checkmark$ & $\checkmark$ & -            & 1.5 & 0 & $\checkmark$ & ACT-CL J0102$-$4915 \\
MACS J0025.4$-$1222 &   6.3724 & $-$12.3770 & 0.584 & $\checkmark$ & $\checkmark$ & -            & 3.7 & 1 & -            & MCXC J0025.4$-$1222  \\
MACS J0257.6$-$2209 &  44.4179 & $-$22.1628 & 0.322 & -            & $\checkmark$ & -            & 3.2 & 1 & $\checkmark$ & MCXC J0257.6$-$2209 \\
MACS J0417.5$-$1155 &  64.3942 & $-$11.9089 & 0.440 & $\checkmark$ & $\checkmark$ & -            & 2.9 & 0 & $\checkmark$ & MCXC J0417.5$-$1154 \\
PLCK G200.9$-$28.2  &  72.5871 &  $-$2.9493 & 0.220 & $\checkmark$ & $\checkmark$ & $\checkmark$ & 4.4 & 1 & $\checkmark$ & \\
RXC J0225.1$-$2928  &  36.3750 & $-$29.5000 & 0.060 & $\checkmark$ & $\checkmark$ & -            & 5.1 & 2 & -            & MCXC J0225.1$-$2928  \\
RXC J0510.7$-$0801  &  77.6846 &  $-$8.0200 & 0.220 & $\checkmark$ & $\checkmark$ & -            & 5.2 & 1 & $\checkmark$ & MCXC J0510.7$-$0801 \\
RXC J0520.7$-$1328  &  80.1967 & $-$13.5022 & 0.336 & $\checkmark$ & $\checkmark$ & -            & 7.7 & 2 & $\checkmark$ & PSZ1 G215.29$-$26.09 \\
RXC J1314.4$-$2515  & 198.5988 & $-$25.2558 & 0.249 & $\checkmark$ & $\checkmark$ & -            & 4.2 & 1 & $\checkmark$ & MCXC J1314.4$-$2515 \\
RXC J2351.0$-$1954  & 357.7704 & $-$19.9133 & 0.248 & -            & $\checkmark$ & -            & 3.1 & 1 & $\checkmark$ & \\
\multicolumn{2}{l}{\textit{\qquad X-ray-selected sample}}\\
J0014.3$-$6604      &   3.5767 & $-$66.0775 & 0.155 & $\checkmark$ & $\checkmark$ & -            & 2.5 & 0 & -             & Abell 2746 \\
J0027.3$-$5015      &   6.8388 & $-$50.2511 & 0.145 & $\checkmark$ & $\checkmark$ & -            & 2.6 & 0 & $\checkmark$ & Abell 2777 \\
J0051.1$-$4833      &  12.7967 & $-$48.5597 & 0.187 & $\checkmark$ & $\checkmark$ & -            & 2.6 & 0 & -             & Abell 2830 \\
J0108.5$-$4020      &  17.1383 & $-$40.3500 & 0.143 & $\checkmark$ & $\checkmark$ & -            & 2.6 & 0 & -             & Abell 2874 \\
J0117.8$-$5455      &  19.4604 & $-$54.9239 & 0.251 & $\checkmark$ & $\checkmark$ & -            & 2.4 & 0 & -             & RXC J0117.8$-$5455 \\
J0145.0$-$5300      &  26.2596 & $-$53.0139 & 0.118 & $\checkmark$ & $\checkmark$ & -            & 2.6 & 1 & $\checkmark$ & Abell 2941 \\
J0145.2$-$6033      &  26.3196 & $-$60.5650 & 0.184 & $\checkmark$ & $\checkmark$ & -            & 2.3 & 0 & $\checkmark$ & PSZ1 G291.34$-$55.32 \\
J0212.8$-$4707      &  33.2246 & $-$47.1328 & 0.115 & $\checkmark$ & $\checkmark$ & -            & 3.1 & 1 & -             & Abell 2988 \\
J0216.3$-$4816      &  34.0796 & $-$48.2731 & 0.163 & $\checkmark$ & $\checkmark$ & $\checkmark$ & 3.1 & 3 & $\checkmark$ & Abell 2998 \\
J0217.2$-$5244      &  34.3025 & $-$52.7469 & 0.343 & $\checkmark$ & $\checkmark$ & -            & 2.8 & 1 & $\checkmark$ & ACT-CL J0217$-$5245 \\
J0225.9$-$4154      &  36.4775 & $-$41.9097 & 0.220 & $\checkmark$ & $\checkmark$ & -            & 2.7 & 1 & $\checkmark$ & Abell 3017 \\
J0232.2$-$4420      &  38.0700 & $-$44.3475 & 0.284 & $\checkmark$ & $\checkmark$ & $\checkmark$ & 2.6 & 0 & $\checkmark$ & PSZ2 G259.98$-$63.43 \\
J0303.7$-$7752      &  45.9433 & $-$77.8692 & 0.274 & $\checkmark$ & $\checkmark$ & -            & 2.9 & 0 & $\checkmark$ & PSZ1 G294.68$-$37.01 \\
J0314.3$-$4525      &  48.5825 & $-$45.4242 & 0.073 & $\checkmark$ & $\checkmark$ & $\checkmark$ & 2.5 & 0 & $\checkmark$ & Abell 3104 \\
J0317.9$-$4414      &  49.4938 & $-$44.2389 & 0.075 & $\checkmark$ & $\checkmark$ & -            & 3.0 & 2 & -             & Abell 3112 \\
J0328.6$-$5542      &  52.1563 & $-$55.7128 & 0.086 & $\checkmark$ & $\checkmark$ & $\checkmark$ & 2.9 & 1 & -             & Abell 3126 \\
J0336.3$-$4037      &  54.0779 & $-$40.6222 & 0.062 & $\checkmark$ & $\checkmark$ & -            & 3.5 & 1 & -             & Abell 3140 \\
J0342.8$-$5338      &  55.7246 & $-$53.6353 & 0.060 & $\checkmark$ & $\checkmark$ & -            & 3.4 & 0 & $\checkmark$ & Abell 3158 \\
J0351.1$-$8212      &  57.7871 & $-$82.2167 & 0.061 & $\checkmark$ & $\checkmark$ & $\checkmark$ & 2.8 & 0 & $\checkmark$ & Abell S405 \\
J0352.4$-$7401      &  58.1229 & $-$74.0308 & 0.127 & $\checkmark$ & $\checkmark$ & $\checkmark$ & 2.6 & 0 & $\checkmark$ & Abell 3186 \\
J0406.7$-$7116      &  61.6908 & $-$71.2750 & 0.229 & $\checkmark$ & $\checkmark$ & -            & 3.0 & 1 & -             & \\
J0416.7$-$5525      &  64.1871 & $-$55.4189 & 0.365 & $\checkmark$ & $\checkmark$ & -            & 2.7 & 0 & -             & \\
J0431.4$-$6126      &  67.8504 & $-$61.4439 & 0.059 & $\checkmark$ & $\checkmark$ & $\checkmark$ & 4.5 & 1 & $\checkmark$ & Abell 3266 \\
J0449.9$-$4440      &  72.4800 & $-$44.6781 & 0.172 & $\checkmark$ & $\checkmark$ & -            & 2.6 & 0 & -             & Abell 3292 \\
J0510.2$-$4519      &  77.5575 & $-$45.3211 & 0.200 & $\checkmark$ & $\checkmark$ & -            & 3.0 & 0 & $\checkmark$ & Abell 3322 \\
J0516.6$-$5430      &  79.1583 & $-$54.5142 & 0.297 & $\checkmark$ & $\checkmark$ & $\checkmark$ & 3.1 & 1 & $\checkmark$ & Abell S520 \\
J0525.8$-$4715      &  81.4650 & $-$47.2506 & 0.191 & $\checkmark$ & $\checkmark$ & -            & 3.0 & 1 & -             & Abell 3343 \\
J0528.9$-$3927      &  82.2346 & $-$39.4628 & 0.284 & $\checkmark$ & $\checkmark$ & -            & 2.6 & 0 & $\checkmark$ & PSZ2 G244.37$-$32.15 \\
J0540.1$-$4050      &  85.0263 & $-$40.8422 & 0.036 & $\checkmark$ & $\checkmark$ & -            & 4.1 & 1 & -             & Abell S540 \\
J0540.1$-$4322      &  85.0417 & $-$43.3822 & 0.085 & $\checkmark$ & $\checkmark$ & $\checkmark$ & 3.4 & 1 & -             & Abell 3360 \\
J0542.8$-$4100      &  85.7117 & $-$41.0014 & 0.640 & $\checkmark$ & $\checkmark$ & -            & 2.4 & 0 & -             & CL J0542.8-4100 \\
J0543.4$-$4430      &  85.8517 & $-$44.5053 & 0.164 & $\checkmark$ & $\checkmark$ & -            & 3.6 & 1 & -             & \\ 
J0545.5$-$4756      &  86.3775 & $-$47.9406 & 0.130 & $\checkmark$ & $\checkmark$ & -            & 2.9 & 1 & -             & Abell 3363 \\
J0600.8$-$5835      &  90.2013 & $-$58.5872 & 0.037 & $\checkmark$ & $\checkmark$ & -            & 2.5 & 0 & -             & Abell S560 \\
J0607.0$-$4928      &  91.7558 & $-$49.4833 & 0.056 & $\checkmark$ & $\checkmark$ & $\checkmark$ & 2.8 & 1 & -             & Abell 3380 \\
J0610.5$-$4848      &  92.6333 & $-$48.8072 & 0.243 & $\checkmark$ & $\checkmark$ & -            & 2.8 & 0 & -             & \\
J0616.8$-$4748      &  94.2233 & $-$47.8050 & 0.116 & $\checkmark$ & $\checkmark$ & -            & 3.0 & 0  & -            & PSZ1 G255.64$-$25.30 \\
J0625.2$-$5521      &  96.3179 & $-$55.3517 & 0.121 & $\checkmark$ & $\checkmark$ & -            & 5.3 & 1 & -             &  \\
J0626.3$-$5341      &  96.5950 & $-$53.6956 & 0.051 & $\checkmark$ & $\checkmark$ & -            & 4.4 & 2 & -             & Abell 3391 \\
J0627.2$-$5428      &  96.8100 & $-$54.4700 & 0.051 & $\checkmark$ & $\checkmark$ & $\checkmark$ & 7.8 & 2 & $\checkmark$ & Abell 3395 \\
J0631.3$-$5610      &  97.8363 & $-$56.1722 & 0.054 & $\checkmark$ & $\checkmark$ & -            & 2.7 & 0 & $\checkmark$ &  \\
J0637.3$-$4828      &  99.3288 & $-$48.4783 & 0.203 & $\checkmark$ & $\checkmark$ & $\checkmark$ & 3.0 & 0 & $\checkmark$ & Abell 3399 \\
J0638.7$-$5358      &  99.6938 & $-$53.9717 & 0.233 & $\checkmark$ & $\checkmark$ & $\checkmark$ & 3.4 & 1 & $\checkmark$ & Abell S592 \\
J0645.4$-$5413      & 101.3721 & $-$54.2189 & 0.167 & $\checkmark$ & $\checkmark$ & -            & 3.4 & 1 & $\checkmark$ & Abell 3404 \\
J0658.5$-$5556$^\dagger$ & 104.6296 & $-$55.9469 & 0.296 & $\checkmark$ & $\checkmark$ & -       & 3.2 & 0 & $\checkmark$ & Bullet\\
J0712.0$-$6030      & 108.0225 & $-$60.5017 & 0.032 & $\checkmark$ & $\checkmark$ & -            & 2.7 & 1 & -             & \\
J0738.1$-$7506      & 114.5375 & $-$75.1067 & 0.111 & $\checkmark$ & $\checkmark$ & -            & 2.6 & 0 & -             & PSZ1 G287.05$-$23.21 \\
J0745.1$-$5404      & 116.2900 & $-$54.0789 & 0.074 & $\checkmark$ & $\checkmark$ & $\checkmark$ & 3.1 & 0 & $\checkmark$ & CIZA J0745.1$-$5404 \\
J0757.7$-$5315      & 119.4438 & $-$53.2636 & 0.043 & $\checkmark$ & $\checkmark$ & -            & 3.2 & 0 & -             & Abell S606 \\
J0812.5$-$5714      & 123.1263 & $-$57.2350 & 0.062 & $\checkmark$ & $\checkmark$ & $\checkmark$ & 2.9 & 0 & -             & PSZ2 G271.60$-$12.50 \\
J0820.9$-$5704      & 125.2483 & $-$57.0797 & 0.061 & $\checkmark$ & $\checkmark$ & $\checkmark$ & 2.9 & 2 & $\checkmark$ & PSZ1 G272.08$-$11.51 \\
J0943.4$-$7619      & 145.8542 & $-$76.3325 & 0.199 & $\checkmark$ & $\checkmark$ & -            & 4.4 & 2 & -             & CIZA J0943.4$-$7619 \\
J0948.6$-$8327      & 147.1642 & $-$83.4656 & 0.198 & $\checkmark$ & $\checkmark$ & $\checkmark$ & 3.1 & 0 & -             & \\
J1040.7$-$7047      & 160.1867 & $-$70.7969 & 0.061 & $\checkmark$ & $\checkmark$ & $\checkmark$ & 4.3 & 1 & -             & CIZA J1040.7$-$7047 \\
J1130.0$-$4213      & 172.5233 & $-$42.2297 & 0.155 & $\checkmark$ & $\checkmark$ & -            & 2.9 & 1 & $\checkmark$ & PSZ1 G287.22$+$18.13 \\
J1145.6$-$5420      & 176.4108 & $-$54.3414 & 0.155 & $\checkmark$ & $\checkmark$ & -            & 3.1 & 1 & -             & PSZ1 G293.32$+$07.33 \\
J1201.0$-$4623      & 180.2642 & $-$46.3906 & 0.118 & $\checkmark$ & $\checkmark$ & -            & 3.2 & 0 & -             & CIZA J1201.0$-$4623 \\
J1240.2$-$4825      & 190.0571 & $-$48.4328 & 0.152 & $\checkmark$ & $\checkmark$ & -            & 3.3 & 1 & -             & \\
J1248.7$-$4118      & 192.1996 & $-$41.3078 & 0.011 & $\checkmark$ & $\checkmark$ & -            & 7.1 & 2 & -             & Abell 3526 \\
J1358.9$-$4750      & 209.7371 & $-$47.8386 & 0.074 & $\checkmark$ & $\checkmark$ & $\checkmark$ & 3.6 & 1 & -             & CIZA J1358.9$-$4750 \\
J1410.4$-$4246      & 212.6188 & $-$42.7769 & 0.049 & $\checkmark$ & $\checkmark$ & -            & 3.6 & 1 & -             & CIZA J1410.4$-$4246 \\
J1423.7$-$5412      & 215.9304 & $-$54.2033 & 0.300 & $\checkmark$ & $\checkmark$ & -            & 3.5 & 1 & $\checkmark$ & CIZA J1423.7$-$5412 \\
J1518.3$-$4632      & 229.5950 & $-$46.5403 & 0.056 & $\checkmark$ & $\checkmark$ & $\checkmark$ & 5.5 & 1 & -             & CIZA J1518.3$-$4632 \\
J1535.1$-$4658      & 233.7879 & $-$46.9792 & 0.036 & $\checkmark$ & $\checkmark$ & -            & 4.4 & 3 & -             & CIZA J1535.1$-$4658 \\
J1539.5$-$8335      & 234.8913 & $-$83.5922 & 0.073 & $\checkmark$ & $\checkmark$ & $\checkmark$ & 2.7 & 0 & $\checkmark$ &  \\
J1601.7$-$7544      & 240.4446 & $-$75.7461 & 0.153 & $\checkmark$ & $\checkmark$ & $\checkmark$ & 3.7 & 1 & $\checkmark$ & PSZ2 G313.88$-$17.11 \\
J1645.4$-$7334      & 251.3592 & $-$73.5817 & 0.069 & $\checkmark$ & $\checkmark$ & $\checkmark$ & 4.8 & 2 & -            & PSZ2 G317.58$-$17.82 \\
J1653.0$-$5943      & 253.2533 & $-$59.7331 & 0.048 & $\checkmark$ & $\checkmark$ & $\checkmark$ & 3.6 & 3 & -            & PSZ1 G329.36$-$09.88 \\
J1705.1$-$8210      & 256.2929 & $-$82.1739 & 0.074 & $\checkmark$ & $\checkmark$ & $\checkmark$ & 2.8 & 0 & -            & Abell S792 \\
J1840.6$-$7709      & 280.1550 & $-$77.1556 & 0.019 & $\checkmark$ & $\checkmark$ & -            & 19.0 & 1 & $\checkmark$ & \\
J2023.4$-$5535      & 305.8500 & $-$55.5917 & 0.232 & $\checkmark$ & $\checkmark$ & $\checkmark$ & 2.7 &  1 & $\checkmark$ & PSZ1 G342.33$-$34.92 \\
J2104.9$-$8243      & 316.2446 & $-$82.7228 & 0.097 & $\checkmark$ & $\checkmark$ & $\checkmark$ & 2.6 & 0 & -            & Abell 3728 \\
J2222.2$-$5235      & 335.5579 & $-$52.5869 & 0.174 & $\checkmark$ & $\checkmark$ & -            & 4.0 & 1 & -            & Abell 3870\\
J2319.2$-$6750      & 349.8000 & $-$67.8400 & 0.029 & $\checkmark$ & $\checkmark$ & $\checkmark$ & 3.5 & 1 & -            & Abell 3990 \\
J2340.1$-$8510      & 355.0429 & $-$85.1783 & 0.193 & $\checkmark$ & $\checkmark$ & $\checkmark$ & 3.0 & 1 & -            & Abell 4023 \\
\hline
\end{longtable}
\tablefoot{$^\dagger$ Observed as part of the X-ray-selected sample; data products can be found under the MCXC designation, J0658.5$-$5556.}
}

\subsection{Radio-selected sub-sample}
The radio-selected sub-sample consists of 41 southern targets which have been previously searched for diffuse cluster radio emission by other studies. Targets were selected from published radio studies or reviews, namely \citet{1999NewA....4..141G, Feretti2012, Lindner2014, Kale2015EGRHS, Shakouri2016ARDES, 2017MNRAS.470.3465B, 2017MNRAS.467..936G, 2017ApJ...841...71G, 2017MNRAS.472..940K, 2017MNRAS.464.2752P}, and \citet{Golovich2019}, and include both systems with and without previous diffuse emission detections. These previous radio studies were restricted to high mass systems, $M_{500} \gtrsim 6 \times 10^{14}$\,M$_\odot$,  derived from X-ray or Sunyaev-Zel'dovich effect \citep{1972CoASP...4..173S} data. Thus, the radio-selected sub-sample contains only high mass clusters. It covers a redshift range of $0.018 < z < 0.87$, with median $z = 0.22$.

Targeting systems of this nature ensured a high scientific return in terms of diffuse emission studies. However, due to the selection, this sub-sample is strongly biased towards clusters with radio halos and relics. The radio-selected clusters are listed in the first panel of Table \ref{tab:sample}, using their {common} names. Where available, an alternate name is provided in the final column of the table. In cases where multiple alternate names exist, the Meta-Catalogue of X-ray-detected Clusters \citep[MCXC catalogue,][]{MCXC} designation, if available, is given.

\subsection{X-ray-selected sub-sample}
The X-ray-selected sub-sample, making up 64\% of the MGCLS, was selected from the MCXC catalogue, in order to create a sample with no direct prior biases towards or against cluster radio properties. From the list of clusters in the MCXC catalogue which were south of $-39^\circ$, we selected MGCLS targets as needed to fill gaps in MeerKAT's observing schedule. 

The X-ray-selected clusters, which cover a redshift range of $0.011 < z < 0.640$ with a median of $z = 0.13$, are listed in the second panel of Table \ref{tab:sample} using their MCXC catalogue designations. Where relevant, common alternate names are also listed. {The X-ray-selected sample covers a luminosity range of $L_{\rm X} \sim (0.1 - 30) \times 10^{44}$\,erg\,s\per, with $\sim$\,60\% of clusters in the range $10^{44}$--$10^{45}$\,erg\,s\per.}

\section{Observations and Data Reduction}\label{sec:initialproc}
\subsection{Observations}
The MGCLS observations were carried out between June 24, 2018 and June 16, 2019 using the full MeerKAT array, with a minimum of 59 antennas per observation. The MGCLS clusters were observed using MeerKAT's L-band receiver (with nominal radio frequency band of 900--1670\,MHz) in the 4k correlator mode (4096 channels across the digitised band of 856--1712\,MHz) with 8~second integrations. 

Data consists of all combinations of the two orthogonal linearly
polarised feeds. 
Each dataset contains observations of the flux
density/delay/bandpass calibrators PKS~B1934$-$638 and/or PMN~J0408$-$6545.  
These were observed for 10~minutes every hour with the remaining time
cycling between the target cluster (10~min) and a nearby astrometric\kk{/phase}
calibrator (1~min). 
These observations spanned 8--12~hours, cycling between the target
cluster and the various calibrator sources, and typically consisted of
$\sim$\,5.5--9.5~hours on source integration, sometimes divided into
multiple sessions.
These were
{scheduled as ``fillers'' during observing schedule gaps}.

\subsection{Initial processing}
All datasets were calibrated and imaged with a simple procedure, described in \citet{DEEP2}, which also verified
 the data quality. 
All calibration and imaging used the \textsc{Obit}
package\footnote{\url{http://www.cv.nrao.edu/~bcotton/Obit.html}}
\citep{Obit}. 

\subsubsection{Calibration and editing}\label{sec:calandedit}
Various processes as described in \cite{DEEP2} were used to identify
data affected by interference and/or equipment failures {which
were then edited out}, typically resulting in $\sim$\,50\% of the frequency/time samples being removed. 
The {remaining} data were calibrated in group delay, bandpass,
and amplitude and  phase. 
The reference antenna was picked on the basis of the best 
{signal-to-noise ratio (SNR)} in the bandpass solutions.  
Our flux density scale is based on the spectrum of PKS~B1934$-$638
{\citep{Reynolds94}}:   
\begin{equation}
\begin{split}
  \log\left(S\right) = & -30.7667 + 26.4908 \log\left(\nu\right) -7.0977 \log\left(\nu\right)^2  \\
   & \qquad + 0.6053 \log\left(\nu\right)^3,
\end{split}
\end{equation}
where $S$ is the flux density in Jy and $\nu$ is the frequency in MHz. The uncertainty in the flux-density scale is estimated to be $\sim$\,5\%.

Small errors in both time and frequency tagging were discovered after the observations had started and were subsequently corrected.  These errors have a small effect on the images, and are more fully described in Section~\ref{sec:issues-time_err}.  The majority of images were made after the errors were fixed.

\subsubsection{\kk{Stokes-I} imaging}\label{sec:stokesi}
We created the maps using the \textsc{Obit} wide-band, wide-field imager \texttt{MFImage}. 
\texttt{MFImage} uses facets to correct for the curvature of the sky, and
multiple frequency bins, which are imaged independently and
{deconvolved} jointly, to allow for the
antenna gain{s} and {a} sky brightness distribution
that vary with frequency.  
A frequency dependent taper was used to obtain a resolution which remained approximately constant over our $\sim$2:1 range in frequency. 
\texttt{MFImage} is described in more detail in \cite{SourceSize}. 

The sky within a radius of 0.8$^\circ$ to 1$^\circ$ of the pointing centre was fully imaged, with outlying facets added to cover sources from the Sydney University Molonglo Sky Survey \citep[SUMSS,][]{SUMSS}
843 MHz catalogue \citep{SUMSScat} {brighter than 5 mJy} 
within 1.5$^\circ$. 
Two iterations of {phase-only self-calibration, with a} 30 second solution {interval}, were used.  Amplitude and phase self-calibration were added 
if the image contained a pixel with a brightness in excess of 0.3\,Jy\,beam\per. 
For Stokes~I \kk{imaging} we used \kk{a maximum of} 30,000 components, a loop gain of 0.1, and fields were typically CLEANed to a depth of \editout{10}\kk{$\sim$\,5}0\,$\mu$Jy\,beam\per. 
No direction dependent corrections were applied; such corrections may be useful for followup studies of individual fields, but do not affect the science results presented here.  

Robust weighting ($-1.5$ in \textsc{AIPS/Obit} usage) was used to down-weight the
very densely sampled inner portion of the \textit{uv}-plane. 
The resulting {FWHM} resolution was in the range 7.5--8.0{\arcsec}.  
{We made images consisting of} 14
{frequency bins, each with a 5\% fractional ($\Delta \nu/\nu$)
  bandpass.} 
When the imaging {was}
complete, a spectrum {was}
fitted in each pixel of the resulting cube.  
Off-source {noise levels (RMS)} in images
which were not dynamic range limited {ranged
over} $\sim$3--5\,$\mu$Jy\,beam$^{-1}$. 
\kk{This is close to the expected thermal noise, with RMS confusion expected to be on the order of 1\,$\mu$Jy\,beam$^{-1}$ \citep[][]{DEEP2}. The local RMS noise varies over the field of view due to contributions from (multiple) strong sources, and is a strong function of the target pointing.}
{Primary beam corrections were only applied in the ``Enhanced
  products'' (see Section~\ref{sec:prod-enhanced}).}  

\subsubsection{Reprocessing with polarimetry}
Changes from the standard procedure described in Section \ref{sec:calandedit} were needed for polarisation calibration.  {The} MGCLS observations did not contain
observations of a polarised calibrator to calibrate the polarisation
response of the array. 
However, each MeerKAT observing session is begun with a calibration using noise signals injected into each antenna which can
be used to calibrate the bulk of the phase and delay difference
between the two recorded orthogonal linear polarisations. 
The remainder of the signal path is stable enough that, after this
initial calibration, appropriate polarisation calibration 
is possible using calibration parameters derived from other,
  properly polarisation-calibrated, data. 
This procedure is discussed in detail in \cite{MKPoln}, however we outline the basic steps here for clarity. 

\lr{Prior to any calibration derived from the data, the initial calibration of the ``X'' and ``Y'' linear feeds from the injected noise signals was performed; this removes most of the phase and delay difference between the X and Y systems.  The remainder is sufficiently stable that a ``standard'' calibration using the same reference antenna corrects it.  In the parallel hand calibration the two bandpass calibrators, PKS~B1934$-$638 and PMN~J0408$-$6545, are sufficiently weakly polarised that they can be considered unpolarised.  After bandpass calibration subsequent gain calibration solved only for Stokes I terms to avoid disturbing the relative X/Y gain ratio.}
   
\lr{Following the parallel-hand calibration, our polarisation calibration procedure was as follows. We required the selected calibration reference antenna to have a set of averaged polarisation calibrations derived from other MeerKAT datasets which had adequate polarisation calibration (including a polarised calibrator).  This ``standard'' set of calibration tables was used to complete the X-Y phase calibration and to correct for on-axis instrumental polarisation via the feed ellipticity and orientation}\footnote{These standard sets of calibration tables are not part of the MGCLS, but can be provided on reasonable request.}. \lr{Since the antennas are equipped with linear feeds, the fundamental reference for the polarisation angle is the nominal orientation of the feeds.} 
   
\lr{The standard polarisation calibration source 3C286 was used to verify and make final corrections to the polarisation calibration.  This source has a polarisation angle of $-$33$^{\circ}$ and a rotation measure (RM) equal to zero \citep{2013ApJS..206...16P}.  A correction of several degrees in polarisation angle and about 1 rad m$^{-2}$ in RM is needed to reproduce the assumed polarisation.  These corrections are stable over several years and have been applied to all polarisation corrected data.}

\editout{Our polarisation calibration procedure was as follows. 
First, the initial calibration of the ``X'' and ``Y'' linear feeds
from the injected noise signals was performed. 
The two bandpass calibrators
PKS~B1934$-$638 and PMN~J0408$-$6545 are sufficiently weakly polarised
that they can be considered unpolarised.  
After bandpass calibration, we solved for Stokes I gains only,
 to avoid disturbing the ratio of the X and Y gains. 
We selected a reference antenna such that it had a set of averaged
polarisation calibrations derived from other datasets which
had adequate polarisation calibration using 3C286. Footnote:(These data, not part of the MGCLS, can be provided on reasonable request).
This polarisation calibration was applied to the data after the parallel-hand
calibration.   }  

A selected subset of the clusters were re-calibrated and
imaged in \texttt{MFImage} to produce Stokes I, Q, U, and V images; these
clusters are indicated in the ``Poln'' column in Table
\ref{tab:sample}.  
Due to the lack of internal polarisation calibration, the Stokes V
images are not sufficient to detect weakly 
{circular polarised sources, but can work as an overall
check} of the quality of the calibration. 
{Strongly circularly or linearly polarised sources ($>$1\%) should be easily
  detectable.} 

\subsubsection{Polarisation imaging}
The imaging in full polarisation was similar to the initial imaging in Stokes~I \kk{(Section~\ref{sec:stokesi})},
but wider and deeper. 
We used the same 5\% frequency bins and the total bandwidth used by the Stokes~I imaging, which allows the recovery of RMs up to
{$\pm$}\,100\,rad\,m$^{-2}$ at full sensitivity, {with decreasing sensitivity beyond this range}. 
The field of view fully imaged has a radius of 1.2$^\circ$. 
In Stokes I, we cleaned to a depth of $\sim$\,80\,$\mu$Jy\,beam$^{-1}$, using up to 500,000 components. Stokes Q and U were CLEANed to a depth
of $\sim$\,30\,$\mu$Jy\,beam\per\ with up to 50,000 components. 
Off-source {noise} values in images which were not
dynamic range limited were $\sim$3\,$\mu$Jy\,beam\per.

\section{MGCLS Data Products\label{sec:products}}
The first MGCLS data release (DR1) is made public with this paper,\footnote{When using DR1 products, this paper should be cited, and the MeerKAT telescope acknowledgement included. See \mgclsdoi\ for details.} which consists of the MGCLS visibilities, the basic
data products (described in Section~\ref{sec:prod-raw}), and a set of enhanced products (described in Section~\ref{sec:prod-enhanced}). 
All DR1 legacy products are available through a DOI\footnote{\mgclsdoi}, and the raw visibilities are accessible through SARAO's Archive Server\footnote{\url{https://archive.sarao.ac.za/}} with project ID ``SSV$-$20180624$-$FC$-$01''. We highlight the primary capabilities of the MGCLS data products in Section~\ref{sec:capabilities}. For issues relating to the scientific usability of the various products, see Section~\ref{sec:issues}.

\subsection{Basic products}\label{sec:prod-raw}
The basic MGCLS product consists of
the standard \texttt{MFImage} output, the structure and description of which are given in
\cite{MFImage}. 
The basic image product is a cube consisting of 16 planes:
\begin{description}
\item 1) the brightness at the reference frequency (typically 1.28\,GHz, although there are slight variations depending on the observation), from a pixel-by-pixel least-squares fit to the brightness, $I$, in each frequency channel; 
\item 2) the spectral index, $\alpha_{908}^{1656}$, from the above fit, or a default value of $-0.6$, as described below;
\item 3--16) images in frequency channels centred at 908, 952, 996, 1044, 1093, 1145, 1200, 1258, 1318, 1382, 1448, 1482, 1594, and 1656 MHz; the 1200 and 1258\,MHz channels are totally blanked for radio frequency interference (RFI).
\end{description}
Note that none of these have primary beam corrections, thus
  the brightness values and spectral index estimates are biased by the frequency-variable primary
  beam shape, and are not suitable for quantitative scientific use. 
These basic products are useful, however,  for full-field
visual searches and source-finding. Images in Stokes Q, U, and V are provided where available.

\subsection{Enhanced products}\label{sec:prod-enhanced}

\subsubsection{Primary beam-corrected image/spectral index cubes}\label{sec:prod-pbcubes}

The basic images were corrected for the primary beam at each frequency, as described in \cite{DEEP2},   
 both at the full resolution of the
image, typically 7.5--8{\arcsec}, and at a convolved 15{\arcsec} resolution
to help recover low surface brightness features. The primary beam-corrected images show the inner 1.2$^\circ\,\times\,1.2^\circ$ portion of the MGCLS pointing, as primary beam corrections are unreliable beyond this region. 
Stokes Q, U, and V cubes are provided where available.

The final enhanced image data products are five-plane cubes (referred to as the \textit{5pln} cubes in the following) in which the first plane is the brightness at the reference
frequency, and the second is the spectral index, $\alpha_{908}^{1656}$, both determined by a least-squares fit to $\log(I)$ vs. $\log (\nu)$ at each pixel.
The third plane is the brightness uncertainty estimate, 
fourth is the spectral index uncertainty, 
and fifth is the $\chi ^2$ of the least-squares fit.  
Uncertainty estimates are only the statistical noise component and do not
include calibration or other systematic effects. As described in more detail below (see Section \ref{sec:issues-spix}), a default value of $-0.6$ is given for the spectral index when the SNR is too low for an accurate fit.

\subsubsection{Primary beam-corrected frequency cubes}\label{sec:prod-freqcubes}
We also provide primary beam-corrected frequency cubes \kk{at full- and 15\arcsec-resolutions}. These cubes consist of the 12 non-blanked
frequency planes with centre frequencies as listed in Section \ref{sec:prod-raw}. 
To account for the unreliability of primary beam corrections far from the pointing centre, 
pixels are blanked as for the \textit{5pln} cubes discussed above. 
Stokes Q, U, and V cubes are provided when available.

\subsection{Primary use cases}\label{sec:capabilities}
The MGCLS legacy products described in Section~\ref{sec:products} provide powerful datasets for a range of scientific enquiry. Here we highlight the main use cases for the MGCLS data.

\subsubsection{Sensitivity to a range of scales} \label{sec:uc-scales}
\begin{figure*}
    \centering
    \includegraphics[width=0.98\textwidth]{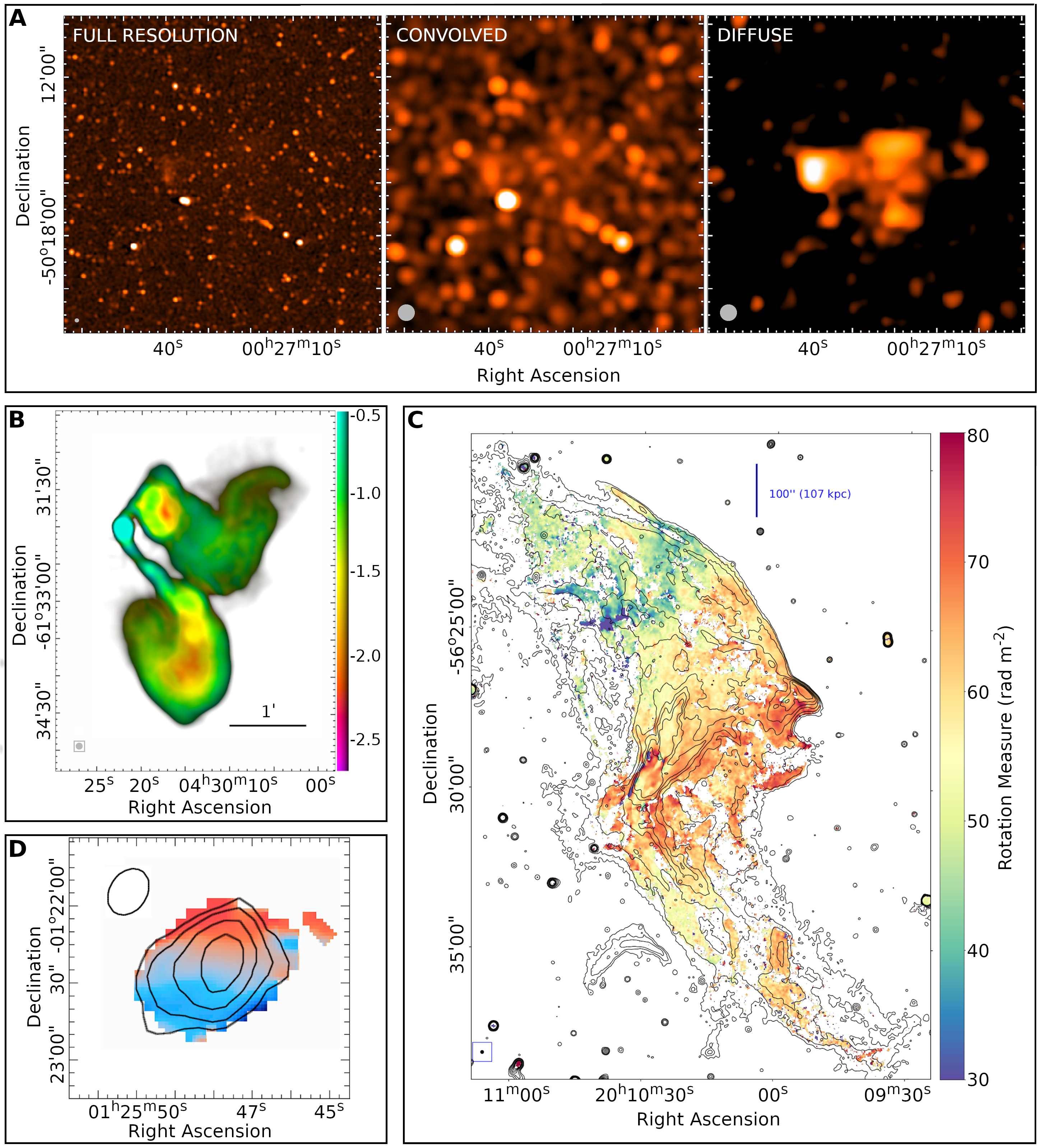}
    \caption{Capabilities of the MGCLS data. \textbf{Panel~A:} Brightness cutouts from the MCXC~J0027.3$-$5015 field, showing MeerKAT's instantaneous sensitivity to a range of angular scales. \textit{From left to right:} full resolution (7.4\arcsec\,$\times$\,7.0\arcsec), convolved 25\arcsec\ resolution, and filtered ``diffuse emission'' at 25\arcsec\ resolution. The synthesised beam is shown in grey at lower left of each image. The colour scale is in square root scaling in each case, with a minimum and maximum brightness of $-10$ and 600\,$\mu$Jy\,beam\per\ \textit{(left)}, 6 and 400\,$\mu$Jy\,beam\per\ \textit{(middle)}, and 25 and 150\,$\mu$Jy\,beam\per\ \textit{(right)}, respectively. See Section~\ref{sec:uc-scales} for further details. 
    \textbf{Panel~B:} Example of an in-band spectral index map of a bent tailed source in the MCXC~J0431.4$-$6126 field, with total intensity coloured by spectral index. The host galaxy, 2MASX~J04302197$-$6132001, is coincident with the radio core (seen in cyan). The colour scale indicates spectral index, and the brightness gradient of the colourbar indicates intensity, with a burned out maximum of 10~mJy\,beam\per. Regions where the SNR was too low to determine a spectral index have been left white. The synthesised beam (7.1\arcsec\,$\times$\,6.7\arcsec) is shown by the grey ellipse at lower left. 
    \textbf{Panel~C:} Example of a rotation measure (RM) map of a complex MGCLS source in the Abell~3667 field. Contours are Stokes~I intensity with levels of (5, 10, 20, 40, 60, 80, 100)\,$\times\,\sigma$, where $\sigma\,=\,6.7\,\mu$Jy\,beam\per. To avoid including spurious RM values, pixels with Stokes~I intensity below 8$\sigma$ have been masked.
    \textbf{Panel~D:} \textup{H\,\textsc{\lowercase{i}}} velocity map of Minkowski's object in Abell~194, at a resolution of 19\arcsec\,$\times$\,15\arcsec (beam shown at top left). Contours show the integrated \textup{H\,\textsc{\lowercase{i}}} flux density at levels of (0.35, 0.7, 1.4, 2.5)~mJy\,beam\per. Colours indicate the relative velocity from $-22$\,\kms\ (blue) to +22\,\kms\ (red) from the central velocity of 5553\,\kms. }
    \label{fig:spixeg}
    \label{fig:confusion}
    \label{fig:A3667RM}
    \label{fig:HI}
\end{figure*}

The configuration of the MeerKAT array, with its dense 1\,km-diameter core of antennas and maximum 7.7\,km baseline, allows for exceptional instantaneous sensitivity to a wide range of angular scales. The full resolution maps have synthesised beam sizes of $\sim$7.5--8{\arcsec} and RMS image noise levels of $\sim$\,3--5\,$\mu$Jy\,beam\per, and are sensitive to extended structures up to tens of arcminutes in extent. An example of the central region of one of the MGCLS fields, MCXC~J0027.3$-$5015, is shown in panel~A of Figure~\ref{fig:confusion}. The left figure of panel~A shows the full resolution ({7.4\arcsec\,$\times$\,7.0\arcsec}) image, dominated by compact sources with faint extended structure at the centre. To increase the sensitivity to the larger scale structure, the typical procedure is to convolve to a lower resolution. The middle figure shows the convolved 25\arcsec-resolution map of the same patch of sky, which is badly ``confused'' due to blending of the compact sources, masking the underlying diffuse emission. 

To exploit MeerKAT's sensitivity to large scale structures, without the problem of source confusion, we filter out all small scale structure with the technique of \citet{2002PASP..114..427R} using a box size of 19\,pixels (23.75\arcsec), and convolve the resulting ``diffuse emission'' image to 25\arcsec. Using figure~3 in that paper, we can roughly quantify what percentage of the flux will be in the diffuse emission image as a function of the characteristic size of any structure. For a 60\arcsec\ structure, $\sim$82\% of the flux will be included, with higher percentages for structures of increasing sizes. Smaller scale features will be heavily suppressed, with only 5--10\% of the flux remaining at 15\arcsec, and $\sim$0\% at 8\arcsec. The result of this process is shown in the right figure of panel~A in Figure~\ref{fig:confusion}, where the structure of the diffuse emission is readily visible. The filtered  25\arcsec-resolution or ``diffuse emission'' maps referred to in the following sections are made using the above filtering technique. \editnew{Note that these filtered maps are not included in the legacy products.}

\subsubsection{In-band spectral index maps}\label{sec:uc-spix}
MeerKAT's wide 0.8\,GHz bandwidth at L-band allows for in-band spectral index studies, with primary beam-corrected spectral index and associated uncertainty maps being part of the legacy products. Panel~B of Figure~\ref{fig:spixeg} shows an example spectral index map for a MGCLS radio galaxy with diffuse lobes. As per the caveats discussed in Section~\ref{sec:issues-spix}, a reliable spectral index can only be fit for pixels with SNR $\gtrsim 10$. Spectral index uncertainty maps contain only the statistical uncertainty from the fit, with constrained spectral indices typically having per-pixel statistical uncertainties between 0.05 and 0.2. 

\subsubsection{Polarisation studies}\label{sec:uc-pol}
All of the MGCLS targets were observed in full polarisation, with 44 of the clusters being mapped in polarisation for DR1 (see Table~\ref{tab:sample} at the end of the paper for the full list). \kk{Allowing for the caveats mentioned in Section~\ref{sec:issues-pol}, t}he sensitivity of the MGCLS polarisation maps will allow for the detection and determination of RMs for a large population of radio sources. Such detections will allow statistical studies of cluster magnetic fields. The determination of RMs of extended sources at high spatial sensitivity will also allow a detailed study of magnetic field strengths and structure across various source morphologies (e.g., radio galaxies and relics). Panel~C of Figure~\ref{fig:A3667RM} shows an RM map for one such extended source in the Abell~3667 field. This map is discussed in more detail in Section~\ref{sec:de-pol}. \kk{}



\subsubsection{\textup{H\,\textsc{\lowercase{i}}} capabilities}\label{sec:uc-hi}
In addition to the continuum and polarimetric use cases, the MGCLS visibilities can also be used for \textup{H\,\textsc{\lowercase{i}}} studies. 
\kk{The MGCLS frequency resolution of $209\,{\rm kHz}$ corresponds to an \textup{H\,\textsc{\lowercase{i}}} velocity resolution of $44.1\,{\rm km}\,{\rm s}^{-1}$ (at $z\,=\,0$). The survey is therefore suitable to approximately resolve the 
velocity structure of galaxies with a rotational amplitude 
of $\gtrsim 100\,{\rm km}\,{\rm s}^{-1}$ (depending on their inclination), which can be seen as a rough threshold dividing dwarf galaxies from more massive objects \citep{lelli_evolution_2014}.}

\begin{figure}
   \centering
 \includegraphics[width=0.95\columnwidth]{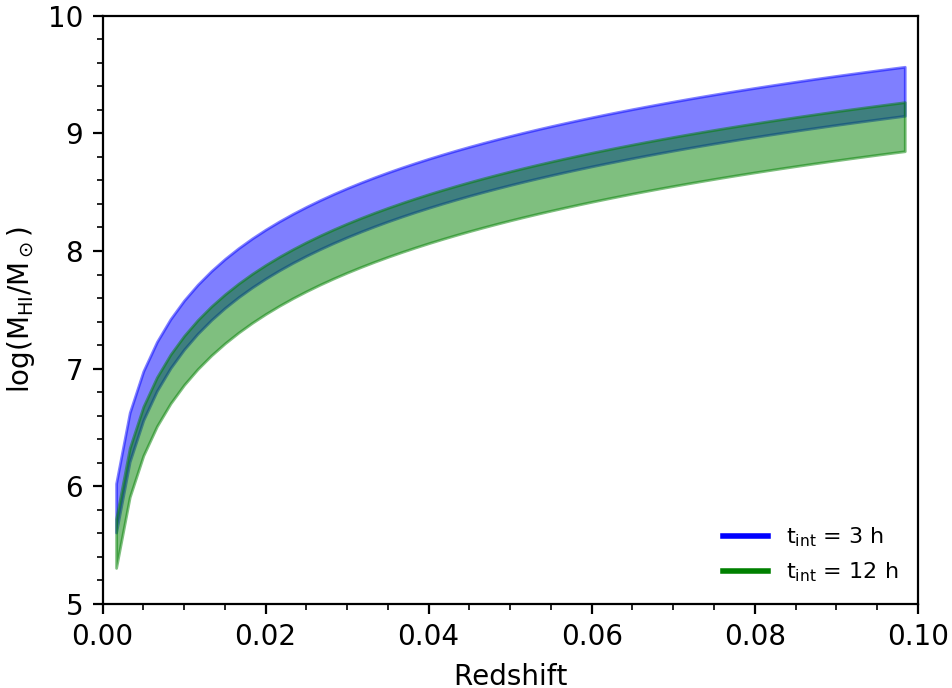}
 \includegraphics[width=0.99\columnwidth,clip=True,trim=0 0 0 40]{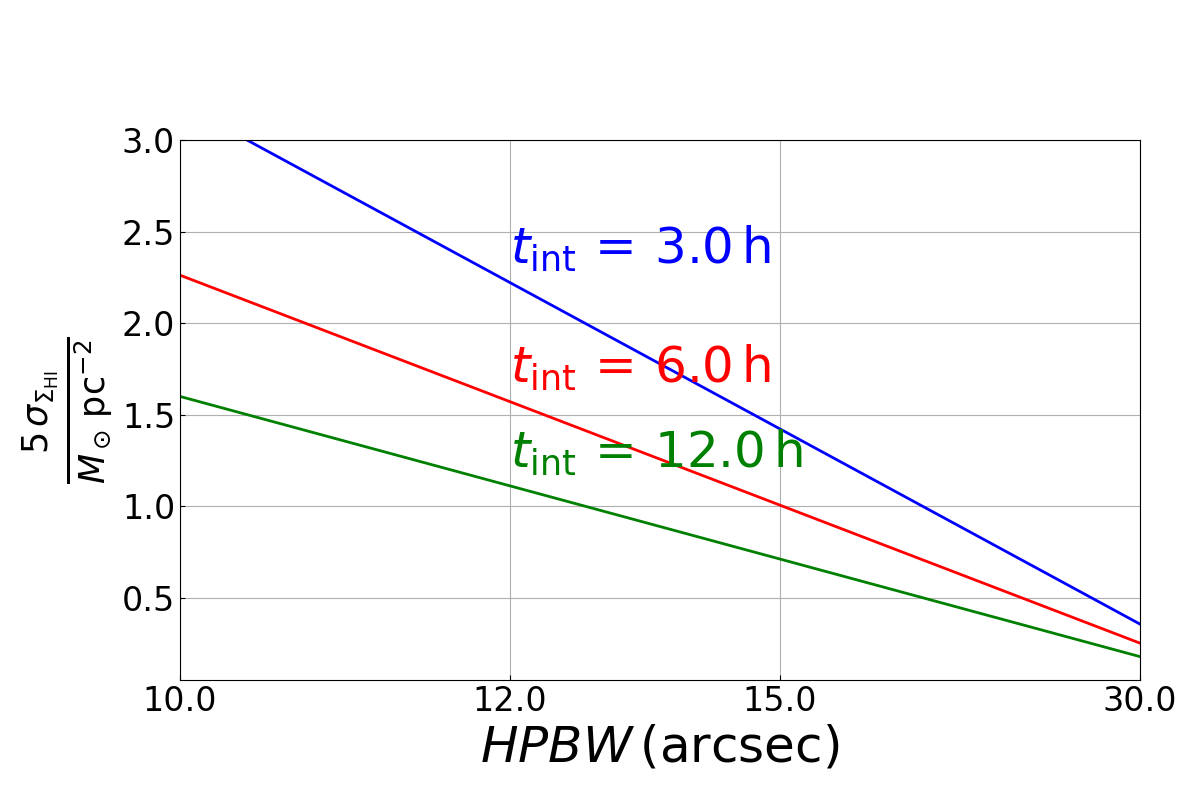}
    \caption{{\bf Top:} Logarithm of the total \textup{H\,\textsc{\lowercase{i}}} mass sensitivity as a function of redshift for integration times between 3 hours and 12 hours (not all MGCLS datasets have the same usable on-source time for \textup{H\,\textsc{\lowercase{i}}} analysis). The shaded areas indicate the \textup{H\,\textsc{\lowercase{i}}} mass limits of the MGCLS data assuming a 5$\sigma$ detection for a galaxy with line width ranging from 44\,\kms\ to 300\,\kms. {\bf Bottom:} Column-density sensitivity versus angular resolution at $z=0.03$ for different integration times present in the MGCLS. The horizontal axis scale is proportional to ${\rm HPBW}^{-2}$. }
    \label{fig:hi-sensitivity}
 \end{figure}
\kk{The top panel of Figure~\ref{fig:hi-sensitivity} shows the \textup{H\,\textsc{\lowercase{i}}} mass sensitivity as a function of redshift for an integration time of 3 and 12 hours (the range of usable on-source time for \textup{H\,\textsc{\lowercase{i}}} analysis in the MGCLS, with most clusters having 6--10 hours on-source), demonstrating that the MGCLS observations are able to detect galaxies below the ``knee'' of the \textup{H\,\textsc{\lowercase{i}}} mass function 
($\log(M^\ast_{\rm \textup{H\,\textsc{\lowercase{i}}}} /{\rm M}_\odot) = 9.94$, \citealt{2018MNRAS.477....2J})
out to $z\lesssim0.1$.}
\kk{Moreover, the angular resolution provided by the survey, shown in the bottom panel of Figure~\ref{fig:hi-sensitivity}, enables one to resolve the structure of the \textup{H\,\textsc{\lowercase{i}}} in galaxies with a resolution between 
10\arcsec\ and 30\arcsec, corresponding to a spatial resolution of 
$\sim4$--$20\,{\rm kpc}$ at redshifts of $0.02 < z < 0.1$. This means that for clusters in this range, larger galaxies will be spatially resolved and we can conduct studies of resolved \textup{H\,\textsc{\lowercase{i}}} galaxies while simultaneously probing extended extragalactic structures in the wide field of view --- a key discriminating feature with respect to single-dish \textup{H\,\textsc{\lowercase{i}}} surveys of rich environments. }
\kk{Panel~D in Figure~\ref{fig:HI} shows an example of a resolved \textup{H\,\textsc{\lowercase{i}}} detection, covering two consecutive channels, in the Abell~194 field. Examples of \textup{H\,\textsc{\lowercase{i}}} science from the MGCLS datasets are given in Section~\ref{sec:hiscience}.}

\subsection{Data quality issues\label{sec:issues}}
There are a number of issues with data quality in the DR1 release that affect the visibilities and images. The images have been corrected for a number of these effects, as described below.  However, they still affect the visibilities, and impose limitations on the accuracy of the images. Users of the data products should take these into consideration when scientific analyses are performed. 

\subsubsection{Dynamic range}\label{sec:issues-DR}
Fields with very strong sources ($I >$ few 100s
mJy\,beam$^{-1}$) are typically limited by residual artefacts from the
brighter sources. 
This is especially true if there are several widely separated bright
sources, as the self-calibration cannot correct direction dependent
effects (DDEs).  
The DDEs which are thought to dominate are asymmetries in the antenna
pattern, pointing errors, and ionospheric refraction. 

\subsubsection{Flux density and spectral index}\label{sec:issues-spix}
\emph{Uncertainties in the primary beam pattern} affect both the derived brightnesses and spectral indices. Near the centre of each field, the total array pattern is very close to that of the individual antennas \citep{DEEP2}. However, application of this individual antenna pattern to
spectra of sources in the outer parts of cluster fields produces non-physical results. This is expected because the array power pattern is broadened by pointing errors in the individual antennas.
Derived brightnesses, and especially spectral indices, 
are therefore not reliable beyond a radius of 36\arcmin. 

We compared the flux densities of the MCGLS compact sources with those from other radio surveys, as discussed in Section~\ref{sec:fluxcheck}, and found them to agree to within about 6\%, with the MeerKAT flux densities being, on average, $6\% \pm 4\%$ lower than those of the other two catalogues.  No corrections to the flux densities have been made in either the catalogues or the maps. 

While MeerKAT's \emph{very wide fractional bandwidth} improves the sensitivity and gives us the ability to derive in-band spectral indices, it also creates uncertainties regarding the effective central frequency at each pixel. 
Simple averaging in frequency space would result in effective reference
frequencies which depend on the spectra of the sources in question and their
position in the field.  When we can adequately fit the spectrum, typically requiring a SNR\,$\geq 10$, the brightness is calculated correctly at the well-defined reference frequency of 1.28\,GHz.  
The spectral fit was accepted if the $\chi^2$ per degree of freedom was
less than 1.5 times the $\chi^2$ per degree of freedom based on assuming
the default value of $-0.6$; otherwise the default value is reported and the central frequency brightness is calculated with that spectral assumption. This results in an underestimate of the brightness by $\sim$\,6\% (10\%) for a true spectral index of $-1$ ($-2$). Typically, these are less than the random errors in these low brightness cases.

Since the flux density threshold for switching to the default spectral index depends on the local noise properties of each cluster image, no global value can be specified. However, inspection of the spectral indices as a function of flux density for a selected region reveals the appropriate local threshold below which the default spectral index has been assigned. Caution is still advised when examining regions of maps where spectral indices very close to the default value are encountered.

\subsubsection{Largest angular size}
MeerKAT has very good short baseline coverage, allowing recovery of
extended emission. 
However the minimum baseline length of 29\,m does restrict the maximum
size of structure which is properly imaged. 
This maximum size scales inversely with
frequency. 
Angular scales less than 10{\arcmin} should be fully recovered. 
Larger scales are less well imaged, especially at the higher
frequencies. 
This effect leads to negative holes around bright extended
structures and an artificial apparent steepening of the
spectrum. 
Very steep fitted spectra of extended emission should be treated with
caution.   

\subsubsection{Astrometry}\label{sec:issues-astrometry} \label{sec:issues-time_err}
A number of instrumental issues can affect the accuracy of the astrometry.  As detailed in this section, we correct the image astrometry in each field by matching the positions with those of the respective optical hosts. This results in a final accuracy better than $0.3\arcsec$ in the high SNR limit. However, the astrometric errors are still present in the visibility data, and users that reprocess raw data need to take these into consideration.

Data in the earliest observations suffered from a 2-second 
\emph{time offset} in the labeling of the data, and a half-channel \emph{frequency error}, which propagate into errors in the
$u,v,w$ coordinates. The column ``Fix'' in Table~\ref{tab:sample} indicates cluster fields
where this issue was corrected in the visibilities. Fields that were not fixed will contain rotation and scaling errors in the positions of sources that depend on position in the field.
These can be as large as 2{\arcsec} at the edge of the fields. 
 
\emph{Calibrator position errors} were present in the initial MeerKAT calibrator list, affecting several calibrators up to a level of several arcseconds. This results in an approximately constant position offset of sources in the affected cluster fields. Corrections were made in the images when incorrect calibrator positions were discovered, but the final corrections were made at the end, in any case, through optical cross-matching.

A \emph{low-accuracy \lr{delay model in the correlator}} in use during the observations can cause a similar, but more subtle, problem \lr{when calibrator and target are widely separated.  An additional bias is
possible when the calibrator does not sufficiently dominate  the
visibility intensities in the field.} The model had insufficient accuracy to
reliably transfer phases measured on the calibrator to those of the target, especially since many of the astrometric calibrators were 10$^\circ$ or more from the
target. This resulted in  approximately constant position offsets of up to
several arcseconds in individual fields \lr{which had to be corrected}.  

\emph{Astrometric corrections} were made for each field centre, removing or reducing the effects noted above. We matched compact radio components to a large number of background quasars, radio galaxies, and star forming galaxies in each field, using the optical/infrared catalogues from the Dark Energy Camera Legacy Survey \citep[DECaLS,][]{Dey_2019}. 
A flux density-weighted average correction was determined for every
cluster field, as indicated in column ``Posn'' in  Table~\ref{tab:sample}, and
appropriate corrections made to each corresponding image. 
Residual systematic errors should be under 0.1\arcsec\ for
clusters as a whole, although errors may be larger for individual sources.

Additional \emph{astrometric checks} were carried out after the above correction by cross-matching with
sources from the International Celestial Reference Frame (ICRF) catalogue \citep{2020A&A...644A.159C}. 
Since the ICRF constitutes the most accurately known set of
astronomical positions, they are an ideal check of the astrometry in
the MGCLS catalogue described in Section~\ref{subsec:compactcat}. 
There are eight ICRF sources in our cluster fields. They are bright (0.3 to 1.8~Jy), and the statistical uncertainty in their MeerKAT position
determinations are therefore small. 
They were chosen for the ICRF for being compact on milliarcsecond
scales, and generally do not have structure on MeerKAT's
$\sim$\,8{\arcsec} scale which might affect the position determination.  
We therefore compared our catalogue positions with the
ICRF3\footnote{\url{http://hpiers.obspm.fr/icrs-pc/newwww/icrf/index.php}}
ones for the eight sources that were included in our fields:
ICRF~J010645.1$-$403419,
J025612.8$-$213729, 
J031757.6$-$441417, 
J033413.6$-$400825,
J060031.4$-$393702, 
J062552.2$-$543850, 
J124557.6$-$412845,
and 
J133019.0$-$312259. 

We found that most of the MGCLS catalogue positions differed from the
ICRF3 ones by less than 1\arcsec\ in either coordinate. 
The single exception was ICRF~J025612.8$-$213729 (QSO~B0253$-$218) for
which the MGCLS catalogue position was $\sim$2.3\arcsec\ north of the
ICRF3 position. 
This source is an exception to the above generalization about lack of
structure, in that \citet{ReidKP1999} show it to be somewhat extended
in a N--S direction at 5\,GHz, with several components spread out over
$\sim\,10$\arcsec. 
This field, MACS~J0257.6$-$2209, is also one of the few for which there were uncorrected timing and frequency errors (column~5 in Table~\ref{tab:sample}, discussed earlier in this section) which will cause position errors far from the field centre. In addition, since the source is resolved, the MGCLS position at 1.28~GHz and $\sim$\,8\arcsec-resolution could differ from the ICRF3 position, which is that of the milliarcsecond core only at 5~GHz.

Ignoring this source, we find that for the remaining seven sources the difference between the MGCLS catalogue positions and the ICRF3 ones
were $-0.04\,\pm\,0.34$\,{\arcsec} in R.A. and $-0.02\,\pm\,0.15$\,{\arcsec} in Dec. 
This check gives us confidence that the positions in the MGCLS
catalogue are accurate, and adopt the uncertainty of 0.36\arcsec\ based on the ICRF3 comparisons.

\subsubsection{Polarisation}\label{sec:issues-pol}
The first plane of the Stokes~Q and U \textit{5pln} cubes (see Section~\ref{sec:prod-pbcubes}) provide a good indication
of where significant polarisation is present, but should not be used
quantitatively on their own.  
Each image was created by a noise-weighted sum of the frequency planes
in the full cube, which is strictly correct only when both the RM and the
spectral index are zero, and when no depolarisation is present. 
At any RM\,$\neq$\,0, the amplitude of Q and U will be reduced, reaching a
factor of two reduction at $|\rm RM| \sim 25$\,rad\,m$^{-2}$, depending on the
source spectral index and the noise in the different frequency
channels.  
The first plane of the Stokes~Q and U \textit{5pln} cubes should therefore be used quantitatively with caution.

Note that polarisation leakage affects the upper half of the band, with the residual polarisation leakage increasing with distance from the field centre within the half-power region of the beam \citep[de Villiers \& Cotton, priv. comm.;][]{2021ITAP...69.6333D}. The instrumental polarisation may reach up to 10\% in the upper part of the band, while it is typically less than 2\% at the lower frequencies. Users should evaluate how the leakage affects their particular science case.

\section{MGCLS Source Catalogues}\label{sec:srccats}
We have produced source catalogues for all fields in the MGCLS, based on the intensity plane of the full-resolution enhanced products described in Section~\ref{sec:prod-enhanced}. Here we detail our source finding method, as well as the various catalogues being provided with the legacy data products. Limitations to the accuracy of the results are discussed in Section~\ref{sec:issues}.  It is very important to note that these catalogues are not complete, and any statistical analyses must consider their limitations. In particular, the sensitivity depends on the distance from the respective field centres, and regions around bright sources are excluded, as discussed below.

\subsection{Source detection}

We used the Python Blob Detection and Source Finder \citep[\textsc{pybdsf},][]{PYBDSF} software to create individual source catalogues for all MGCLS fields, using the full resolution, primary beam-corrected data products. \textsc{pybdsf} searches for islands of emission and attempts to fit models consisting of one or more elliptical Gaussians to them. Gaussians are then grouped into sources, and there may be more than one source per island. Each source is given a code: `S' for single-Gaussian sources which are the only source on its island, `C' for single-Gaussian sources that share an island with other sources, and `M' for sources comprised of two or more Gaussian fits. We used the default 3\,$\sigma_{\rm img}$ island boundary threshold and 5\,$\sigma_{\rm img}$ source detection threshold, where $\sigma_{\rm img}$ is the local image RMS. As many images have variable image noise levels across the field, we allow \textsc{pybdsf} to calculate the 2D RMS map during the source finding. 

For sources in regions of high image noise, for example those near bright sources with strong sidelobes, the typical statistical uncertainty in peak source brightness is a factor of $\sim$\,2 larger than for sources elsewhere in the same field. Spurious source detections are common around very bright sources, with \textsc{pybdsf} sometimes cataloguing sidelobes as sources. To mitigate spurious detections in our DR1 catalogues, we excised all catalogue entries around bright sources. A source was considered bright if its peak brightness was higher than the bright source limit for that field, $I^{\rm bs}_{\rm lim}$. This limit is connected to the image dynamic range, such that
\begin{equation}
	I^{\rm bs}_{\rm lim} = 10^{-4} \times \frac{ I_{\rm max} }{\sqrt { \sigma_{\rm global} } }\,{\rm Jy\,beam^{-1}},
\end{equation}
where $I_{\rm max}$ is the maximum source brightness in the image, and $\sigma_{\rm global}$ is the median image RMS, both in units of Jy\,beam\per. The region around a bright source within which catalogue entries were excised, $r_{\rm cut}$, scales with the source brightness:
\begin{equation}
	r_{\rm cut} = 0.005 \times \left(1 + \log_2 \frac{I^{\rm bs}_{\rm peak}}{I^{\rm bs}_{\rm lim}}\right)\,{\rm deg},
\end{equation}
where both $I^{\rm bs}_{\rm lim}$ and $I^{\rm bs}_{\rm peak}$, the peak brightness of the bright source, are in Jy\,beam\per. A median of 2.6\% of nominally detected sources were removed per field through this process. 

\subsection{Compact source catalogue}\label{subsec:compactcat}

From our \textsc{pybdsf} results, we compiled a single MGCLS compact source catalogue from all fields, only including sources which could be fit with a single Gaussian component (source codes `S' or `C'), after the spurious source excision. The full DR1 catalogue\footnote{Available at \mgclsdoi.} contains $\sim$626,000 sources from the 115 cluster fields, with an excerpt shown in Table~\ref{table:ptsrcs}. The catalogue columns, described in the Table caption, contain standard radio source information including the integrated flux density, peak brightness, and source size, with catalogue source positions provided in decimal degrees. The source identifier (first column of the catalogue) uses an IAU classification, with the designation MKTCS~JHHMMSS.ss$\pm$DDMMSS.s, where the decimal positional information is truncated, rather than rounded. 

\begin{table*}
\caption{Excerpt of the MGCLS compact source catalogue at 1.28\,GHz. The full catalogue, which includes all cluster fields, is available online at \mgclsdoi. Cols: (1) MGCLS source ID using the IAU designation of the form MKTCS~JHHMMSS.ss$\pm$DDMMSS.s, where the decimals are truncated; (2--5) J2000 R.A. and Dec., and associated $1\,\sigma$ \kk{uncertainty}, respectively; (6--7) Total integrated Stokes~I flux density and associated $1\,\sigma$ uncertainty at the reference frequency, respectively; (8--9) Peak Stokes~I brightness and associated 1$\sigma$ uncertainty, respectively; (10--12) Source size: FWHM of the major and minor axes of the source, and source p.a.; (13) Cluster field of the source. \kk{All uncertainties are statistical only, and are determined from the images as per \citet{1997PASP..109..166C}.}}
\label{table:ptsrcs}      
\centering
\small
\begin{tabular}{ccccccccccccc} 
\hline\hline               
(1) & (2) & (3) & (4) & (5) & \multicolumn{1}{c}{(6)} & \multicolumn{1}{c}{(7)} & \multicolumn{1}{c}{(8)} & \multicolumn{1}{c}{(9)} & \multicolumn{1}{c}{(10)} & (11) & (12) & (13) \\
Src. Name & R.A.$_{\rm J2000}$ & Dec.$_{\rm J2000}$ & $\Delta$R.A. & $\Delta$Dec. & \multicolumn{1}{c}{$S^{\rm 1.28\,GHz}_{\rm tot}$} & \multicolumn{1}{c}{$\Delta S_{\rm tot}$} & \multicolumn{1}{c}{$I^{\rm 1.28\,GHz}_{\rm peak}$} & \multicolumn{1}{c}{$\Delta I_{\rm peak}$} & \multicolumn{1}{c}{$s_{\max}$} & $s_{\min}$ & $s_{\rm p.a.}$ & Field\\    
MKTCS & (deg) & (deg) & (deg) & (deg) & \multicolumn{1}{c}{(mJy)} & \multicolumn{1}{c}{(mJy)} & \multicolumn{1}{c}{(mJy/b)} & \multicolumn{1}{c}{(mJy/b)} & \multicolumn{1}{c}{(\arcsec)} & (\arcsec) & \multicolumn{1}{c}{($^\circ$)}\\
\hline                        
J001059.77$-$190940.3 & 2.7491 & $-$19.1612 & 0.0000 & 0.0000 & 0.695 & 0.020 & 0.669 & 0.011 & 7.6  & 7.5  & 6   & Abell\_13 \\
J001059.94$-$190654.9 & 2.7498 & $-$19.1153 & 0.0001 & 0.0002 & 0.113 & 0.028 & 0.081 & 0.013 & 9.8  & 7.8  & 21  & Abell\_13 \\
J001059.14$-$195204.2 & 2.7464 & $-$19.8679 & 0.0002 & 0.0003 & 0.106 & 0.027 & 0.056 & 0.010 & 14.2 & 7.3  & 143 & Abell\_13 \\
J001059.23$-$194540.7 & 2.7468 & $-$19.7613 & 0.0002 & 0.0003 & 0.077 & 0.024 & 0.045 & 0.010 & 11.1 & 8.4  & 153 & Abell\_13 \\
J001059.50$-$192405.3 & 2.7479 & $-$19.4015 & 0.0001 & 0.0001 & 0.050 & 0.012 & 0.060 & 0.007 & 8.0  & 5.8  & 179 & Abell\_13 \\
J002318.34$-$254121.6 & 5.8264 & $-$25.6894 & 0.0001 & 0.0002 & 0.053 & 0.016 & 0.061 & 0.010 & 7.4  & 6.5  & 10  & Abell\_22 \\
J002317.08$-$253627.2 & 5.8212 & $-$25.6076 & 0.0000 & 0.0000 & 0.939 & 0.021 & 0.827 & 0.011 & 8.3  & 7.6  & 36  & Abell\_22 \\
J002318.08$-$253621.5 & 5.8253 & $-$25.6060 & 0.0002 & 0.0001 & 0.073 & 0.021 & 0.067 & 0.011 & 8.5  & 7.1  & 62  & Abell\_22 \\
J002317.47$-$252111.6 & 5.8228 & $-$25.3532 & 0.0001 & 0.0001 & 0.242 & 0.030 & 0.198 & 0.015 & 8.6  & 7.8  & 143 & Abell\_22 \\
J002317.15$-$261532.9 & 5.8215 & $-$26.2592 & 0.0003 & 0.0001 & 0.573 & 0.095 & 0.156 & 0.021 & 19.6 & 10.3 & 80  & Abell\_22 \\
\hline                                   
\end{tabular}
\end{table*}

\subsubsection{Comparison with previous radio catalogues}\label{sec:fluxcheck}

To verify the MGCLS compact source flux densities, we compare them to those from other radio surveys. To cover all MGCLS pointings, we use catalogues from both the 1.4\,GHz NRAO VLA Sky Survey \citep[NVSS, north of $-40^\circ$ declination,][]{NVSS}  and the 843\,MHz SUMSS survey \citep[south of $-30^\circ$ declination,][]{SUMSS}. 
To scale the NVSS and SUMSS flux densities to the MGCLS reference frequency of 1.28\,GHz, we assume a power law $S_\nu \propto \nu^\alpha$ with a fiducial spectral index of $\alpha\,=\,-0.7$ \citep{2017A&A...602A...1S}.

To avoid incompleteness effects due to differences in sensitivity between the three surveys, we consider only sources with SNR\,$\geqslant$\,50 in all three surveys. This high limit also minimises effects from  additional faint MGCLS sources within the larger NVSS and SUMSS beams, as noted below. To cross-match MGCLS compact sources with their NVSS and SUMSS counterparts, we use a 5{\arcsec} radius. This radius is a compromise between maximizing the number of real counterparts and minimizing the number of spurious matches. By shifting the MGCLS sources by 1{\arcmin} and repeating the cross-matching, we determine that the expected percentage of spurious matches is 4.1\% and 3.8\% for NVSS and SUMSS, respectively. These have a negligible effect on the flux density comparisons. 

\begin{figure}
   \centering
   \includegraphics[width=0.9\columnwidth]{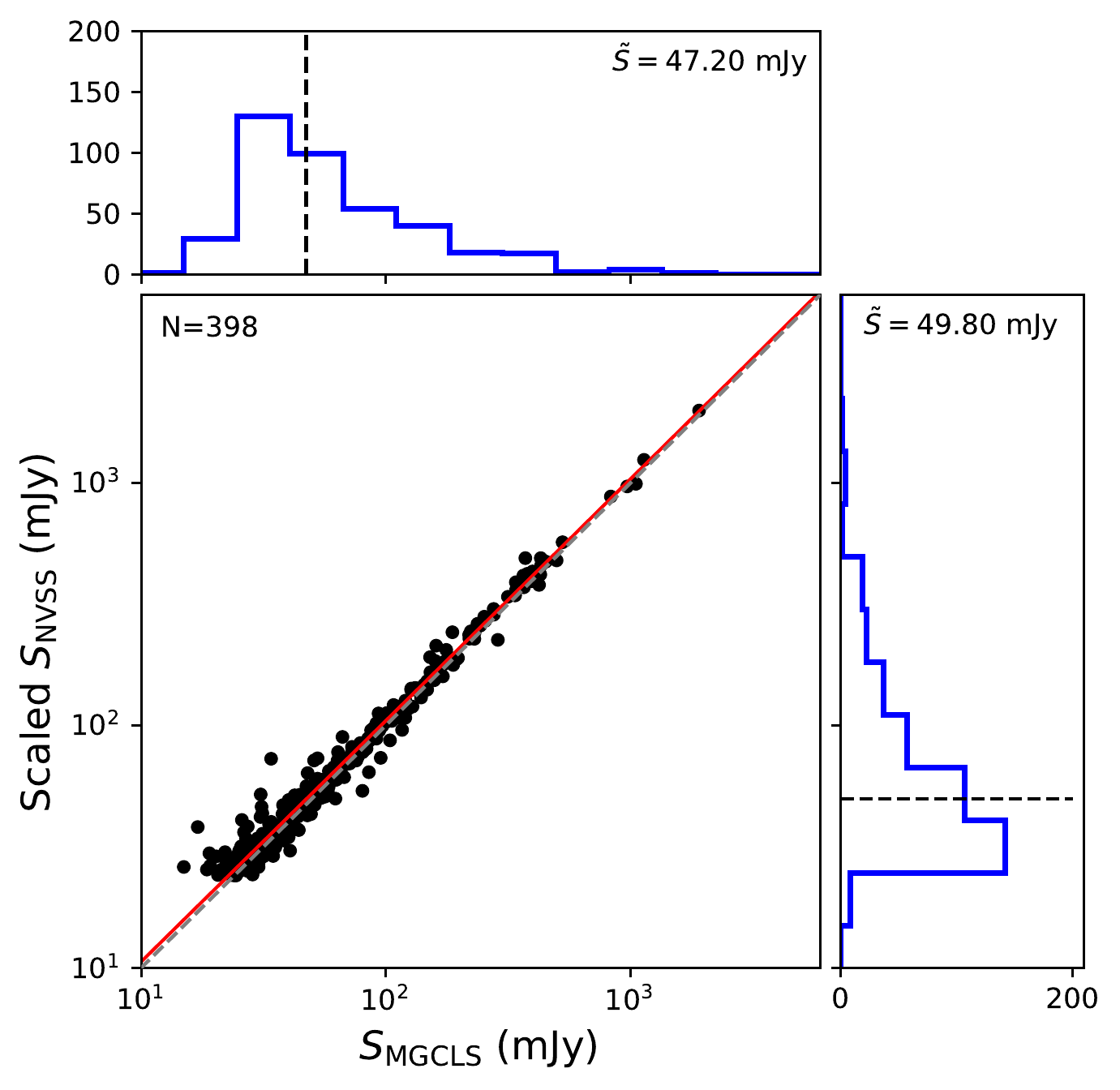}\\
   \includegraphics[width=0.9\columnwidth]{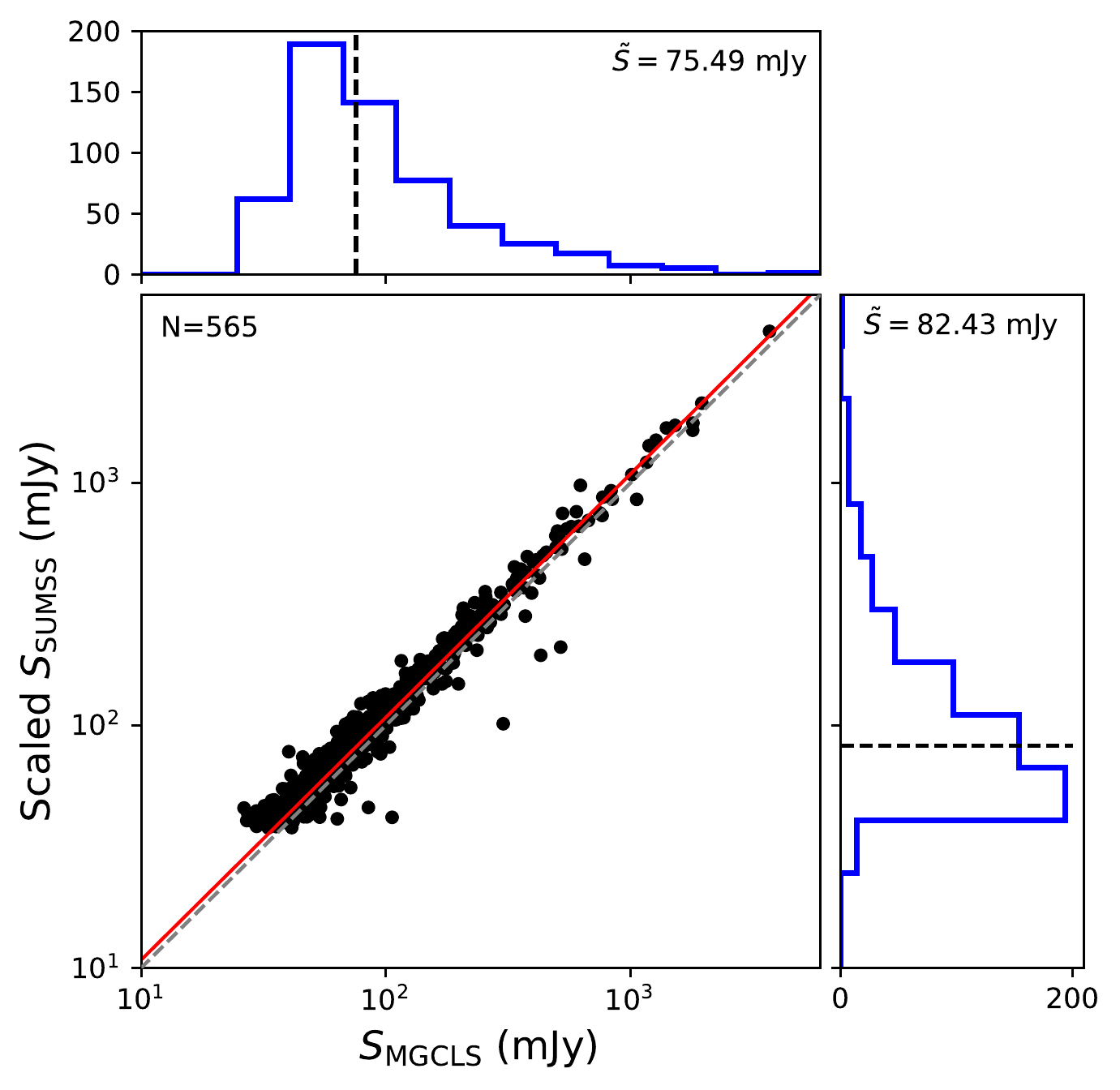}
   \caption{Comparison between the integrated flux density of MGCLS compact sources with their counterparts in the NVSS \textbf{(top)} and SUMSS \textbf{(bottom)} catalogues, with the latter two being scaled to the MGCLS frequency of 1.28\,GHz. The best-fit relation in each case is shown by the solid red line and is consistent with a linear relationship, with the number of sources in the fit shown upper left. The grey dashed line shows the exact one-to-one relationship. The scaled sky survey flux densities are typically 6\% higher than their MGCLS counterparts, so this represents a possible small bias in the MGCLS flux scale.
   Histograms show the relevant flux density distributions, with the black dashed lines indicating the respective median values, $\Tilde{S}$. 
   }
   \label{fig:fluxcheck}%
\end{figure}

Our cross-matching yields a total of 398 and 565 compact MGCLS sources with NVSS and SUMSS counterparts, respectively. Figure~\ref{fig:fluxcheck} shows the flux density scale comparison between the MGCLS compact sources and the scaled NVSS (top panel) and scaled SUMSS (bottom panel) sources. The MGCLS flux densities are in good agreement with those of both NVSS and SUMSS. We fit a power law of the form 
\begin{equation}
S_{\rm scaled\_cat}=\kappa \times S_{\rm MGCLS}^\gamma,
\end{equation}
to the MGCLS and scaled NVSS and SUMSS flux densities. Values of 1 for both $\kappa$ and $\gamma$ would indicate an exact one-to-one correspondence. We obtain fit values of $\kappa_{\rm NVSS} = 1.06 \pm 0.02, \gamma_{\rm NVSS} = 1.00 \pm 0.01$, and $\kappa_{\rm SUMSS} = 1.06 \pm 0.04, \gamma_{\rm SUMSS} = 1.00 \pm 0.01$ for the NVSS and SUMSS comparisons, respectively. This is consistent with a linear relation between the MGCLS and scaled fluxes, with the scaled NVSS and SUMSS fluxes being marginally higher than those from MGCLS. The MGCLS sources were not chosen to be isolated, so the poorer resolution of the sky surveys ($\sim$\,45\arcsec) may lead to a single NVSS or SUMSS source having contributions from additional faint MGCLS sources; at the lowest fluxes, this could result in a small bias in opposite spectral index directions for the two surveys. However, the equality of the fitted flux densities show that any such effects are negligible. 

\begin{figure}
   \centering
   \includegraphics[width=\columnwidth]{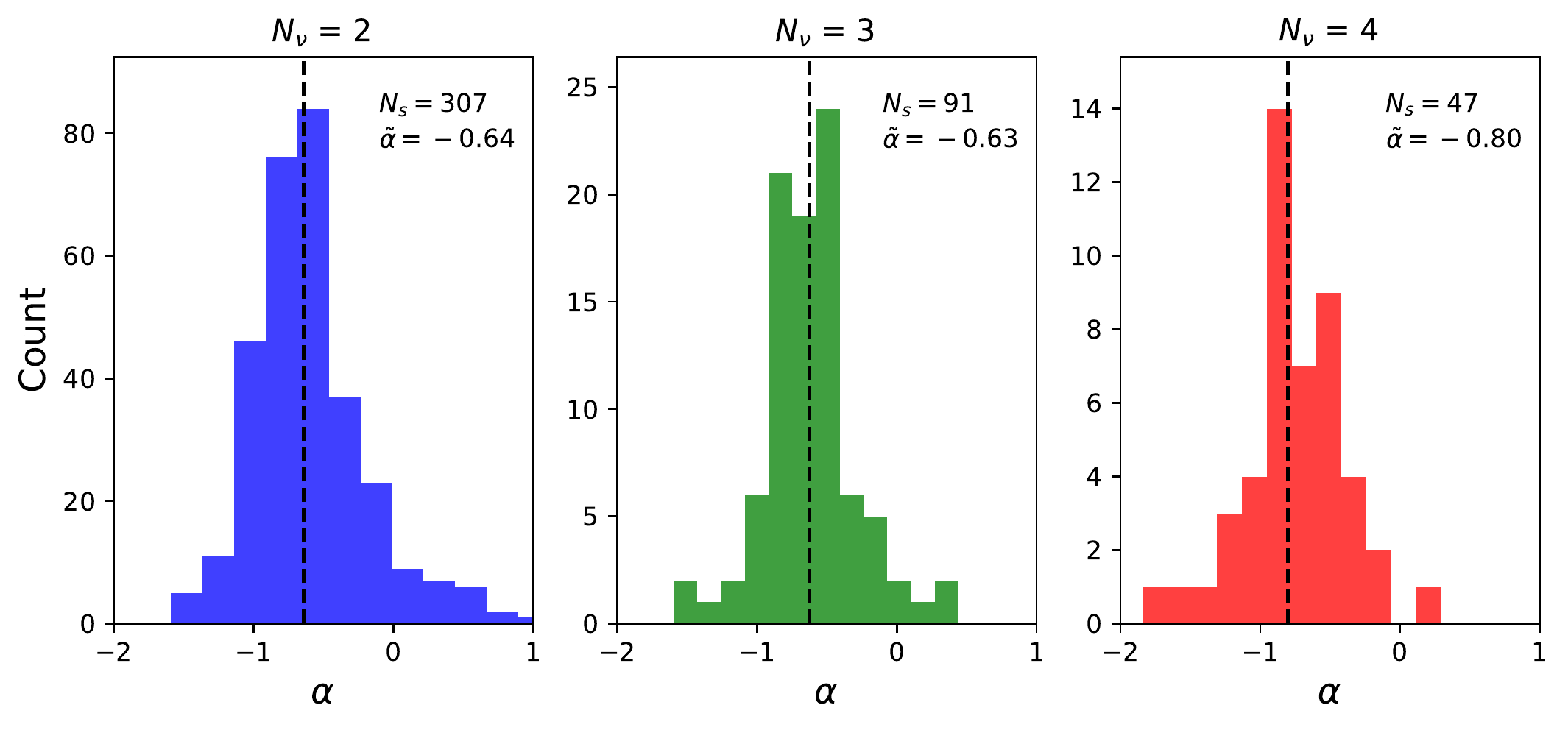}
   \caption{Spectral index distributions for the cross-matched MGCLS compact sources, using data from the MGCLS (1.28~GHz), NVSS (1.4\,GHz), SUMSS (843\,MHz), and TGSS (148~MHz) catalogues. \textbf{Left to Right:} Distributions for MGCLS sources with flux density measurements at 2, 3, and 4 frequencies, respectively. The number of compact sources, $N_s$, and the median spectral index, $\Tilde{\alpha}$, are indicated upper right of each panel, and the respective median spectral indices are indicated by  vertical dashed lines.}
   \label{fig:spixcheck}%
\end{figure}

We also examine the spectral index distribution of the radio cross-matched MGCLS compact sources. For this purpose, we include 148\,MHz data from the TIFR GMRT Sky Survey \citep[TGSS, north of $-53^\circ$ declination,][]{TGSS} in addition to the NVSS and SUMSS surveys. 
Given the different sky coverages, some MGCLS sources have flux densities at only one additional frequency, whereas others have three. We fit a single $S_\nu \propto \nu^{\alpha}$ power-law and inspected each fit to discard both spurious matches and sources showing spectral curvature. Figure~\ref{fig:spixcheck} shows the resulting spectral index distributions. The distributions of spectral indices for sources with two, three, and four frequency points are similar, although the $N_\nu = 4$ distribution has a lower median spectral index. Subtle differences may reflect differences in the spectral populations of the different surveys, but are beyond the scope of the current work.

\subsubsection{Optical cross-matching}\label{sec:optxmatch}

We have created optical cross-match catalogues for the compact sources in the \object{Abell~209} and \object{Abell~S295} fields, using data from DECaLS \citep{Dey_2019}. These fields were selected due to their existing DECaLS coverage and their decent MGCLS dynamic range. Cross-match catalogues for other MGCLS fields will be compiled in follow-up works.

To identify DECaLS counterparts for the MGCLS compact sources, we used the likelihood ratio method \citep{1992MNRAS.259..413S, 2009ApJS..180..102L, 2011MNRAS.416..857S}, and we use \emph{LR} to denote the likelihood ratio. The \emph{LR} {here} is defined as the ratio of the probability that an optical source (at a given distance from the radio position and with a given optical magnitude) is the true counterpart, to the probability that the same source is a spurious alignment, i.e., \emph{LR}\,$= (q(m) \times f(r)) / n(m)$, where $q(m)$ is the expected number of true optical counterparts with magnitude $m$, $f(r)$ is the probability distribution function of the positional uncertainties in both the radio and the optical source catalogues, and $n(m)$ is the background density of optical galaxies of magnitude $m$ in the DECaLS $r$-band (or $g$- or $z$-band). The magnitudes are AB magnitudes from DECaLS DR8.

\begin{figure}
   \centering
   \includegraphics[width=0.9\columnwidth,clip=True,trim=0 0 0 0]{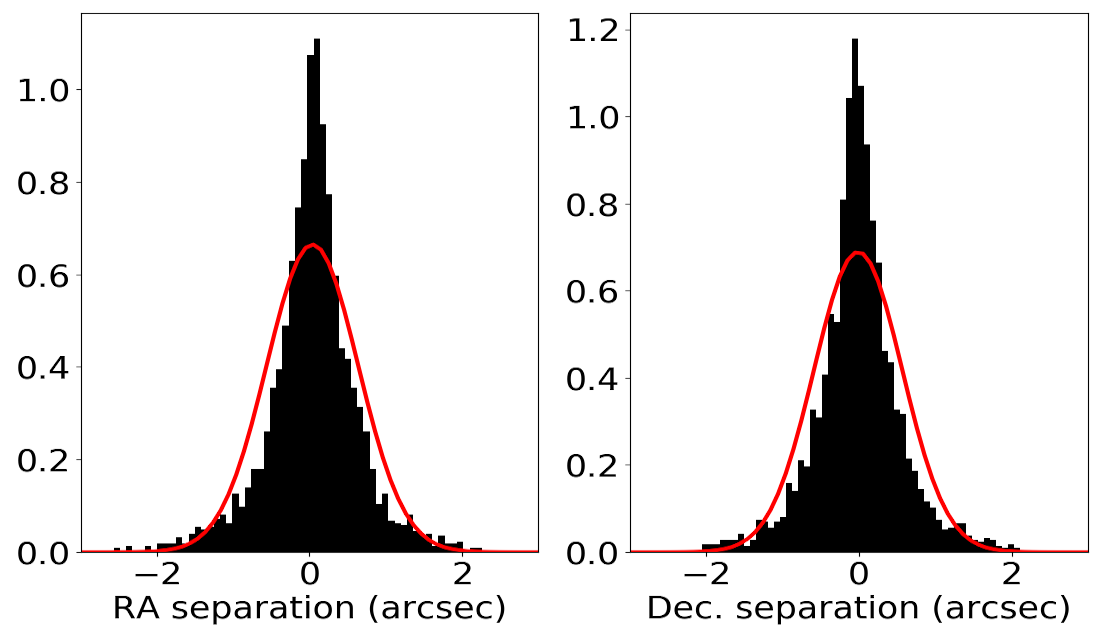}
   \caption{Histograms of the R.A. and Dec.\ angular separations between the positions of radio sources in the Abell~209 compact source catalogue and their optical counterparts, for \emph{LR} $>$ 0.5. In each panel, the red line shows the normalised Gaussian distribution. 
   }
   \label{fig:A209optdist}%
\end{figure}

The \textit{a priori} probability $q(m)$ is determined as follows. First, the radio and optical source catalogues are matched by finding the closest counterpart within a fixed search radius of 4\arcsec. We choose this radius based on our distributions of cross-matches, as shown in Figure~\ref{fig:A209optdist}, where the distributions (\emph{LR} > 0.5) of the R.A. and Dec. angular separations between the positions of radio sources can be approximated by Gaussians. The search radius of 4\arcsec\ is the optimal radius where we detect the most true sources. Radii greater than 4\arcsec\ will sharply increase the number of spurious matches. The number of spurious matches (of magnitude $m$) is estimated by scaling $n(m)$ to the area of 4\arcsec\ radius within which we search for counterparts. This is then subtracted from the number of counterparts (as a function of magnitude) to determine the number of true associations, $q(m)$. 

In our case, the probability distribution $f(r)$ is a two-dimensional Gaussian distribution: 
\begin{equation}
f(r) = \frac{1}{2 \pi \delta^2} \exp\left({-\frac{r^2}{2 \delta^2}}\right). 
\end{equation}
Here, $r$ is the angular distance (in arcsec) from the radio source position, and $\delta$ is the combined positional error given by $\sqrt{\delta_{\rm decals}^2 + \delta_{\rm mgcls}^2}$, {where ${\delta_{\rm decals}}$ is the positional uncertainty from the DECaLS catalogue, and ${\delta_{\rm mgcls}}$ is the positional uncertainty from the MGCLS compact source catalogue}. For each source in the MGCLS DR1 compact source catalogue, we adopted an elliptical Gaussian distribution for the positional errors, with the uncertainties in R.A. and Dec. on the radio position reported in the radio catalogue. We assume a systematic optical position uncertainty of 0.2{\arcsec} in both R.A. and Dec. for the DECaLS catalogue \citep{Dey_2019}.

\begin{table*}
\caption{Excerpt of the compact source catalogue for Abell~209 with optical cross-match information from DECaLS. The full catalogues of cross-matched sources for Abell~209 and Abell~S295 are available online. Columns (1)--(11) list the radio source properties and are the same as columns (1)--(11) in the DR1 compact source catalogue presented in Table~\ref{table:ptsrcs}.  Additional columns: (12--13) R.A. and Dec. of DECaLS cross-match; (14) $g$-band magnitude; (15) $r$-band magnitude; (16) $z$-band magnitude; (17) WISE W1-band magnitude; (18) WISE W2-band magnitude; (19) photometric redshift {(see Section 8.2 for description; typical uncertainty $\delta z / (1+z) = 0.03$)}; (20--24) 1$\sigma$ uncertainty on $g$, $r$, $z$, $W1$, and $W2$ magnitudes, respectively; (25) angular separation between MGCLS source and optical match; (26) match flag: 0\,-\,no optical counterpart, 1\,-\,the best optical counterpart according to the probabilities in column 28, 2\,-\,the second-best optical counterpart according to the probabilities in column 28, etc.; (27) Probability that the MeerKAT source has a matched counterpart; (28) Probability that the MeerKAT source has the listed matched counterpart. {In cases where there are more than one optical counterpart detected within 4\arcsec, this column lists the probability of each of the objects being the optical counterpart.} 
See Section~\ref{sec:optxmatch} for details. } 
\label{table:optxmatch}      
\centering       
\begin{tabular}{cccccccccc} 
\hline                 
(1) & (2)--(11) & (12) & (13) & (14) & (15) & (16) & (17) & (18) & (19) \\
Src. name &
& R.A.$_{\rm opt}$ 
& Dec.$_{\rm opt}$ 
& $g$ 
& $r$ 
& $z$ 
& $W1$ 
& $W2$ 
& $z_{\rm p}$ 
\\
MKTCS & & (deg) & (deg) & & & & &\\
\hline                        
J231022.80$-$140627.7 & ... & 23.17312 & $-$14.10746 & 19.469 & 19.01   & 18.772 & 19.115 & 19.448 & 0.22 \\
J231014.16$-$135358.5 & ... & 23.17047 & $-$13.89944 & 21.503 & 20.748  & 20.244 & -      & -      & 0.25 \\
J231001.56$-$133249.2 & ... & 23.16717 & $-$13.54746 & 19.678 & 18.752 & 18.128 & 18.221 & 18.5   & 0.06 \\
J230946.80$-$133547.0 & ... & 23.16301 & $-$13.59645 & 19.76  & 18.499 & 17.597 & 17.12  & 17.34  & 0.3  \\
J230944.28$-$132944.5 & ... & 23.16238 & $-$13.49577 & 20.445 & 19.572 & 18.854 & 18.374 & 18.521 & 0.17 \\
J230939.60$-$135751.1 & ... & 23.16135 & $-$13.96480 & 21.981 & 20.995 & 20.313 & 19.654 & 19.611 & 0.21 \\
J230919.08$-$132915.3 & ... & 23.15536 & $-$13.48751 & 20.279 & 19.534 & 18.984 & 18.898 & 19.151 & 0.51 \\
J230919.08$-$132354.9 & ... & 23.15526 & $-$13.39845 & 22.113 & 21.104 & 20.535 & 20.349 & -      & 0.27 \\
J230916.20$-$132101.0 & ... & 23.15449 & $-$13.35043 & 19.775 & 18.915 & 18.294 & 18.285 & 18.601 & 0.23 \\

\hline                                   

\end{tabular}
\begin{tabular}{ccccccccc} 
\hline                
(20) & (21) & (22) & (23) & (24) & (25) & (26) & (27) & (28) \\
$\Delta g$ 
& $\Delta r$ 
& $\Delta z$ 
& $\Delta W1$ 
& $\Delta W2$ 
& $d$ 
& $F_{\rm xm}$ 
& $P_{\rm has\_xm}$ 
& $P_{\rm this\_xm}$ \\
& & & & & (\arcsec) & \\
\hline                        
0.003 & 0.003 & 0.008 & 0.03  & 0.095 & 1.1 & 1 & 1    & 1 \\
0.008 & 0.005 & 0.01  & -     & -     & 0.8 & 1 & 1    & 1 \\
0.004 & 0.002 & 0.003 & 0.013 & 0.039 & 1.6 & 1 & 0.94 & 1 \\
0.005 & 0.002 & 0.003 & 0.006 & 0.017 & 0.1 & 1 & 1    & 1 \\
0.01  & 0.005 & 0.008 & 0.016 & 0.042 & 0.3 & 1 & 1    & 1 \\
0.015 & 0.007 & 0.012 & 0.045 & 0.099 & 2.5 & 1 & 0.8  & 1 \\
0.006 & 0.003 & 0.005 & 0.024 & 0.067 & 0.5 & 1 & 1    & 1 \\
0.019 & 0.01  & 0.016 & 0.087 & -     & 0.7 & 1 & 0.99 & 1 \\
0.003 & 0.002 & 0.003 & 0.013 & 0.04  & 0.4 & 1 & 1    & 1 \\

\hline                                   

\end{tabular}
\end{table*}

The optical cross-match catalogue includes all matches within 4{\arcsec} of a compact MGCLS source detection, along with their \emph{LR} probabilities and flags (no \emph{LR} cutoff imposed). Table~\ref{table:optxmatch} shows an excerpt from the optical cross-match catalogue for the Abell~209 field, with the full catalogues for it and the Abell~S295 fields available online\footnote{\mgclsdoi}. The presence of more than one counterpart for a particular radio source provides additional information to that contained in the \emph{LR} itself, which can then be used to estimate the reliability of the counterpart source, or the probability that a particular source is the correct counterpart. The reliability for radio source $i$, as defined by \citet{1992MNRAS.259..413S}, is calculated as
\begin{equation}
REL_i = \frac{LR_i} {\sum \,LR_{\rm searchradius} + (1 - Q)}, 
\end{equation}
where $\sum LR_{\rm searchradius}$ is the sum of \emph{LR} for all possible DECaLS counterparts to the radio source within our search radius of 4\arcsec, and $Q$ is the fraction of MGCLS compact radio sources with optical counterparts above the DECaLS magnitude limit. Comparison of $\sum REL_i$ with the total number of counterparts with \emph{LR} > $LR_{\rm cutoff}$ provides an estimate of the spurious identification rate, or error rate (ER). The choice of the cutoff in \emph{LR} is a trade-off between maximum completeness and maximum purity. Completeness is defined as the fraction of radio catalogue sources which have an optical counterpart, and purity (given by $1\,-\,$ER) is the fraction of radio-optical source matches which are real. 

\begin{figure}
   \centering
   \includegraphics[width=\columnwidth,clip=True,trim=30 0 15 50]{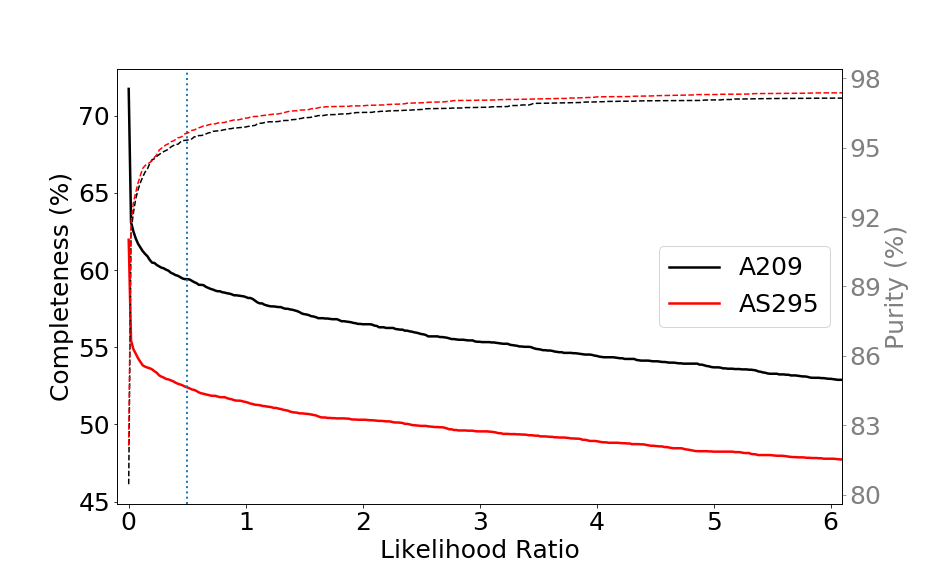}
   \caption{Completeness (left axis, solid lines) and purity (right axis, dashed lines) versus the likelihood ratio (\emph{LR}) for Abell~209 (black) and Abell~S295 (red). Imposing a \emph{LR} cutoff at 0.5 (vertical dotted line) for Abell~209, the estimated spurious identification rate (error rate, or 1 $-$ purity) is 4.5\%, with a completeness of 59\%. }
   \label{fig:cpvlr}%
\end{figure}

Figure~\ref{fig:cpvlr} shows the completeness and purity for both Abell~209 and Abell~S295, with no \emph{LR} cutoff imposed. Our chosen $LR_{\rm cutoff}$ of 0.5 is indicated by the vertical dotted line. A value of $LR_{\rm cutoff} = 0.5$ corresponds to an estimated spurious identification rate of 4.5\% in Abell~209, with 59\% (2723 of 4581) of radio sources in the Abell~209 compact source catalogue having optical counterparts. 

\subsection{Extended sources}\label{sec:srcs-ext}
\begin{figure}
    \centering
      \includegraphics[width=0.98\columnwidth]{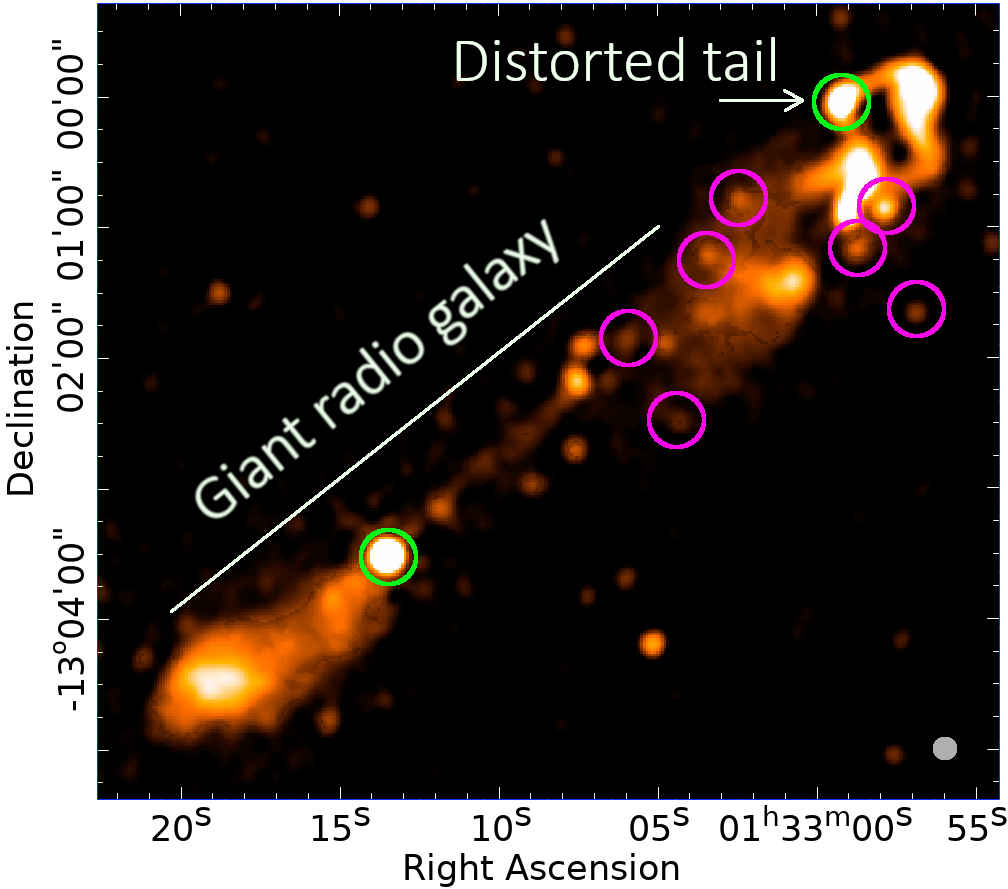}
      \caption{Full resolution (7.6\arcsec\,$\times$\,7.4\arcsec) MGCLS Stokes~I intensity image of a multi-host structure in the field of Abell~209. We indicate the two possible intersecting multi-component radio galaxies (giant radio galaxy and a distorted tailed galaxy). The compact radio components circled in green indicate the positions of the likely optical hosts, while the magenta circles indicate other superposed or nearby compact radio emission with optical counterparts. The brightness scale is logarithmic, and saturates at 1~mJy\,beam\per. The synthesised beam is indicated at lower right.}
    \label{fig:A209gal}
\end{figure}
We do not provide catalogues for the extended sources, {that is, those for which the \textsc{pybdsf} fit required multiple Gaussians, indicated in the \textsc{pybdsf} output with code `M'}. These need to be verified visually, a process that is extremely time consuming. However, to provide an indication of the number of extended sources in the MGCLS, we {performed} this verification for sources in the Abell~209 and Abell~S295 fields. 

Extended sources can be separated into two categories: \textit{blended} sources, i.e., those with overlapping Gaussian components, and \textit{multi-component} sources with multiple, non-overlapping and often visually separable, components. There are 158 and 347 blended sources in the Abell~209 and Abell~S295 fields, respectively, roughly 3--5\% of the number of compact sources in the respective fields. {Their integrated flux densities} range from $85\,\mu$Jy to $105$\,mJy, with the largest blended sources being just over 30\arcsec\ across. Extrapolating these numbers to the full MGCLS sample, there are of the order of 29,000 blended sources in the full survey.

We define the multi-component sources as those with distinct structures such as jets, cores, or lobes. Identifying the different components which comprise a single source can be difficult, and is still typically done through visual inspection. Automated methods, such as those discussed in\editout{Section~} \kk{Appendix~\ref{app:astronomaly}}, will be needed as our datasets become larger and larger. We visually inspected the Abell~209 and Abell~S295 fields, finding 33 and 26 multi-component sources in each field, respectively. The largest of these multiple-component sources spans an angular size of 9.8\arcmin. For each source we measured the integrated flux density, including all components, using pixels above 3$\sigma$, where $\sigma$ is the local RMS. Catalogues of these sources are presented in Appendix \ref{app:extsrccats} in Tables~\ref{tab:extsrcs_A209} and \ref{tab:extsrcs_AS295}, which include the likely optical/IR counterpart from either the DECaLS or Wide-field Infrared Survey Explorer \citep[WISE,][]{Wright_2010} catalogues. The catalogue position for each extended radio source is fixed to that of its likely optical/IR counterpart, if available, or otherwise given by the flux-weighted centroid.

Due to the combination of high sensitivity and resolution, many well-resolved and multi-component radio galaxies have contaminating foreground or background sources that can currently only be separated visually, if at all. Figure~\ref{fig:A209gal} shows one example {of a giant radio galaxy \citep{2016MNRAS.456..512C, 2018ApJS..238....9K} in the Abell~209 field}.  This $\sim$\,7.7\arcmin\ structure appears to have several possible optical/IR counterparts, all of which are associated with compact radio features. The left-most green circle in Figure~\ref{fig:A209gal} indicates the position of WISE~J013313.50$-$130330.5, a candidate quasar \citep{2019arXiv191205614F} at a redshift of $z=0.289$ and the likely counterpart for the giant radio galaxy. At that redshift, the radio galaxy extent would be $>$\,1.5\,Mpc. The right-most green circle is associated with WISE~J013259.20$-$130002.0, which has no available redshift. It appears to be a distorted bent-tail source whose features merge (in projection) with those of the western lobe of the assumed giant radio galaxy. Spectral index studies and redshift measurements will likely to be needed to disentangle such structures \citep{2021MNRAS.508.2910M}, which appear frequently {in the MGCLS images, due to the high sensitivity to extended structures}.

\section{MGCLS Diffuse Cluster Emission}\label{sec:clusterde}

\longtab{
\begin{longtable}{lrrclcrcl} 
\caption{\label{tab:diffuse}Catalogue of the 99 diffuse cluster radio sources detected in the MGCLS. Several of the 62 MGCLS clusters in this catalogue are host to more than one diffuse cluster source. Cols: (1) Cluster name; (2--3) NED cluster position: J2000 R.A. and Dec.; (4) Cluster redshift; (5) Diffuse source classification --- mini-halo (MH), halo (H), relic (R), phoenix (Ph), candidate (c), unknown/unclear (U); see Section~\ref{sec:clusterde} for further details. Elongated halos/halo candidates with embedded bright AGN sources are indicated by $^\dagger$; (6) Indicates whether or not the diffuse source is a new detection; (7) Largest angular size in arcminutes; (8) Largest physical linear size at the cluster redshift in Mpc; (9) Notes on the detection with references, in square brackets, to previous studies. HT -- head-tail galaxy; WAT -- wide-angle tail galaxy.}\\
\hline\hline       
(1) & \multicolumn{1}{c}{(2)} & \multicolumn{1}{c}{(3)} & (4) & (5) & (6) & \multicolumn{1}{c}{(7)} & \multicolumn{1}{c}{(8)} & (9)\\
Cluster Name & \multicolumn{1}{c}{R.A.$_{\rm J2000}$} & \multicolumn{1}{c}{Dec.$_{\rm J2000}$} & $z$ & Class & New? & \multicolumn{1}{c}{LAS} & \multicolumn{1}{c}{LLS} & Notes [Refs] \\    
 & \multicolumn{1}{c}{(deg)} & \multicolumn{1}{c}{(deg)} & & & & \multicolumn{1}{c}{(\arcmin)} & \multicolumn{1}{c}{(Mpc)} & \\
\hline  
\endfirsthead
\endfoot
\caption{continued.}\\
\hline\hline
(1) & \multicolumn{1}{c}{(2)} & \multicolumn{1}{c}{(3)} & (4) & (5) & (6) & \multicolumn{1}{c}{(7)} & \multicolumn{1}{c}{(8)} & (9)\\
Cluster Name & \multicolumn{1}{c}{R.A.$_{\rm J2000}$} & \multicolumn{1}{c}{Dec.$_{\rm J2000}$} & $z$ & Class & New? & \multicolumn{1}{c}{LAS} & \multicolumn{1}{c}{LLS} & Notes [Refs] \\    
 & \multicolumn{1}{c}{(deg)} & \multicolumn{1}{c}{(deg)} & & & & \multicolumn{1}{c}{(\arcmin)} & \multicolumn{1}{c}{(Mpc)} & \\
\hline     
\endhead
\hline
\endlastfoot
Abell 13            &   3.384  &    $-$19.501 &  0.094 & cR  &            &  3.0 & 0.32 & W of centre; [8, 13, 17, 37]\\ 
Abell 22            &   5.161  &    $-$25.722 &  0.142 & cH  &            &  2.7 & 0.41 & [8]\\
Abell 85            &  10.453  &    $-$9.318  &  0.056 & \kk{M}H$^\dagger$ & \checkmark & 4.0 & 0.26 & \\
                    &          &              &         & Ph$^a$ &         &  5.6 & 0.36 & SW of centre; [8, 18]\\
Abell 168           &  18.791  &       0.248  &  0.045 & R   &            & 12.4 & 0.66 & N of centre; [10] \\
Abell 209           &  22.990  &   $-$13.576  &  0.209 & H   &            &  7.9 & 1.62 & Embedded HT; [17, 19, 20, 41, 42]\\
Abell 370           &   39.960 &    $-$1.586  &  0.375 & cH  & \checkmark &  3.1 & 0.96 & [44] \\
Abell 521           &   73.536 &    $-$10.244 &  0.248 &  R  &            &  6.0 & 1.40 & SE of centre; [11, 15, 28, 41]\\
                    &          &              &         & H   &            &  5.3 & 1.23 & [4, 6, 20, 28, 42]\\
                    &          &              &         & R   & \checkmark &  4.0 & 0.93 & NW of centre \\
Abell 545$^b$       &   83.102 &    $-$11.543 &  0.154  & H   &            &  4.7 & 0.75 & [1, 17]\\
Abell 2645          &  355.320 &    $-$9.028 &  0.251  & cH  & \checkmark &  2.7 & 0.64 & Irregular shape; [5]\\
Abell 2667          &  357.920 &    $-$26.084 &  0.232 & H$^\dagger$ &    &  4.6 & 1.02 &  [16] \\
Abell 2744          &    3.567 &    $-$30.383 &  0.307 & H   &            &  7.5 & 2.04 & [8, 13, 17, 21, 32, 42]\\
                    &          &              &         & R   &            &  5.8 & 1.57 & NE of centre; [8, 13, 21, 30, 32, 42]\\
                    &          &              &         & R   &            &  4.6 & 1.25 & SE of centre; [32] \\
Abell 2751          &    4.058 &    $-$31.389 &  0.081 & cPh & \checkmark &  0.9 & 0.08 & NW of centre; [8] (says relic)\\
Abell 2811          &   10.537 &    $-$28.536 &  0.108 & H   &            &  5.2 & 0.62 & [8] \\
Abell 2813          &   10.852 &    $-$20.621 &  0.292 & cH  & \checkmark &  3.1 & 0.81 & \\
Abell 2895          &   19.546 &    $-$26.973 &  0.228 & cPh & \checkmark &  0.8 & 0.18 &  E of centre \\
Abell 3365          &   87.050 &    $-$21.935 &  0.093 & R   & \checkmark &  6.8 & 0.70 &  NE of centre; [45] \\
                    &          &              &         & R   & \checkmark &  3.4 & 0.35 & W of centre; [45] \\ 
Abell 3376          &   90.426 &    $-$39.985 &  0.047 & R   &            & 28.1 & 1.55 & E of centre; [2, 12, 23]\\
                    &          &              &         & R   &            & 18.8 & 1.04 & W of centre; [2, 12, 23]\\
Abell 3558          &  201.978 &    $-$31.492 &  0.048 & H   &            &  4.5 & 0.25 & [7, 39]\\
Abell 3562          &  202.783 &    $-$31.673 &  0.050 & H   &            & 11.5 & 0.67 & [7, 14, 40]  \\
                    &          &              &         & R   &            &  8.6 & 0.50 & SW of centre; [40]\\
Abell 3667          &  303.140 &    $-$56.841 &  0.056 & R   &            & 25.8 & 1.67 & SE of centre; [22, 33]\\ 
                    &          &              &         & R   &            & 35.2 & 2.28 & NW of centre; [22, 33] \\
Abell 4038          &  356.880 &    $-$28.203 &  0.030 & cMH & \checkmark &  4.5 & 0.16 & Embedded HT \\
Abell S295          &   41.400 &    $-$53.038 &  0.300  & H   &            &  4.0 & 1.07 & [46] \\
Abell S1063         &  342.181 &    $-$44.529 &  0.348 & H   &            &  5.5 & 1.62 &  [44] \\
Abell S1121         &  351.284 &    $-$41.212 &  0.190 & U   &            &  2.2 & 0.42 & SE of centre \\
Bullet              &  104.658 &    $-$55.950 &  0.297  & H   &            &  8.5 & 2.26 &  [26, 35]\\
                    &          &              &         & R   &            &  3.8 & 1.01 & E of centre; [36]\\
El Gordo            &   15.719 &    $-$49.250 &  0.870  & H   &            &  3.3 & 1.53 & [3, 27]\\
                    &          &              &         & R   &            &  2.0 & 0.93 & NW of centre; [3, 27]\\
                    &          &              &         & R   &            &  1.0 & 0.46 & SE of centre; [3, 27]\\
                    &          &              &         & R   &            &  0.9 & 0.42 & E of centre; [27]\\
PLCK G200.9$-$28.2  &   72.587 &     $-$2.949 &  0.220  & R   & \checkmark &  5.6 & 1.19 & E of centre  \\
                    &          &              &         & R   & \checkmark &  2.7 & 0.58 & SW of centre; [24]\\
                    &          &              &         & cR  & \checkmark &  1.0 & 0.21 & NW of centre \\
MACS J0257.6$-$2209$^c$ & 44.412 &  $-$22.163 &  0.322 & U   &            &  1.6 & 0.45 & [16]\\
                    &          &              &         & cR  & \checkmark &  1.4 & 0.39 & SW of centre \\
MACS J0417.5$-$1154 &   64.394 &    $-$11.909 &  0.443 & H$^\dagger$ &    &  4.9 & 1.68 & [9, 31, 34]\\
                    &          &              &         & cR  & \checkmark &  1.0 & 0.34 & N of centre \\
                    &          &              &         & cR  & \checkmark &  1.0 & 0.34 & NW of centre \\
RXC J0510.7$-$0801  &   77.685 &    $-$8.020  &  0.220 & cH$^\dagger$ & \checkmark & 4.9 & 1.04 & \\
                    &          &              &         & cR  & \checkmark &  3.0 & 0.64 & N of centre; possible WAT \\
RXC J0520.7$-$1328$^d$ &  80.200 &  $-$13.502 &  0.336 & cR  &            &  4.2 & 1.21 & SE of centre; [29]\\
                    &          &              &         & U   &            &  7.1 & 2.05 & SE of centre; bubble?; [29] \\
RXC J1314.4$-$2525  &  198.599 &    $-$25.256 &  0.244 & H   &            &  4.7 & 1.08 &  [38, 48] \\
                    &          &              &         & R   &            &  4.7 & 1.08 & W of centre; [38, 48] \\
                    &          &              &         & R   &            &  2.7 & 0.62 & E of centre; [38, 48] \\
RXC J2351.0$-$1954  &  357.770 &    $-$19.913 &  0.248 & cR  &            &  9.8 & 2.28 & W of centre; [8] \\
                    &          &              &         & cR  &            &  5.5 & 1.28 & E of centre; [8]\\
                    &          &              &         & cH  & \checkmark &  4.0 & 0.93 & [8] \\
J0027.3$-$5015      &    6.839 &    $-$50.251 &  0.145 & cMH & \checkmark &  2.5 & 0.38 & \\
J0145.0$-$5300      &   26.260 &    $-$53.014 &  0.117 & H   & \checkmark &  4.9 & 0.62 & \\ 
J0145.2$-$6033      &   26.320 &    $-$60.565 &  0.181 & cMH & \checkmark &  1.9 & 0.35 & \\
J0216.3$-$4816      &   34.080 &    $-$48.273 &  0.163 & cMH & \checkmark &  1.4 & 0.24 & \\
J0217.2$-$5244      &   34.303 &    $-$52.747 &  0.343 & cR  & \checkmark &  1.5 & 0.44 & N of centre \\
J0225.9$-$4154      &   36.478 &    $-$41.910 &  0.220 & H   & \checkmark &  2.4 & 0.51 &  \\ 
J0232.2$-$4420      &   38.070 &    $-$44.348 &  0.284 & H   &            &  5.8 & 1.49 & [25]\\
                    &          &              &         & cR  & \checkmark &  3.3 & 0.85 & S of centre \\
                    &          &              &         & cR  & \checkmark &  1.8 & 0.46 & E of centre \\
J0303.7$-$7752      &   45.943 &    $-$77.869 &  0.274 & H   & \checkmark &  3.8 & 0.95 &  \\
J0314.3$-$4525      &   48.583 &    $-$45.424 &  0.072 & cMH & \checkmark &  2.4 & 0.20 & \\
J0342.8$-$5338      &   55.725 &    $-$53.635 &  0.060 & MH$^\dagger$ & \checkmark & 5.2 & 0.36 & \\
J0351.1$-$8212      &   57.787 &    $-$82.217 &  0.061 & U   & \checkmark &      &      &  See Section~\ref{sec:de-j0351}\\
J0352.4$-$7401      &   58.123 &    $-$74.031 &  0.127 & R   & \checkmark & 15.6 & 2.13 & SE of centre \\
                    &          &              &         & H   & \checkmark & 10.7 & 1.46 &  \\
                    &          &              &         & cR  & \checkmark &  6.8 & 0.93 & NW of centre \\
                    &          &              &         & R   & \checkmark &  3.9 & 0.53 & NNW of centre \\
                    &          &              &         & R   & \checkmark &  5.0 & 0.68 & N of centre \\
J0431.4$-$6126      &   67.850 &    $-$61.444 &  0.059 & R   & \checkmark &  9.7 & 0.66 & SE of centre \\
                    &          &              &         & U   & \checkmark &      &      &  Confused tail \\
J0510.2$-$4519      &   77.558 &    $-$45.321 &  0.200  & cMH & \checkmark &  1.3 & 0.26 &  \\
J0516.6$-$5430      &   79.158 &    $-$54.514 &  0.295 & H   & \checkmark &  5.7 & 1.51 & \\
                    &          &              &         & R   & \checkmark &  5.7 & 1.51 & N of centre \\
                    &          &              &         & cR  & \checkmark &  2.3 & 0.61 & S of centre \\
J0528.9$-$3927      &   82.235 &    $-$39.463 &  0.284 & H   &            &  4.0 & 1.06 & [46] \\
J0627.2$-$5428      &   96.810 &    $-$54.470 &  0.051 & R   & \checkmark &  5.4 & 0.32 & W of centre  \\ 
J0631.3$-$5610      &   97.836 &    $-$56.172 &  0.054 & cR  & \checkmark &  7.7 & 0.49 & W of centre \\ 
J0637.3$-$4828      &   99.329 &    $-$48.478 &  0.203 & U   & \checkmark &  7.6 & 1.52 &  \\
                    &          &              &         & cR  & \checkmark &  2.0 & 0.40 & NW of centre \\
J0638.7$-$5358      &   99.694 &    $-$53.972 &  0.227  & H$^\dagger$ &    &  6.4 & 1.40 &  [43] \\
J0645.4$-$5413      &  101.372 &    $-$54.219 &  0.164 & H   & \checkmark &  7.7 & 1.30 &  \\
                    &          &              &         & R   & \checkmark &  3.5 & 0.59 & SW of centre \\
J0745.1$-$5404      &  116.290 &    $-$54.079 &  0.074 & U   & \checkmark &  4.2 & 0.35 & S of centre \\
J0820.9$-$5704      &  125.248 &    $-$57.080 &  0.061 & U   & \checkmark &      &      & S of centre \\
J1130.0$-$4213      &  172.523 &    $-$42.230 &  0.155 & cR  & \checkmark &  2.1 & 0.34 & NE of centre \\
J1423.7$-$5412      &  215.930 &    $-$54.203 &  0.300 & U   & \checkmark &      &      & N of centre \\
J1539.5$-$8335      &  234.891 &    $-$83.592 &  0.073 & MH  & \checkmark &  2.8 & 0.23 & \\
                    &          &              &         & cR  & \checkmark &  2.2 & 0.18 & W of centre \\
J1601.7$-$7544      &  240.445 &    $-$75.746 &  0.153 & H   & \checkmark &  5.9 & 0.94 &  \\
J1840.6$-$7709      &  280.155 &    $-$77.156 &  0.019 & cMH & \checkmark &  2.5 & 0.06 & \\
J2023.4$-$5535      &  305.852 &    $-$55.592 &  0.232 & H   &            &  4.7 & 1.04 & [47]\\
                    &          &              &         & R   &            &  2.9 & 0.64 & [47]\\

\hline                                   
\end{longtable}
\tablefoot{$^a$ Mixed classification in the literature (see Section~\ref{sec:de-a85}). $^b$ Diffuse emission from the H$\alpha$ region in the Orion Nebula is also detected in this field. $^c$ MACS~J0257.6$-$2209 has a published detection of a giant radio halo \citep{2017ApJ...841...71G} but in the higher-quality MGCLS data, the reported giant halo looks like the blending of other sources. $^d$ Diffuse emission in RXC~J0520.7$-$1328 is complex and difficult to classify (see Section~\ref{sec:de-j0520}).

\textbf{References:} \textbf{1}~\cite{2003A&A...400..465B}; \textbf{2}~\cite{2006Sci...314..791B}; \textbf{3}~\cite{2016MNRAS.463.1534B}; \textbf{4}~\cite{2008Natur.455..944B}; \textbf{5}~\cite{2013ApJ...777..141C}; \textbf{6}~\cite{2009ApJ...699.1288D}; \textbf{7}~\cite{2018A&A...620A..25D}; \textbf{8}~\cite{2021PASA...38...10D}; \textbf{9}~\cite{2011JApA...32..529D}; \textbf{10}~\cite{Dwarakanath_2018}; \textbf{11}~\cite{2006A&A...446..417F}; \textbf{12}~\cite{10.1093/mnras/stv1152}; \textbf{13}~\cite{2017MNRAS.467..936G}; \textbf{14}~\cite{2005A&A...440..867G}; \textbf{15}~\cite{2008A&A...486..347G}; \textbf{16}~\cite{2017ApJ...841...71G}; \textbf{17}~\cite{1999NewA....4..141G}; \textbf{18}~\cite{2000NewA....5..335G}; \textbf{19}~\cite{2006AN....327..563G}; \textbf{20}~\cite{2009A&A...507.1257G}; \textbf{21}~\cite{2001A&A...376..803G}; \textbf{22}~\cite{10.1093/mnras/stu1669};  \textbf{23}~\cite{10.1111/j.1365-2966.2012.21519.x}; \textbf{24}~\cite{2017MNRAS.472..940K}; \textbf{25}~\cite{2019MNRAS.486L..80K}; \textbf{26}~\cite{Liang_2000}; \textbf{27}~\cite{Lindner2014}; \textbf{28}~\cite{2013A&A...551A.141M}; \textbf{29}~\cite{2014A&A...565A..13M}; \textbf{30}~\cite{Orr__2007}; \textbf{31}~\cite{2017MNRAS.464.2752P}; \textbf{32}~\cite{Pearce2017}; \textbf{33}~\cite{1997MNRAS.290..577R}; \textbf{34}~\cite{2018Ap&SS.363..133S}; \textbf{35}~\cite{2014MNRAS.440.2901S}; \textbf{36}~\cite{2015MNRAS.449.1486S};  \textbf{37}~\cite{2001AJ....122.1172S}; \textbf{38}~\cite{2019MNRAS.489.3905S};  \textbf{39}~\cite{2000MNRAS.314..594V}; \textbf{40}~\cite{2003A&A...402..913V}; \textbf{41}~\cite{2007A&A...463..937V}; \textbf{42}~\cite{2013A&A...551A..24V}; \textbf{43}~\citet{ASKAPJ0638}; \textbf{44}~\cite{2020A&A...636A...3X}; \textbf{45}~\citet{2011A&A...533A..35V}; \textbf{46}~\cite{Knowles2020}; \textbf{47}~\cite{2020ApJ...900..127H}; \textbf{48}~\cite{2005A&A...444..157F}.
} 
}

A key aspect of radio observations of galaxy clusters is the detection of diffuse cluster-scale synchrotron emission, which carries information about the cluster formation history \citep[see][for observational and theoretical reviews, respectively]{2019SSRv..215...16V,2014IJMPD..2330007B}. There are several different \editout{classifications}\lr{categories} of diffuse cluster radio emission, historically separated into three main classes: radio halos, mini-halos, and radio relics. {All classes are characterised by low surface brightness and\lr{, typically,} steep radio spectra ($\alpha \lesssim -1.1$).} 

Radio halos are diffuse sources which cover scales greater than 500\,kpc, with many spanning Mpc scales. They are typically seen to have morphologies closely linked to those of the X-ray {emitting} ICM. Both individual studies \citep[e.g.,][]{2001MNRAS.320..365B,Lindner2014} and statistical studies of large samples \citep{2013ApJ...777..141C,Kale2015EGRHS,2021A&A...647A..51C} have shown a strong link between radio halos and particle re-acceleration following major cluster mergers{, as well as correlations between source radio power and cluster mass and thermal properties}.

Radio mini-halos are found in the central region of dynamically relaxed, cool-core clusters \citep[see][for a recent update]{2017ApJ...841...71G}. They are \lr{defined to be} smaller than radio halos, with projected sizes ranging from a few tens to a few hundreds of kpc, usually confined within cold fronts at the cluster centre \citep{2008ApJ...675L...9M}.  Mini-halo clusters always have a radio active brightest cluster galaxy (BCG) which is thought to provide at least {a fraction} of the seed electrons necessary to produce the diffuse emission \citep[see e.g.,][]{2020MNRAS.499.2934R}. Particle re-acceleration induced by gas sloshing is possibly the driving mechanism for mini-halos \citep[e.g.,][]{2011ApJ...743...16Z,2013ApJ...762...78Z}.

Radio relics are elongated Mpc-scale structures located at the periphery of merging galaxy clusters. Their observed properties, which include a high degree of polarisation \citep[see for instance the case of the Sausage cluster,][]{2010Sci...330..347V} are consistent with the idea that they are related to the presence of merger-induced shocks in the ICM. Double radio relics are found in a number of clusters, the prototype case being Abell~3667 \citep{1997MNRAS.290..577R}, and in some cases a radio halo is detected as well \citep[see e.g,][]{2012MNRAS.426...40B,Lindner2014}. Radio phoenix sources are a sub-class of relics thought to be related to revived fossil emission from radio AGN in the cluster region \citep{2019SSRv..215...16V}.

Radio halos and relics have been detected in an increasing number of merging clusters, over a broad range of cluster masses \citep[see][for recent updates]{2019SSRv..215...16V,2021A&A...647A..50C} and over a wide range of redshifts \citep{Lindner2014,2021NatAs...5..268D}. The detection of mini-halos, by contrast, remains limited, mainly due to observational constraints. \lr{In addition to the three main classes, more clusters} with very steep-spectrum filaments have been detected recently \citep[e.g., Abell\,2034,][]{2016MNRAS.459..277S}, \lr{requiring further investigation of the connection between the structures and particle reservoirs radio galaxies deposit into the ICM, and the effects of cluster merger events} \citep[see e.g.,][]{2017NatAs...1E...5V,2017SciA....3E1634D}. 

Of the 115 clusters observed in this Legacy Survey, 62 have some form of \editout{detected} diffuse cluster emission, with several clusters hosting more than one diffuse source. Table \ref{tab:diffuse}, found at the end of the paper, presents the list of all 99 diffuse cluster structures {or candidates} detected in the survey. Fifty-six of these are new. For each diffuse source we list the emission classification, as well as angular and physical projected sizes, and position relative to cluster centre (where relevant). {Classifications are based on a combined interrogation of the full- and 15{\arcsec}-resolution MeerKAT data products, a 25{\arcsec}-resolution filtered image (see Section~\ref{sec:uc-scales} for details), and any available optical and X-ray imaging.} {
Candidate structures are those with either marginal detections or where the classification is uncertain. Where a diffuse source does not fit into any of the current classes, but is not clearly associated with an individual radio galaxy, 
we classify it as `unknown'.} \editout{A full analysis of all diffuse cluster sources, including flux densities, spectral index, and radio power measurements, will be presented in a follow-up paper (Kolokythas et al., in prep.).} Our diffuse cluster emission detections can be summarised as follows: \editout{two}\kk{three} new mini-halos and seven new mini-halo candidates, \kk{27}\editout{26} halo detections and six candidates (of which 13 are new), 28 relics and 18 candidates (of which 26 are new), one known phoenix source and two new candidates, and nine diffuse sources, six of which are new, with ambiguous or unknown classifications. 

The galaxy clusters observed in the MGCLS provide just a glimpse of the many diffuse cluster emission discoveries likely to be made in the Square Kilometre Array era. In the following, we present a few examples to show the much improved images compared to previous observations, opening up new areas of investigation, as well as \editout{new} discoveries \lr{highlighting specific interesting science issues}. Flux densities for the diffuse emission in these systems have been measured by integrating the surface brightness {within the 3$\sigma$ contour} in the 15{\arcsec}{-resolution} image and then subtracting the compact source contributions, which are determined from the full resolution image.
{In some cases the 25{\arcsec}-resolution filtered image shows a greater extent to the diffuse structures than the 15{\arcsec}-resolution products; the flux densities quoted here may be considered as lower limits to the total amount of diffuse emission.} 


\subsection{\lr{New insights into known sources}}

\kk{Twenty-eight of the MGCLS targets are hosts of previously known diffuse cluster emission. In many cases the MGCLS data provide an additional frequency or deeper detections of low surface brightness emission, yielding new insights into well-known sources (e.g., El~Gordo, Abell~3376). Here we provide three examples.}
\editout{to emphasise the potential of this survey to improve our understanding of well-studied non-thermal sources.}

\subsubsection{Abell 85: A new type of halo?}\label{sec:de-a85}
\begin{figure}
   \centering
   \includegraphics[width=0.95\columnwidth]{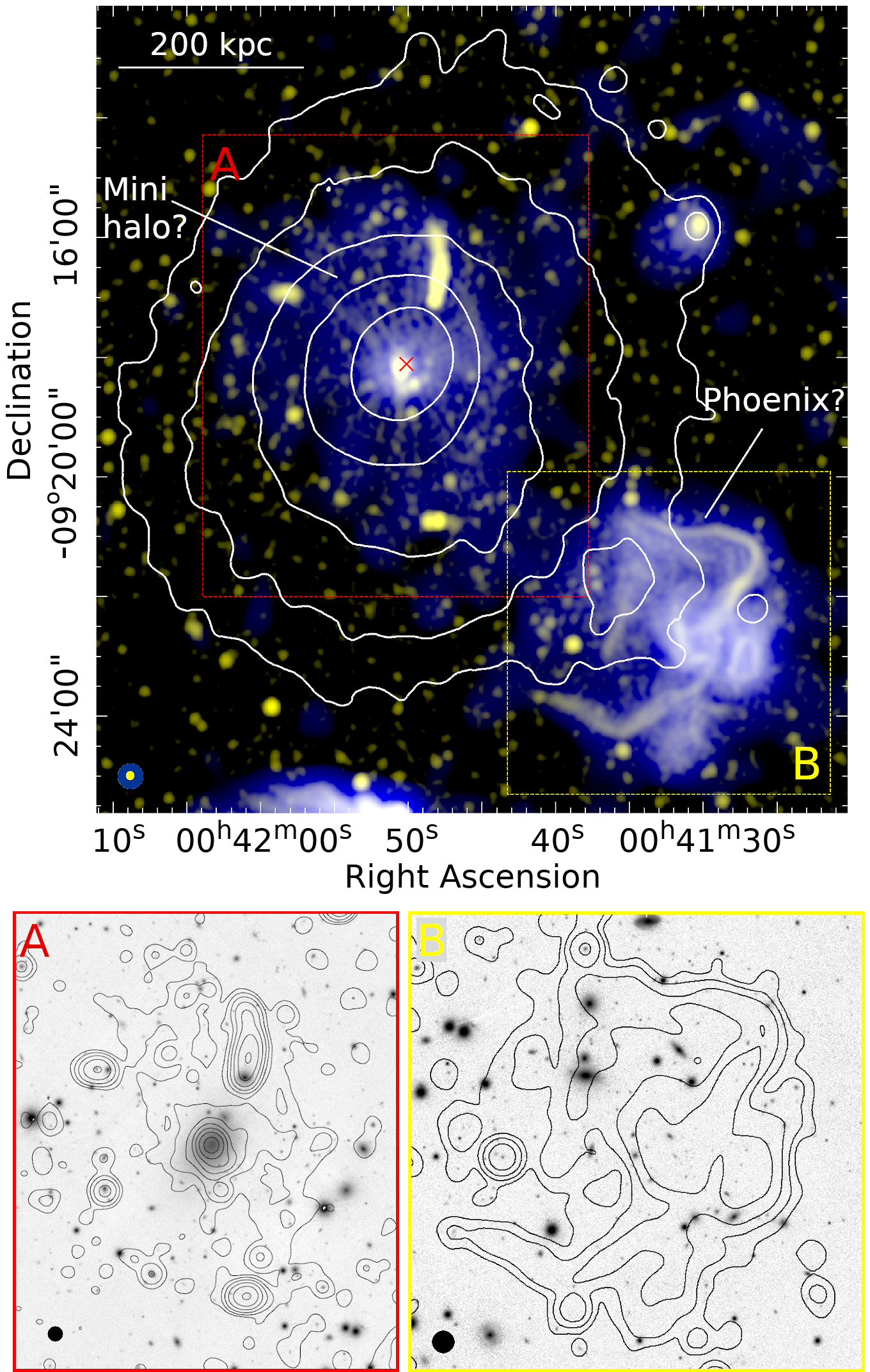}
   \caption{Multi-resolution MGCLS view of Abell~85. \textbf{Top:} Full-resolution (7.7\arcsec\,$\times$\,7.1\arcsec; yellow, logscale between 0.006 and 2\,mJy\,beam\per) and filtered 25\arcsec-resolution (blue, logscale between 0.06 and 3\,mJy\,beam\per) Stokes~I intensity image of Abell~85, \lr{with the  bright BCG and tailed radio galaxy emission filtered out}. Smoothed archival \textit{Chandra} \kk{0.5--7\,keV} X-ray contours (levels: 1.0, 1.7, 3.4, 6.8, 13.6\,$\times\,10^{-7}$\,counts\,cm$^{-2}$\,s\per) \lr{are} overlaid in white. The respective synthesised MGCLS beams are shown at lower left. The physical scale at the cluster redshift is indicated at upper left, and the red $\times$ marks the NED cluster position. A newly detected elongated \kk{mini-}halo (largest linear size\,$\sim$\,370\,kpc) is seen around the cluster BCG. Filamentary structures are seen in the known phoenix/revived fossil plasma source SW of the cluster\editout{, with the blue cross indicating the position of the likely optical counterpart}. \textbf{Bottom:} PanSTARRS $r$-band images of the boxed regions from the top panel, with MGCLS 15\arcsec-resolution contours overlaid. Contour levels start at 3\,$\sigma = 30\,\mu$Jy\,beam\per\ and increase by factors of 3. \editout{The blue circle in box B marks the likely optical counterpart for the phoenix source.}}
   \label{fig:abell85}%
\end{figure}

The MGCLS multi-resolution view of the \object{Abell~85} system, a cool-core cluster at $z=0.0556$ and part of the Pisces-Cetus supercluster, is shown in the top panel of Figure~\ref{fig:abell85}. We detect two diffuse sources\editout{in this system}: a complex phoenix or revived fossil plasma source southwest of the cluster centre \kk{(previously known)}, and a newly \editout{detected}\kk{discovered} elongated radio \kk{mini-}halo \kk{which may be an example of a new class of sources which represents an evolutionary bridge between mini-halos and halos}.

\kk{The newly detected diffuse source in Abell~85 surrounds the BCG, with contours shown in panel~A of Figure~\ref{fig:abell85}, and has a 1.28~GHz flux density of $\sim$\,5.5\,mJy measured from the 15{\arcsec}-resolution image. The 25{\arcsec}-resolution image, also shown in Figure~\ref{fig:abell85}, reveals a much greater extent to this source, with a largest angular size of 5.6\arcmin\ ($\sim$\,370\,kpc).} \editout{The bright emission of the BCG and tailed radio galaxy north of it has been filtered out before convolving to 25{\arcsec} resolution.}\kk{ This}\editout{new candidate halo} 
\kk{structure has not been detected in any of the low resolution radio data available in the literature at other frequencies \citep[see][and references therein]{2021PASA...38...10D}, possibly due to its low surface brightness ($\sim$\,1.1\,$\mu$Jy\,arcsec$^{-2}$ in the diffuse emission filtered  25\arcsec-resolution image) and its proximity to the 27\,mJy BCG and 13\,mJy tailed source.}
\kk{Given its size and location in a cool-core system, w}e classify this new detection\editout{, elongated in the NE-SW direction and which fills the inner core of the X-ray emission,} as a \kk{mini-halo. We note, however, that the MGCLS contains discoveries of several other examples of elongated diffuse halos with embedded central radio-loud BCGs (indicated by $^\dagger$ in Table \ref{tab:diffuse} at the end of the paper), some of which are on Mpc scales. We may be observing a new type of diffuse structure, bridging both halo and mini-halo classifications. A handful of cool-core clusters have been found to host larger scale radio halos \citep{Bonafede2014,2017MNRAS.466..996S}, and in each case the cluster shows signs of merger activity in addition to the cool X-ray core, as is the case for Abell~85 \citep{2016ApJ...831..156Y}. }

The phoenix source \editout{is well-known and }has been studied at multiple frequencies \citep{1984PASAu...5..516S, 1994AuJPh..47..145S, 2000NewA....5..335G, 2001AJ....122.1172S, 2021PASA...38...10D}, \editout{with the classification varying between}\lr{variously classified as }a revived fossil plasma source, a phoenix, and a relic. At full resolution (yellow in Figure~\ref{fig:abell85}), MeerKAT images this source in unprecedented detail, resolving a large number of filamentary structures extending from a brighter complex core region. \kk{At the filtered  25\arcsec-resolution (blue in Figure~\ref{fig:abell85}) we recover the full angular extent of the source as seen at lower frequencies \citep{2021PASA...38...10D}}. The diffuse emission extends far from the filaments, suggesting that some form of distributed re-acceleration is necessary.

The source's complex morphology could result from recurrent episodes of AGN activity, \kk{although determining the host galaxy(ies) for these structures is not trivial. Several potential hosts can be seen in the PanSTARRS $r$-band image \editout{in the bottom right panel of} in Figure~\ref{fig:abell85}. Sub-arcsecond resolution imaging is needed to identify the radio core(s), if present, in order to cross-match with optical counterpart(s)}.\editout{The first indications of the filamentary structure were seen in, but only for the brightest inner regions of the source.} The toroidal structure in the western region of the source may indicate late stage radio-mode feedback in a non-central cluster galaxy source \citep[such as in e.g., M87,][]{2013ApJ...775..118N,2018ApJ...855..128W}\kk{, possibly}\editout{. For example, the structure could be} caused by an evolved radio lobe that has been transformed into a ``smoke-ring''-like feature \kk{\citep{2002MNRAS.331.1011E}}. \editout{In this evolved AGN picture, the presumed host would be WISEA~J004130.92$-$092233.1, indicated by the blue cross in the top panel and the blue circle in the PanSTARRS $r$-band image in the bottom right panel of Figure~}
Such vortex structures can travel much further than amorphous plasma structures can, as shown by hydrodynamical simulations \citep{doi:10.1098/rspa.1957.0022}, and the radio features here {could therefore be} at a more advanced stage than the radio emission in M87. The buoyancy produced by this off-centre bubble is sufficiently powerful to uplift the ICM, visible in the \textit{Chandra} X-ray contours in Figure~\ref{fig:abell85}. 
\editout{On either side of the presumed host, f}\kk{F}ilamentary wings of radio emission trail the bubble for about 200\,kpc (at the cluster redshift). The filaments are quite narrow (10--20\,kpc), which constrains transport parameters for the enclosed cosmic-ray electrons, as well as ambient turbulent motions that would disrupt such structures \citep{2017SciA....3E1634D}. Such filaments can survive in turbulent motions provided that there are Reynolds stresses of magnetic fields that thread the wings \citep{2018MNRAS.473.3454B}. Determining the Alfv\'{e}n scale --- the physical scale at which the magnetic tension starts to play a role in turbulent dynamics --- may assist in clarifying their role. \editout{We also recover the full angular extent of the source from the diffuse emission filtered  25\arcsec-resolution MeerKAT image, as has been seen at lower frequencies  The diffuse emission is indicated in blue in the top panel of Figure~}
\editout{. The diffuse emission extends far from the filaments, suggesting that some form of distributed re-acceleration is necessary.}
\editout{The complex morphology of the phoenix source might also indicate recurrent episodes of AGN activity.} 

\editout{A spectral index study could help constrain the ages of the cosmic-ray population.} \lr{Spectral information can help constrain the history of the cosmic-ray population in the phoenix source.}  From the full source volume, we measure a flux density of $S_{\rm 1283\,MHz} = 242.4 \pm 12.1$\,mJy. Comparing to a flux density of $S_{\rm 148\,MHz} = 9.82 \pm 0.98$\,Jy within the same solid angle from TGSS \citep{TGSS}, we obtain a \editout{substantially flatter, but still }steep integrated spectrum of $\alpha^{1283}_{148} = -1.7 \pm 0.3$ for the full source. This latter value is in agreement with the 148--300\,MHz spectral index of $-1.85 \pm 0.03$ derived by \citet{2021PASA...38...10D}. The brightest regions of the complex structure have sufficient SNR across the MeerKAT bandwidth to determine in-band spectral indices, and we find values\footnote{Statistical uncertainties on the in-band spectral index are less than 0.05 in these regions.} ranging from $-2.9$ to $-3.5$. The steep spectra of the diffuse and filamentary structures indicate old electron populations, however the cause of the extreme steep spectrum in the central regions is unclear. There may be an artificial steepening effect from the frequency-dependent \emph{uv}-coverage; \lr{ although this should not be important in the brightest fine scale structures, this requires further investigation.}


\subsubsection{RXC J0520.7$-$1328: \kk{Revealing a diffuse multiplex}}\label{sec:de-j0520}

\begin{figure}
   \centering
   \includegraphics[width=0.98\columnwidth]{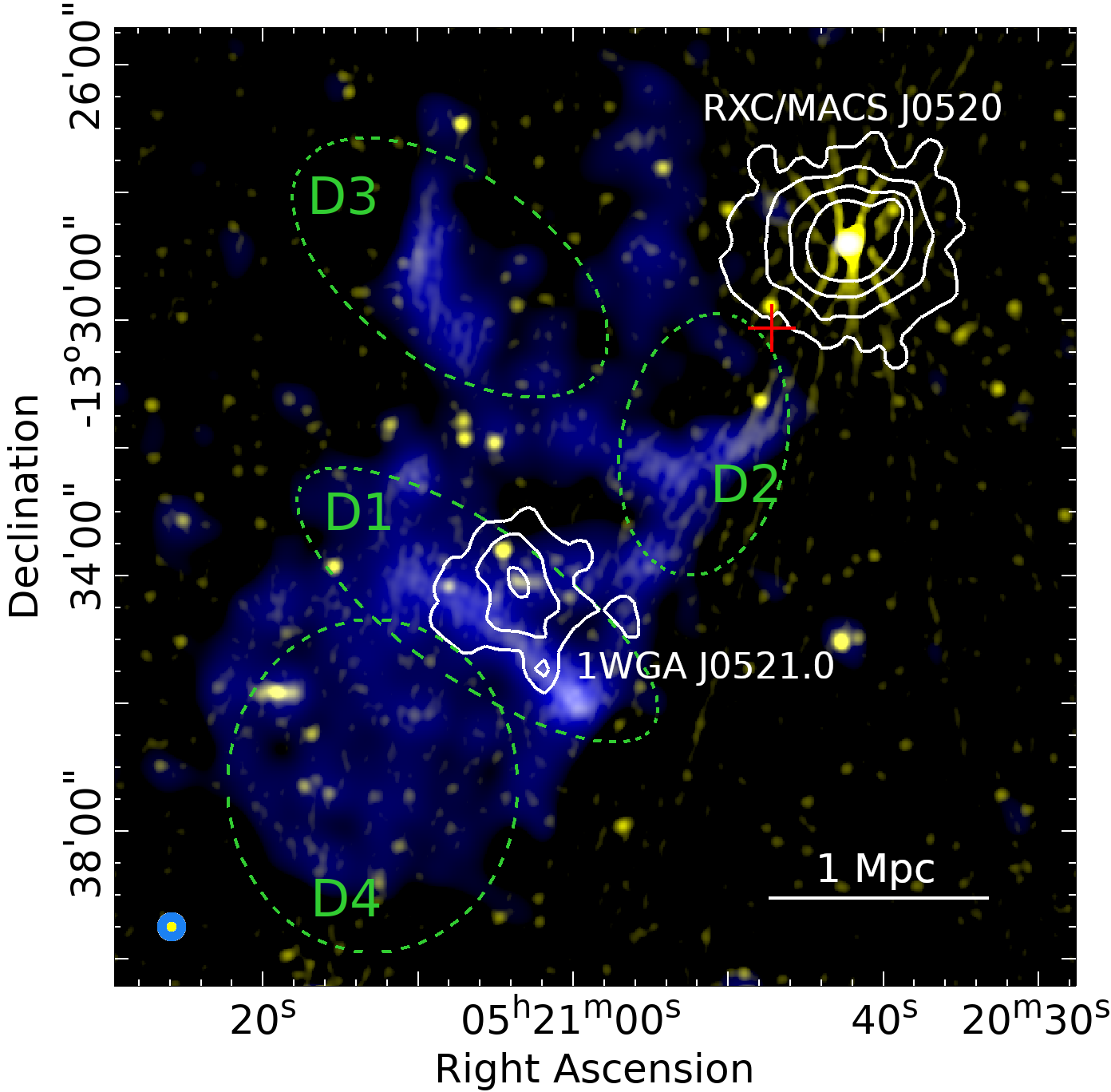}
   \caption{MGCLS full-resolution (8.6\arcsec; yellow) and diffuse emission filtered  25\arcsec-resolution (blue) Stokes~I intensity image of a portion of the RXC~J0520.7$-$1328 field, showing complex extended emission within the region. The corresponding synthesised beams are shown at lower left by the filled yellow and blue ellipses, respectively. The colour scales are in square root scaling between 10--500\,$\mu$Jy\,beam\per\ (yellow), and 20--500\,$\mu$Jy\,beam\per\ (blue). The physical scale at the cluster redshift is indicated at lower right, and the cross marks the NED cluster position. Smoothed archival \textit{Chandra} \kk{0.5--7\,keV} X-ray contours (levels: 3, 6, 12, 24\,$\times\,10^{-8}$\,counts\,cm$^{-2}$\,s\per) are overlaid in white, with RXC~J0520.7$-$1328 and its companion cluster 1WGA~J0521.0$-$1333 labelled. Dashed green ellipses indicate the regions of diffuse emission, D1 to D4, identified in \citet{2014A&A...565A..13M}. The full-resolution MGCLS image contains artefacts which radiate from the bright compact source in RXC~J0520.7$-$1328.  }
   \label{fig:j0520}%
\end{figure}

\lr{The MGCLS reveals complex structures in the previously studied} cluster \object{RXC\,J0520.7$-$1328} \citep[also known as MACS\,J0520.7$-$1328,][]{MACS} \kk{ that do not obviously fit any of the existing paradigms.}  RXC\,J0520.7$-$1328, at $z=0.336$, is part of a possible cluster pair, with its companion, 1WGA\,J0521.0$-$1333, lying to the southeast at a redshift of $z=0.34$ \citep{2014A&A...565A..13M}. Multi-resolution MGCLS images of this pair, presented in Figure~\ref{fig:j0520}, show a large, complex, diffuse source southeast of RXC\,J0520.7$-$1328, coincident in part with 1WGA\,J0521.0$-$1333. The brightest portion of the radio emission is a bar-like feature seen at both full (8.6\arcsec; yellow) and 25{\arcsec} (blue) resolution, at the southeast edge of 1WGA\,J0521.0$-$1333's X-ray emission. This feature was observed with the GMRT at 323\,MHz \citep{2014A&A...565A..13M}, along with three other distinct regions of faint diffuse emission \citep[labels D2 to D4 in Figure~\ref{fig:j0520}, taken from][]{2014A&A...565A..13M}. Although characterised as a relic by \citeauthor{2014A&A...565A..13M}, the bar's southwest end is significantly brighter in the full-resolution MeerKAT image than the rest of the feature, and could instead be a head-tail galaxy. Although no MeerKAT polarisation maps are available for this system, spatial spectral index studies may be able to disentangle these two possibilities. 

The filtered  25\arcsec-resolution MGCLS image shows that the bar-like feature is connected to two other low-surface-brightness structures: a filled circular `bubble' overlapping the southeast region of 1WGA\,J0521.0$-$1333, coincident with the 323\,MHz D4 source, and a partial ring or arc of radio emission lying between the two clusters (incorporating sources D2 and D3). MeerKAT reveals much more detail and extent to the complex emission seen at lower frequency. It also reveals a new faint structure ($\sim$\,0.8\,$\mu$Jy\,arcsec$^{-2}$ surface brightness) to the east of RXC\,J0520.7$-$1328. Despite the improved sensitivity, due to the location and morphology of the structures, it remains difficult to classify the majority of the emission in terms of typical cluster-related nomenclature. A more detailed analysis, including deeper X-ray data, polarimetry, and spectral index studies, will be required to resolve the nature of this emission.

\subsubsection{\kk{Abell 3667: Polarisation with MGCLS}}\label{sec:de-pol}

\begin{figure}
   \centering
   \includegraphics[width=0.95\columnwidth]{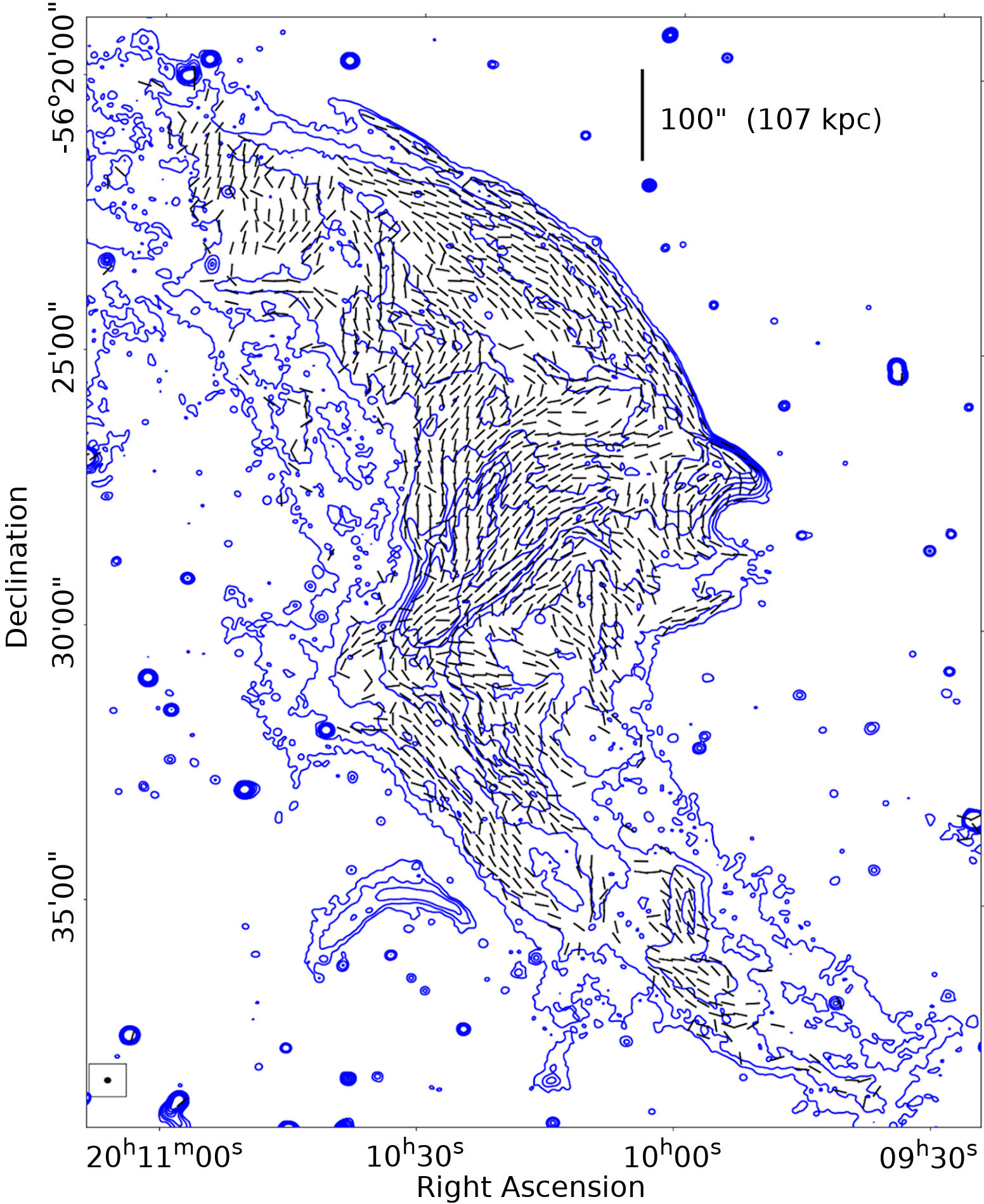}
   \caption{Inferred magnetic field angles for the northwest relic in Abell~3667, after correction for Faraday rotation (which is shown in Figure \ref{fig:A3667RM}). Blue contours show the full-resolution (7.5\arcsec\ $\times$ 7.4\arcsec) Stokes~I MGCLS intensity, with the synthesised beam shown in the boxed region at lower left. Contour levels are 10, 20, 40, 60, 80, 100\,$\times\,1\sigma$, where $\sigma = 6.7\,\mu$Jy\,beam\per. Magnetic field vectors (angles only) are shown where the linearly polarised intensity is $>$\,8$\sigma$. The magnetic field is highly ordered along the NW edge of the relic.
   }
   \label{fig:A3667PA}%
\end{figure}



Radio relics are among the most polarised radio sources known, reaching 70\% polarisation levels in localised regions \citep{2010Sci...330..347V, 2017MNRAS.472.3605L}. Here we \kk{showcase the MGCLS dataset's ability to probe polarisation structure in diffuse sources on $\sim$\,10\arcsec\ scales by} present\kk{ing} initial \editout{MGCLS }results from polarised observations of the north-west radio relic in the Abell~3667 galaxy cluster ($z = 0.056$). This \kk{relic} is one of the most impressive examples of radio emission generated by a merger shock wave \citep{1997MNRAS.290..577R, 2003PhDT.........3J}.  \editout{Here we find that the magnetic field structures are consistent with those expected from merger shock waves.} \editout{A complete discussion on this source will be presented in de Gasperin et al. (in prep.). } The shock wave \editout{that generates the relic source} propagates in the ICM with a speed of 1200\,\kms, at roughly two times the sound speed \citep{2010ApJ...715.1143F}. The polarisation is maximal close to the shock front, as well as in other regions of the radio relic where the magnetic field is likely compressed and aligned by the action of the shock wave \citep{1998A&A...332..395E}. 

In Figure~\ref{fig:A3667PA} we see that the magnetic field appears aligned with the NW edge of the radio relic at the location of the shock; this is seen in other radio relics as well \citep[see, e.g.,][]{2010Sci...330..347V,2015PASJ...67..110O}. We also observe a patchy, disordered structure to the magnetic field away from the edge; this is seen in other relics, and is expected as turbulence develops post-shock \citep{2017JKAS...50...93K}. In Abell~3667, we can rule out disorder due to Faraday rotation in the ICM, since we have removed these effects using the RMs shown in Panel~C of Figure \ref{fig:A3667RM}. 
At the same time, the RM is seen to \editout{decrease from the central regions of the relic towards the north}\lr{increase from the north towards the central regions of the relic}; 
this could be due to the subcluster identified in that region with X-ray observations \citep{2010ApJ...715.1143F}, if it is located in front of the radio emitting plasma.

\subsection{\kk{New clusters with diffuse emission}}

\kk{There are 34 MGCLS clusters in which we detect cluster diffuse emission for the first time. }\lr{Here we present three examples which raise interesting science issues and showcase the capabilities of the MGCLS.}
\subsubsection{J0352.4$-$7401: Multiple relics in a massive merger}\label{sec:de-j0352}

\begin{figure}
   \centering
   \includegraphics[width=0.9\columnwidth]{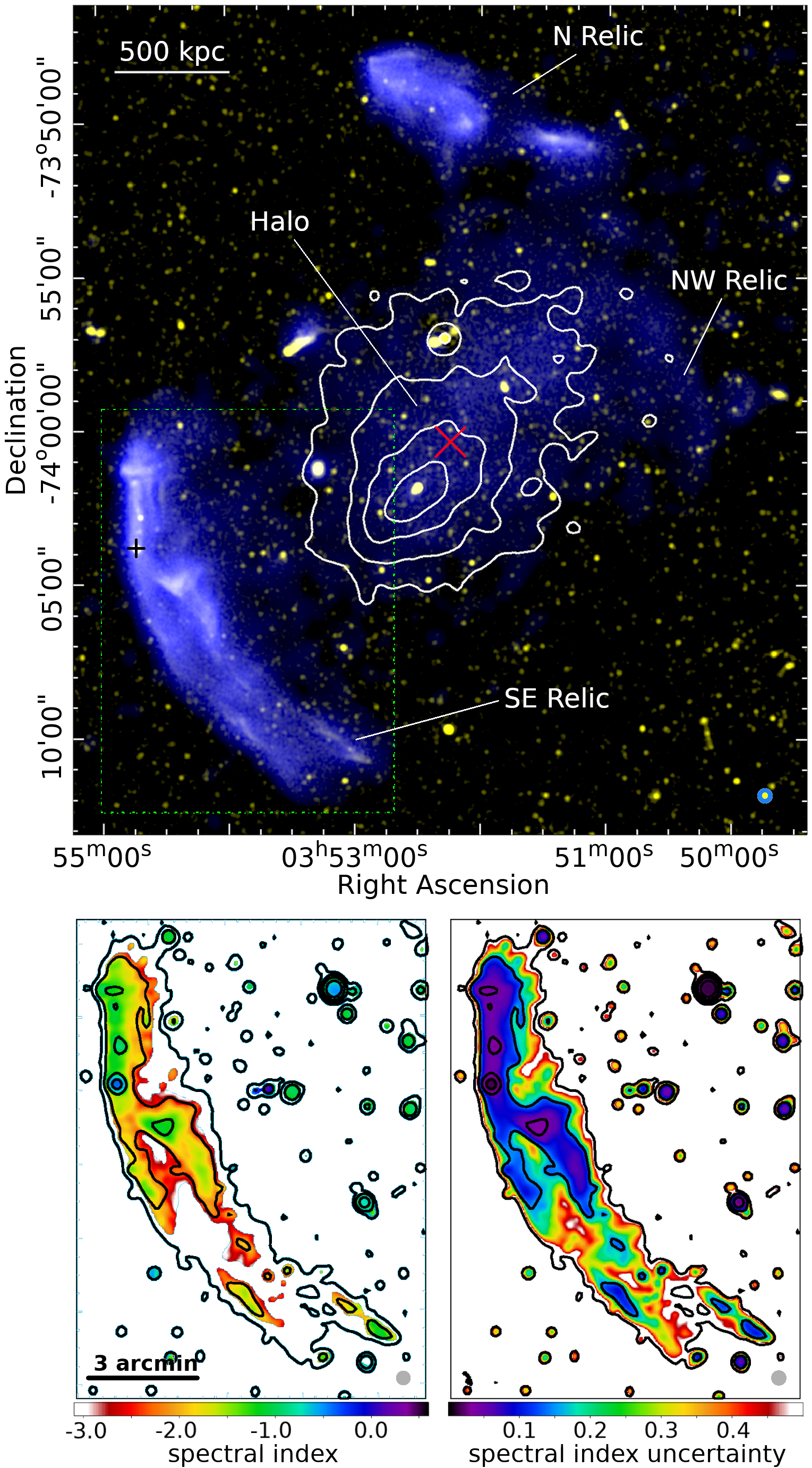}
   \caption{MGCLS view of the newly detected multiple-relic and radio halo system in MCXC~J0352.4$-$7401. \textbf{Top:} MGCLS full-resolution (7.0\arcsec$ \times$ 6.8\arcsec; yellow) and filtered  25\arcsec-resolution (blue) Stokes~I intensity images. The colour scales are in square root scaling between 50--500\,$\mu$Jy\,beam\per\ (yellow), and 5--300\,$\mu$Jy\,beam\per\ (blue). Synthesised MGCLS beams are shown lower right, and the physical scale at the cluster redshift is shown upper left. The red $\times$ marks the NED cluster position, and the black cross indicates the position of a 3\,mJy compact source coincident with the SE relic. Smoothed archival \textit{Chandra} \kk{0.5--7\,keV} contours (levels: 4, 8, 16, 30\,$\times$\,10$^{-7}$\,counts\,cm$^{-2}$\,s\per) are overlaid in white. \textbf{Bottom:} In-band spectral index (left) and spectral index uncertainty (right) maps of the dashed region from the top panel, with 15\arcsec-resolution MGCLS intensity contours overlaid. Contours start at $5\sigma = 40\,\mu$Jy\,beam\per\ and increase in factors of 4. The synthesised beam is indicated by the filled grey circle at lower right of each panel.}
   \label{fig:j0352}%
\end{figure}

The MGCLS has detected several new relics, many with large scale filamentary structure. One such example is the newly detected SE relic in \object{MCXC~J0352.4$-$7401} (Abell~3186, $z=0.1270$) shown in Figure~\ref{fig:j0352}. The relic, located approximately 1.2\,Mpc SE of the cluster centre, is very large, \editout{being}2\,Mpc long and 340\,kpc wide. \editout{The MGCLS full-resolution image, shown in yellow in the top panel of Figure~}
\editout{, reveals several notable features.} The brightest region of the SE relic is a straight bar, aligned N--S and spatially connected to a 3\,mJy compact source (marked by the black cross).  No optical counterpart is visible in the Digitized Sky Survey (DSS). The in-band spectral index map of the SE relic, \editout{presented in the bottom left panel of}\lr{also in} Figure~\ref{fig:j0352}, shows this compact source to have a significantly flatter spectrum ($\alpha \sim -0.4$) than that of the bar ($\alpha \sim -1.0$ to $-1.2$). \editout{In contrast to this linear region, t} The overall shape of the relic is curved, with the southern part consisting of two parallel filaments with complex structure. The filaments have steeper spectra than the northern bar, with in-band spectral indices of $-1.5$ at the centre of the filaments, steepening to $-2.5$ at the edges. The trailing edge of the relic is much wider in 25{\arcsec}-resolution imaging of the field (shown in blue in the top panel of Figure~\ref{fig:j0352}), with a maximum width of $\sim$\,580\,kpc. The SE relic has a measured 1.28~GHz flux density of $\sim$\,65\,mJy, which corresponds to a total radio power of $P_{\rm1.28\,GHz} \sim 2.8 \times 10^{24}$\,W\,Hz\per. 

This system also hosts a smaller 1.1~Mpc relic north of the cluster, as well as a large radio halo which fills the X-ray emitting region. The N relic has two components in the full resolution image and its morphology is strikingly linear, reminiscent of revived radio galaxies, although no obvious optical counterpart is seen in DSS images. Both the SE and N relics lie far beyond the observed thermal ICM (shown by white \textit{Chandra} contours in Figure~\ref{fig:j0352}). The large-scale radio halo, with a 1.28~GHz flux density of $\sim$\,20\,mJy, lies at the centre of the cluster, and its SE--NW elongated morphology closely follows that of the X-ray emission. This giant radio halo, with a largest linear size of 1.5\,Mpc, has a faint ($\sim 0.2\,\mu$Jy\,arcsec$^{-2}$) diffuse arc-like protrusion off its NW edge, mirroring the SE relic. As the shape and position of the arc-like structure may hint at the presence of another shock front, we classify it as a candidate relic.

The presence of the multiple relics, and their location with respect to the asymmetric X-ray emission, are indicative of a major cluster merger. Recent numerical simulations of galaxy cluster mergers suggest that a merger shock may gradually detach from the dense merging clumps \citep{2020MNRAS.494.4539Z}, and propagate to the cluster outskirts. There is observational radio evidence for the presence of such merger shocks in the periphery of clusters, for example in the Coma cluster \citep{1991A&A...252..528G,2011MNRAS.412....2B} and in Abell~2744 \citep{Pearce2017,2013A&A...551A..24V}. Given that the N and SE relics lie far beyond the detectable X-ray emitting region, MCXC~J0352.4$-$7401 \lr{may be in an advanced merger state}, with the merger-induced shocks having propagated into the periphery of the system. \editout{If this is indeed the case, we are currently observing a cluster in an advanced merger state.} We note that simulations also predict a weaker secondary shock front closer to the centre, now seen in the Coma cluster \citep{2021A&A...651A..41C}. The NW candidate relic seen in MCXC~J0352.4$-$7401 may be evidence for such a shock.

There are only a handful of double-relic clusters which also host a radio halo: e.g., El~Gordo \citep{Lindner2014}; CIZA~J2242.8$+$5301 \citep{2010Sci...330..347V,2018ApJ...865...24D}; MACS~J1752.0$+$4440 \citep{2012MNRAS.425L..36V}; MACS~J1149.5$+$2223 \citep{2012MNRAS.426...40B}. Double-relic clusters are an important subclass of merging clusters, as the merger is observed close to the plane of the sky. Systems such as MCXC~J0352.4$-$7401 can therefore provide insights into the dynamical state of the halo emission, along with morphological and spectral properties of the transition from diffusive shock acceleration to second order turbulent re-acceleration. In addition to MCXC~J0352.4$-$7401, there are six halo$+$double-relic systems in the MGCLS: known systems (El~Gordo, RXC~J1314$-$2525), and four new ones (Abell~521, MCXC~J0516.6$-$5430, MCXC~J0232.2$-$4420, and RXC~J2351.0$-$1954).

\subsubsection{J0351.1$-$8212: ``Boomerangs'' and an off-centre mini-halo?}\label{sec:de-j0351}
\begin{figure}
   \centering
   \includegraphics[width=0.9\columnwidth]{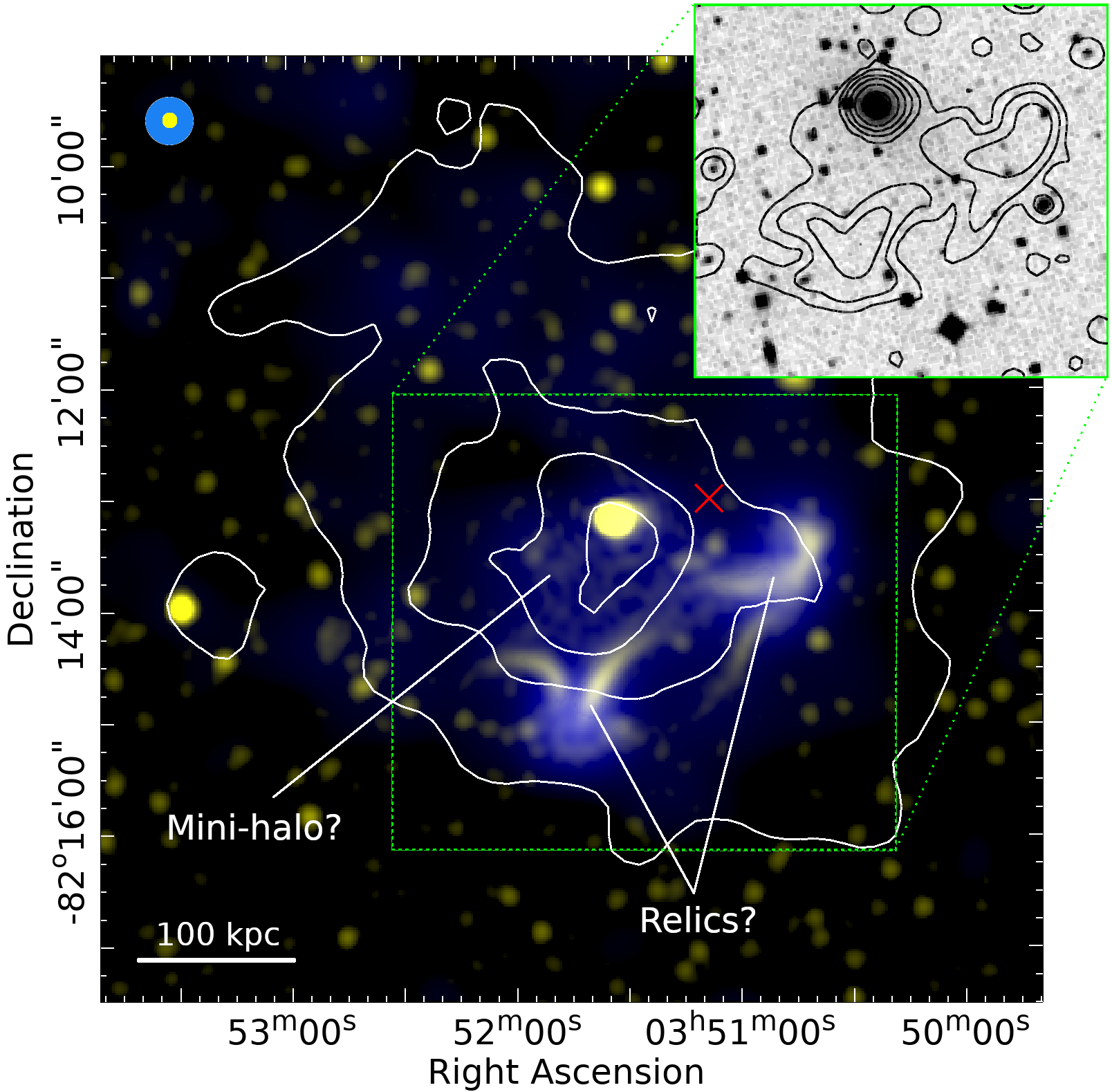}
   \caption{MGCLS full-resolution (7.9\arcsec$ \times$ 7.5\arcsec; yellow) and filtered  25\arcsec-resolution (blue) Stokes~I intensity image of MCXC~J0351.1$-$8212, showing newly detected candidate mini-halo and relic-like structures near the cluster centre. The corresponding synthesised beams are shown at upper left. The colour scale is in square root scaling between 10--500\,$\mu$Jy\,beam\per\ (yellow), and 50--800\,$\mu$Jy\,beam\per\ (blue). Smoothed archival \textit{Chandra} \kk{0.5--7\,keV} X-ray contours (levels: 7, 12, 17, 22\,$\times\,10^{-8}$\,counts\,cm$^{-2}$\,s\per) are overlaid in white. The physical scale at the cluster redshift is indicated at lower left, and the red $\times$ marks the NED catalogue position. \textbf{Inset:} DSS $r$-band image of the dashed region in the main figure, with 15\arcsec-resolution MGCLS contours overlaid. Contours start at $3\,\sigma = 30\,\mu$Jy\,beam\per\ and increase in factors of 3. 
   }

   \label{fig:rg-buoyancy}%
\end{figure}
\kk{\object{MCXC~J0351.1$-$8212}, a nearby system at $z = 0.0613$, is one of the newly discovered hosts of diffuse cluster emission, however the structures seen in this system present a confusing picture.} We detect a new mini-halo-like structure around\kk{, but offset from,} the BCG\editout{of the nearby ($z = 0.0613$) cluster MCXC~J0351.1$-$8212}. The source has a 1.28~GHz flux density of only 2\,mJy and {our preliminary estimate of the maximum linear extent in the sky plane is} 240\,kpc. The multi-resolution MGCLS image in Figure~\ref{fig:rg-buoyancy}, with white \textit{Chandra} contours overlaid, shows the diffuse radio emission to be primarily SW of the BCG and X-ray peak, instead of being centreed thereon. The size, and therefore flux density, of this possible mini-halo are difficult to clearly determine due to the presence of several brighter filamentary ``boomerang''-shaped sources less than 100\,kpc from the BCG, visible in both the full-resolution (7.9\arcsec$ \times$ 7.5\arcsec) Stokes~I intensity map shown in yellow, and at 15\arcsec-resolution, indicated by black contours on the DSS $r$-band image of the region shown in the figure inset. The filaments do not appear to have obvious optical counterparts in DSS, and have very steep spectra ($\alpha < -2.5$) in our initial spectral maps (not shown). Going out from the X-ray peak, there is a significant decrease in the brightness from \textit{Chandra} data at the position of the filaments.  We therefore tentatively classify them as relics, although the small sizes and physical proximity to the BCG make the classification unclear. 

We note the similarity of the filamentary structures to those in Abell~133 \citep{2001AJ....122.1172S,2010ApJ...722..825R}. In that case, the filaments cap a buoyant blob of AGN material from the BCG, which has dragged up cool thermal material from the cluster core. A similar process may be at work in MCXC~J0351.1$-$8212, although why there would be two such filamentary caps is unclear. If this is another case of buoyant lifting of the radio and X-ray plasmas, then the mini-halo-like structure may in fact be the remnant of the radio lobe(s). \kk{Deeper optical imaging and investigation of the spectral shape of the various diffuse and filamentary components may be able to distinguish between the mini-halo/relic and AGN-related scenarios.}

\subsubsection{J0631.3$-$5610: \kk{Distant AGN or} faint relic?}\label{sec:de-relics}
\begin{figure*}
    \centering
    \includegraphics[width=0.98\textwidth]{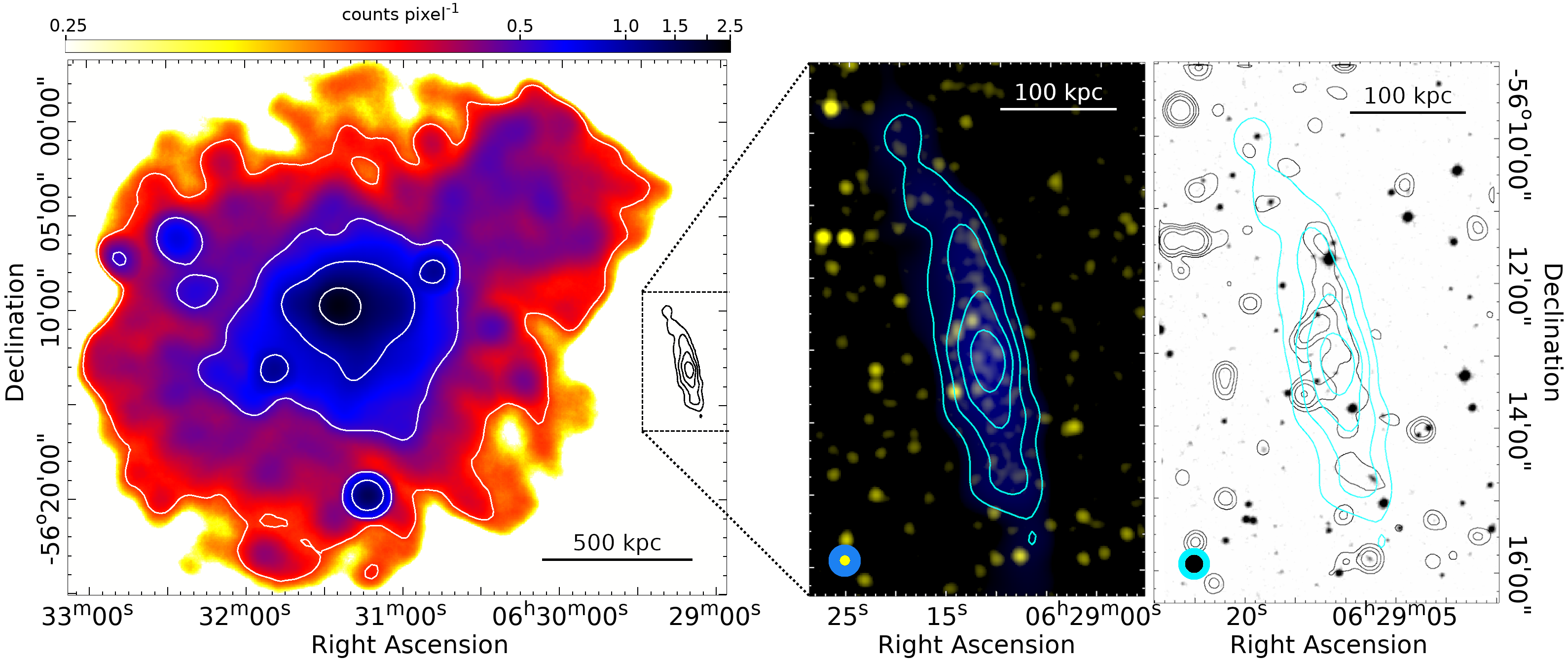}
    \caption{A faint relic-like structure in the MCXC~J0631.3$-$5610 ($z=0.054$) cluster field. \textbf{Left:} Smoothed archival \textit{XMM-Newton} \kk{0.2--12\,keV} counts image of the cluster, with white contours at levels of 0.3, 0.5, 0.9, and 1.7 counts per pixel. The position of the relic-like source relative to the cluster region is indicated, with filtered  25\arcsec-resolution MGCLS contours of the radio source, to the west, in black; the radio source appears to be 1.5\,Mpc from the cluster centre. \textbf{Middle:} Multi-resolution MGCLS image of the radio source (yellow - 7.6\arcsec\ $\times$ 7.5\arcsec resolution; blue - filtered  25\arcsec\ resolution), with cyan contours from the filtered  25\arcsec-resolution image (levels: 20, 40, 60, 80\,$\mu$Jy\,beam\per). The colour scale is in square root scaling between 8--200\,$\mu$Jy\,beam\per\ (yellow), and 5--300\,$\mu$Jy\,beam\per\ (blue). The physical scale at the cluster redshift is shown at upper right, and the synthesised MGCLS beams at lower left. \textbf{Right:} DSS $r$-band image of the same region as the middle panel. Cyan contours are the same as in the left inset, and black contours are from the 15\arcsec-resolution MGCLS image with levels of 3, 6, 10, 20\,$\times\,1\sigma$, where $\sigma = 10\,\mu$Jy\,beam\per. The synthesised MGCLS beams are shown at lower left. There appears to be no obvious optical counterpart for the radio source.}
    \label{fig:faintrelics}
\end{figure*}
One of the strengths of MeerKAT at L-band is its extreme sensitivity to faint extended emission. Figure~\ref{fig:faintrelics} shows an example of one of the fainter relic-like structures detected in this survey, with the left inset panel showing the full-resolution (7.6\arcsec\ $\times$ 7.5\arcsec) and filtered 25\arcsec-resolution MGCLS images in yellow and blue, respectively. The source, with a mean surface brightness of 6\,$\mu$Jy\,beam{\per} in the full-resolution map, has a largest angular size of 5.9\arcmin, revealed in the filtered 25{\arcsec}-resolution image. The source appears to be 1.5\,Mpc west of the \object{MCXC~J0631.3$-$5610} ($z=0.054$) cluster centre, when comparing to an X-ray image of the region shown in Figure~\ref{fig:faintrelics}. The total 1.28\,GHz flux density of the radio source is 2.8\,$\pm$\,0.2\,mJy, measured from the filtered 25\arcsec-resolution image. At the cluster redshift and assuming a conservative spectral index of $-1.0$ with a 10\% uncertainty, this corresponds to a $k$-corrected 1.4\,GHz radio power of ($1.7\,\pm\,0.3)$\,$\times\,10^{22}$\,W\,Hz{\per}.

The source has no obvious optical counterpart in the DSS $r$-band image of the region, shown in the right inset panel of Figure~\ref{fig:faintrelics}, and no redshift information is available for any of the WISE sources in the region. \kk{Although showing no evidence of a radio core, it is possible that this source is a distant dying radio galaxy with the available optical imaging too shallow to identify a counterpart. However, g}iven the morphology and orientation of the source relative to the X-ray emitting gas, it is possible that it is a relic or phoenix source related to the cluster. Its physical size would be $\sim$\,370\,kpc at the cluster redshift. If this is a relic source, its radio power makes it the lowest luminosity relic known, a factor of 5 below the lowest luminosity relic listed by \citet{2014MNRAS.444.3130D}. With the low X-ray-derived cluster mass of M$_{500}=1.3\times10^{14}$\,M$_\odot$ \citep{2019MNRAS.483..540L}, it lies just below the extrapolation of \citeauthor{2014MNRAS.444.3130D}'s luminosity-mass relation.  This reinforces and extends to lower masses the problem that these radio luminosities are significantly larger than those seen in simulations \citep[e.g.,][]{2020MNRAS.493.2306B}. Observational estimates of  the efficiency of shock acceleration \citep[e.g,][]{2020A&A...634A..64B} at relics also indicate improbably high values, given our current understanding.

\section{Illuminating Individual Radio Galaxies}\label{sec:individsrcs}
Among the many thousands of extended radio galaxies in the survey, we highlight a small group selected for the interesting science issues they raise.\editout{We describe some of the more intriguing examples below, while detailed investigations and modeling of these and other such sources are beyond the scope of the current paper.} The science issues include possible missing pieces of radio galaxy physics as well as complex interactions with the external medium that go beyond the simple relative motions that created tailed radio galaxies. For each source below, we indicate if it is a member of its respective target cluster; otherwise it should be considered as a serendipitous detection.

In this section we use optical overlays from the Dark Energy Camera (DECam) Legacy Survey (DECaLS), available through the Dark Energy Legacy Survey site\footnote{\url{https://www.legacysurvey.org}} and from the NOAO Astro Data Lab\footnote{\url{https://datalab.noao.edu/ls/ls.php}}, and overlays from PanSTARRs\footnote{\url{https://outerspace.stsci.edu/display/PANSTARRS/Pan-STARRS1+data+archive+home+page}}. \kk{An investigation of a machine learning tool to automatically identify interesting sources is discussed in Appendix C.}


\subsection{Lateral edge enhancement}

\begin{figure}
   \centering
   \includegraphics[width=0.98\columnwidth]{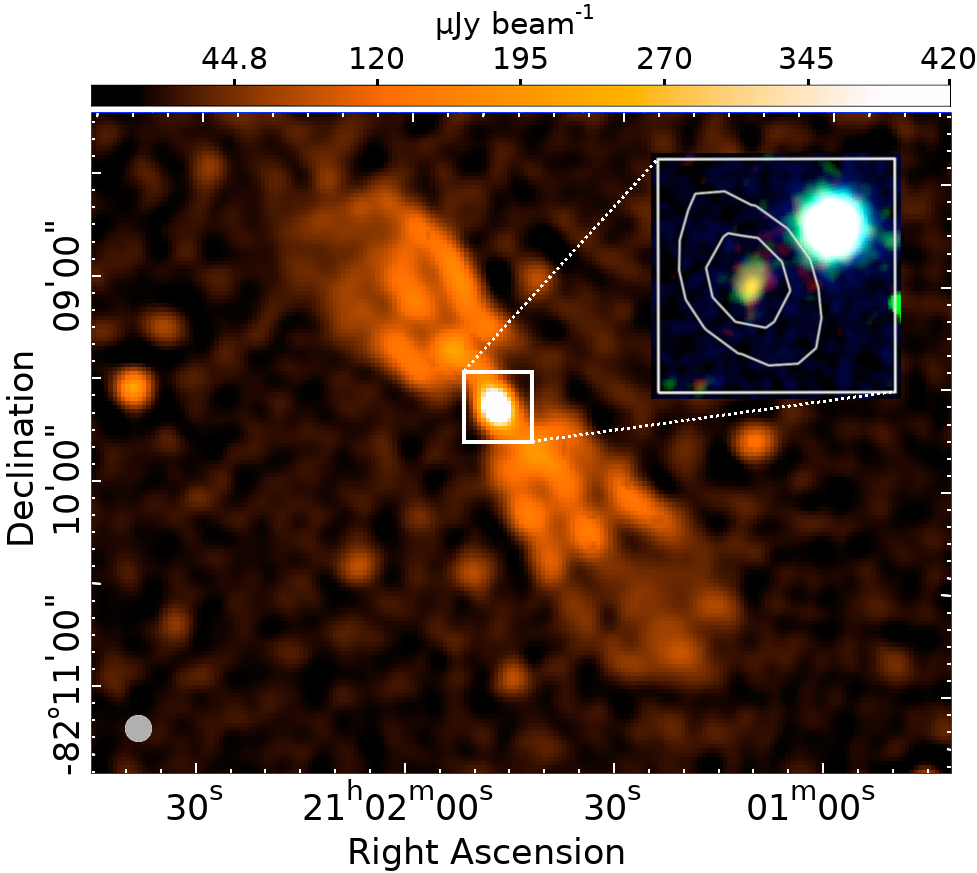}
   \caption{Full-resolution (7.6\arcsec\ $\times$ 7.4\arcsec) MGCLS Stokes~I intensity map of the \emph{LBS} source, a unique double-lobed radio galaxy with lateral edge-brightening. 
   The inset is a DECam RGB colour composite (using the $zir$-bands, respectively), with contours from the full-resolution MGCLS image showing the radio core centreed at $\mbox{R.A.} = \rm21^h01^m47.7^s$, $\mbox{Dec.} = \rm -82^\circ09^\prime36$\arcsec\ on a faint, irregularly shaped, very red object \lr{(which appears yellow-green in the $zir$ inset)}. The MGCLS synthesised beam is indicated at lower left.}

   \label{fig:rg-latedgeridges}%
\end{figure}

\lr{The laterally brightened source shown in Figure~\ref{fig:rg-latedgeridges} (hereinafter \emph{LBS}), is unique in terms of the brightening of its lateral edges, along with the presence of central jet-like features of similar length and brightness in each lobe. While} some radio galaxies have bright hot spot regions at their ends \citep[defined as Fanaroff-Riley type IIs,][]{FR}, brightened \emph{lateral} edges are not observed.  Instead, radio galaxy lateral edges are observed to either cut off sharply, e.g., when pressure confined,  or to fade slowly \citep[see, e.g., Fig. 34 in][]{1991AJ....102..537L}. \lr{\emph{LBS}} was found serendipitously in the field of cluster MCXC~J2104.9$-$8243. Here we describe some aspects of this unique source and some possibilities (none of them very attractive) for explaining the edge brightened features.

In most respects, \emph{LBS} appears normal.  Its host is a faint, red, irregularly shaped object seen in the DECaLS image (see inset in Figure~\ref{fig:rg-latedgeridges}). No redshift is available. The NE and SW lobes have total MGCLS flux densities of $4.0\pm0.8$ and $1.9\pm0.5$\,mJy, with lengths of 80\arcsec\ and 100\arcsec, respectively. The length:width ratio of the lobes is $\sim$2:1.  The source's R-ratio \citep[{$S_{\rm core}/(S_{\rm total}-S_{\rm core})$},][]{OrrBrowne1982}
is $\sim14\%$, within the normal range, and the core has a spectral index of $\sim-0.9$ 
indicating the likely presence of small-scale optically thin jets.  No reliable spectral indices could be determined for the lobes.

The lobes seem to divide abruptly  into three separate lengthwise narrow features (jets and edges) at an angular distance of $\sim$10\arcsec\ from the core. Most of the flux density appears to be in these narrow features, with little indication of more extended emission. Where the jets and enhanced edges are bright, they are each somewhat resolved transversely, with deconvolved widths of $\sim$5--15\arcsec.  Approximately half-way to the ends, the jets and bright edges \emph{all} drop by a factor of $\sim$\,2 in brightness, and the structure becomes less clear. 

Jets typically fade with distance from the core, as seen here, due to expansion or other energy losses. However, the fact that the transverse edges fade at the same distance from the core as the jets do is difficult to explain in  current models.  We briefly considered several possible explanations for this behaviour.  One possibility could be that the bright edges are due to backflow from the terminus of the jet.  However, this does not appear attractive because a) no hotspots are seen, which are expected if there is strong backflow, and b) any edge brightening due to backflows would be brighter near the end away from the host, while we observe the opposite.  Another possibility is that the bright edges are regions of strong magnetic fields, perhaps generated by shear with the external medium. However, once again, there is no reason for them to drop in brightness at the same distance from the host as the jets. A hollow cylinder could perhaps be invoked for the bright edges, similar to what is seen in some bipolar nebulae \citep{1972ApJ...174..583A},  but again, the correspondence with the jet profile appears fortuitous.  In short, we can provide no adequate explanation for this source's unique morphology.  


\subsection{Exceptionally stable bent jets}

\begin{figure}
   \centering
   \includegraphics[width=\columnwidth]{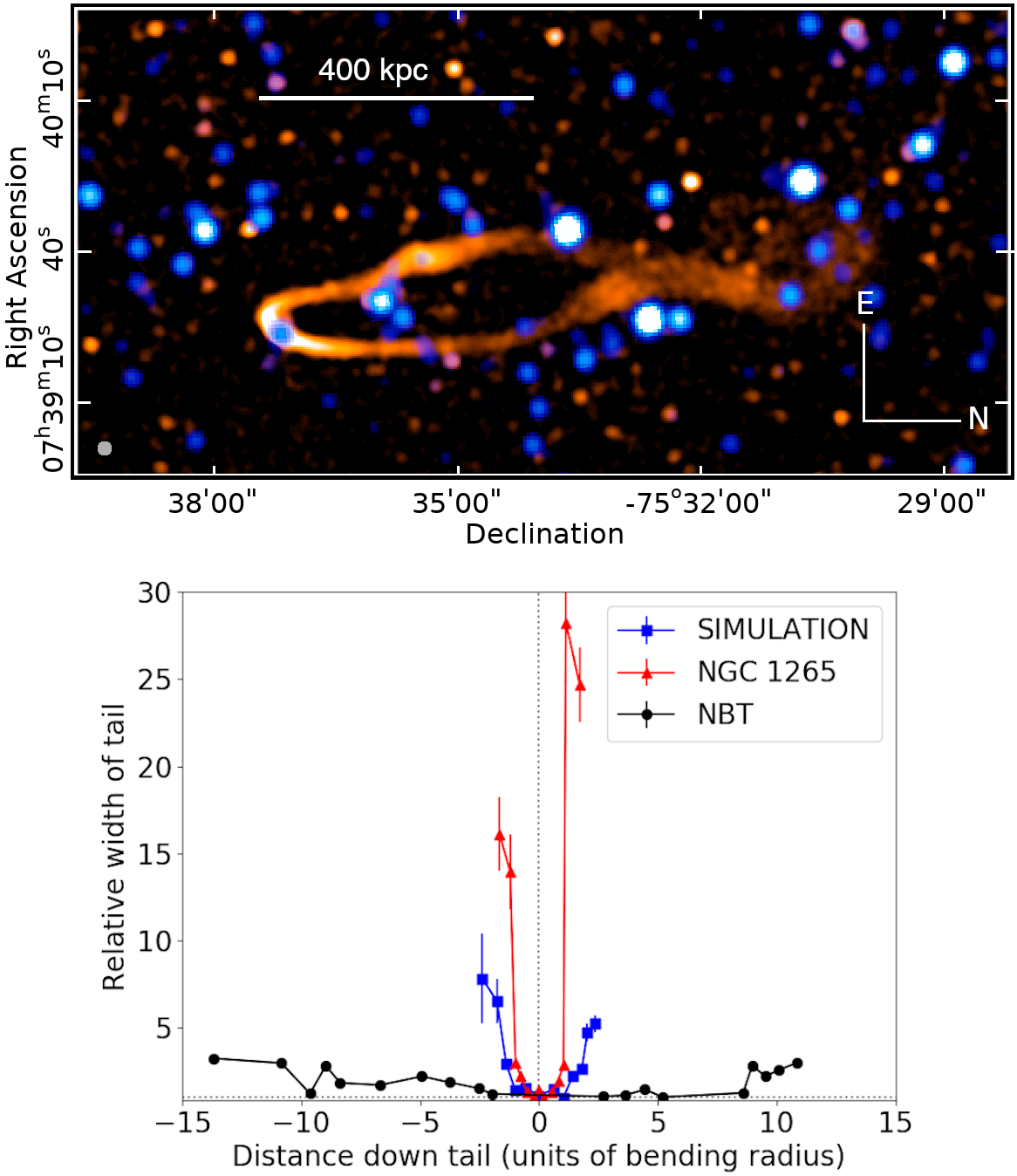}
   \caption{The narrow bent tail source, \emph{NBT}, showing that it remains well-collimated long after it bends, a unique behaviour. \textbf{Top:} MGCLS Stokes~I full-resolution (7.5\arcsec\,$\times$\,7.4\arcsec) intensity image in orange, overlaying the WISE~W1 image in blue. The radio brightness scale is logarithmic, saturating at 0.2 mJy\,beam\per. The host, centreed at the south end of the twin tails, is located at $\rm R.A. = 07^h39^m23.89^s,~Dec. = -75^\circ37^\prime11.3^{\prime\prime}$. The physical scale at the host redshift is indicated, and the MGCLS synthesised beam is shown at lower left. \textbf{Bottom:} The increase in jet widths, in units of the jet radius near the core,  as a function of distance from the core. They show the dramatic difference between \emph{NBT} and a prototypical narrow-angle tail \citep[NGC~1265, using the 3.7{\arcsec} resolution map from][]{Gendron-Marsolais2020} and a numerically simulated tail, from \citet{2019ApJ...884...12O}. 
   }
   \label{fig:rg-impossibletail}%
\end{figure}

The narrow bent-tail source shown in the top panel of Figure~\ref{fig:rg-impossibletail} (hereinafter \emph{NBT}), is unique among bent-jet sources in \editout{maintaining very well-collimated jets}\lr{showing very little lateral expansion of the jets} far beyond where they have bent by $\sim$90 degrees from their original direction. \lr{There are many examples where no expansion may be apparent \citep{2021Natur.593...47C}, but the resolution is insufficient to measure the widths before and after the bend.} \lr{By contrast to the bent sources,} straight jets can be very well-collimated, terminating in small radio and sometimes even optical hot spots subtending $\ll1^\circ$ as seen from the core \citep[e.g.][]{1987ApJ...314...70R}. But bent jets such as narrow- or wide-angle tails always \lr{develop broad cocoon-like structures after the jets bend} \citep[e.g.,][]{1986ApJ...301..841O, 1981ApJ...246L..69R}. As those jets bend due to their motion with respect to the surrounding medium, they tend to become unstable and \lr{are enveloped in much broader structures} \citep{2019ApJ...884...12O}.

\emph{NBT}'s host is \object{WISE~073923.89$-$753711.3}; at a photometric redshift of $z=0.108$ \citep{Dalyaetal18}  it is $\sim$930~kpc long, placing it in the ``giant'' class. It was found serendipitously at the very large projected distance of $\sim$3.7\,Mpc from the nearest known cluster at a similar redshift (MCXC~J0738.1$-$7506 at $z=0.111$). \emph{NBT}'s monochromatic luminosity at 1.28~GHz is $6.4 \times 10^{23}$\,W\,Hz\per, typical for bent-tail sources. Profiles across the eastern tail (top tail in Figure \ref{fig:rg-impossibletail}) show that it broadens and fades 
until it disappears at $\sim$400~kpc from the host, while the western tail extends more than twice that.

In the bottom panel of Figure~\ref{fig:rg-impossibletail} we compare the transverse expansion of the jets in \emph{NBT} to those of NGC~1265, the prototypical head-tail/narrow-angle-tail galaxy \citep{Miley1973}. We also include the results from a simulation of synchrotron radiation from a tailed radio galaxy with jets of internal Mach number $M = 2.5$ bent by a transverse $M=0.9$ wind \citep{2019ApJ...884...12O}. The distances in each case are plotted as the straight line separation from the core, in units of the bending radius ($r_b$) of the jets.  For \emph{NBT} we set $r_b$=35\arcsec, half the distance between the parallel tails; the apparent bending radius appears smaller near the host, but since the jets do not emerge exactly perpendicular to the direction of relative motion, they do not follow a simple circular path \citep[see Appendix of][]{2019ApJ...884...12O}.  The jet widths are measured as the FWHM for Gaussian fits to the jet/tail. In NGC~1265, the jets expand by a factor of 3 at 1.0\,$r_b$, while for the simulated jet, this occurs at $\sim 1.5\,r_b$. By contrast, \emph{NBT} reaches a factor of 3 expansion only at $\sim 10\,r_b$.

It is not clear how to maintain this extraordinary collimation.  In simulated jets, strong eddies form in the uniform surrounding medium, causing the jets to wobble and be partially disrupted \citep{2019ApJ...884...12O}. Pieces of jet thus transfer their momentum through the surrounding medium to the other tail, causing disruption events there. By this time, the extent of the synchrotron emitting region of each tail, mixed with the external medium, can be many times the initial jet width. The situation is even more complex if there are flow inhomogeneities in the surrounding medium.

It might appear attractive to invoke projection effects to make the bending radius appear smaller than it is in the true plane of the jets. However, to mimic the behaviour of the jets shown in Figure~\ref{fig:rg-impossibletail}, we need to assume foreshortening by a factor of $\sim7$, or an angle of $\sim82^\circ$ between the plane of the sky and the plane of the jets.  At this angle, by the time the jets in \emph{NBT} are moving parallel to one another, they would be separated by $\sim$1~Mpc. In comparison, the separation between the tails in narrow-angle tail sources (where the tails are parallel to each other), is typically of order 10s~kpc \citep[e.g.,][]{Gendron-Marsolais2020}.  Therefore, projected or not, \emph{NBT}  is unique in maintaining its collimation, and whether we are looking at a very special environment, or a new type of jet physics, remains a puzzle for the future.

\subsection{\lr{A 300 kpc} ring around interacting spirals}
\begin{figure}
   \centering
   \includegraphics[width=\columnwidth]{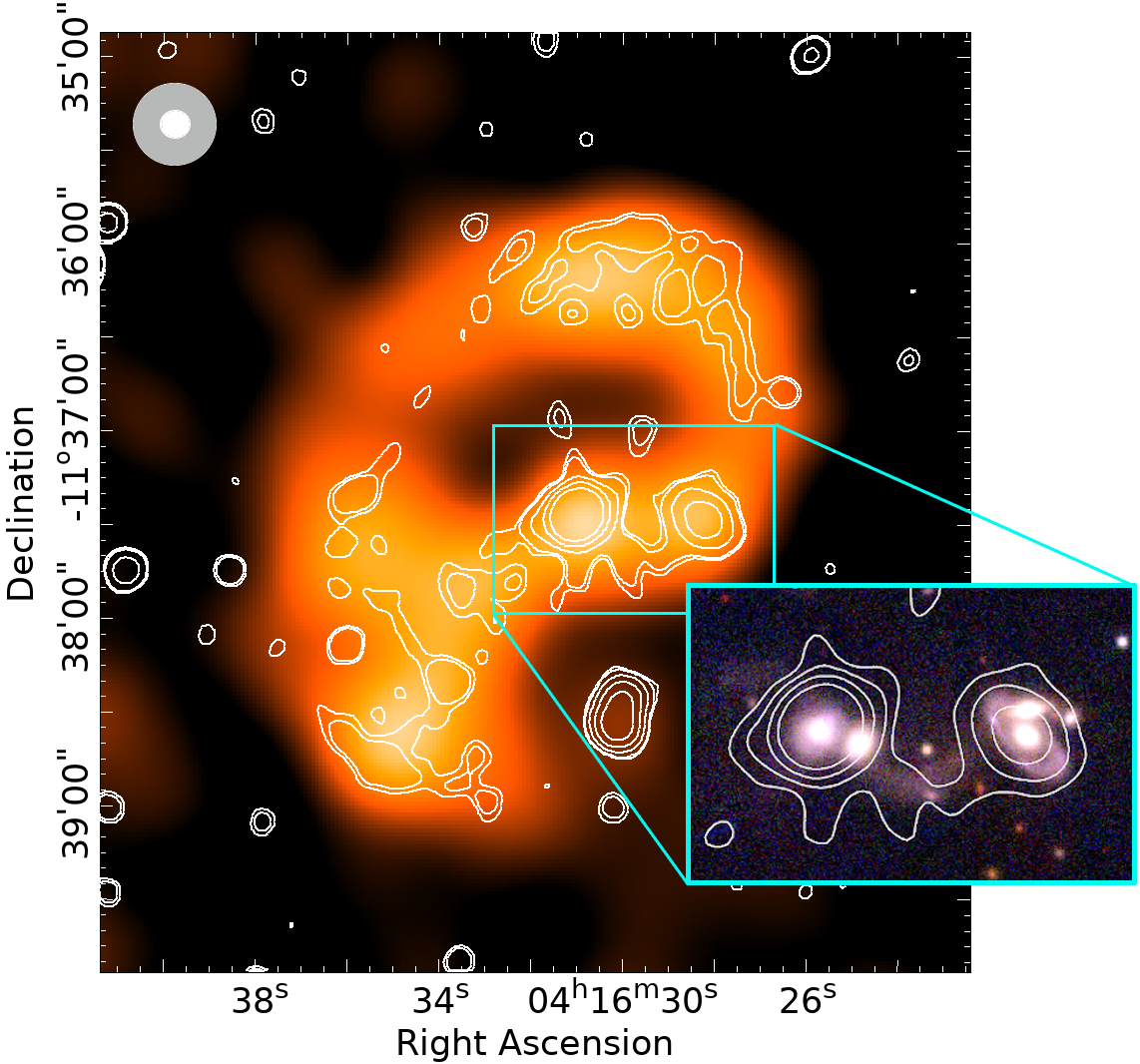}
   \caption{Filtered 25\arcsec-resolution MGCLS Stokes~I intensity image of a large diffuse radio ring surrounding and connecting to an interacting pair of pairs of galaxies in the MACS~J0417.5$-$1155 field. The brightness scale is linear and saturates at 0.2\,mJy\,beam\per. The interacting pairs are seen in the PanSTARRS false colour $gri$-composite inset. White contours show the full-resolution (7.9\arcsec\,$\times$7.8\arcsec) MGCLS Stokes~I intensity, with a 1$\sigma$ local RMS noise of 6\,$\mu$Jy\,beam\per. The first two contours are at levels of 3.5 and 5$\sigma$, thereafter increasing by a factor of three. The same contours, starting from 5$\sigma$, are shown in the inset. The synthesised beams are shown at upper left (grey - filtered  25\arcsec\ resolution, white - full resolution). 
   } 
   \label{fig:interact}
\end{figure}

\lr{We have discovered a $\sim$300~kpc-diameter ring around two pairs of interacting spiral galaxies.  To our knowledge, this structure is unique and not easily explained by currently known processes in such galaxies.} In the full-resolution Stokes~I intensity image of the MACS~J0417.5$-$1155 field, faint emission is seen in a ring-like structure around two compact radio sources $\sim0.4$ degrees NW of and unrelated to the cluster. The ring becomes much more distinct in the filtered 25\arcsec-resolution intensity map, shown in Figure~\ref{fig:interact}. No other such rings are seen around stronger sources in this field or any other field in the MGCLS, so it is unlikely that the ring is an artefact.
   
Each of the bright compact radio sources near the centre overlaps a \emph{pair} of interacting spiral galaxies, as shown in the PanSTARRS $gri$-composite image inset in Figure~\ref{fig:interact}. The eastern radio source is centred on the brighter galaxy of its pair, \object{WISEA~J041630.96$-$113728.0} (also 6dFGS~gJ041631.0$-$113728) at a redshift of 0.086 \citep{2009MNRAS.399..683J}. The radio source is unresolved, has a flux density of 14\,mJy, a spectral index of $-0.76\pm0.02$, and a marginally detected ROSAT PSPCB counterpart. The other galaxy in the pair, 2MASX~J04163041$-$1137306, has a redshift of 0.087. These objects are in the foreground of the target cluster, MACS~J0417.5$-$1155 ($z=0.44$). The western radio source is also associated with the brighter member of its pair, \object{WISEA~J041628.16$-$113729.1}, at a redshift of 0.0846, with its companion 2MASX~J04162816$-$1137236 at a redshift of 0.083. It has a total flux density of 1.3\,mJy, and a spectral index of $-0.9\pm0.05$. Encircling and connecting with these compact components, the extended emission (seen in the low resolution imaging) has a total flux density of $5\pm1$\,mJy and a maximum diameter of $\sim$\,300\,kpc, assuming it is associated with the compact sources.  The luminosity of the diffuse emission and the western compact emission are each $\sim10^{26}$\,W\,Hz\per, while the eastern source's luminosity is an order of magnitude higher. 
All of these are orders of magnitude higher than expected from starburst activity \citep{Condon1992}. 

With several more galaxies, these pairs are part of the Hickson Compact Group 27 \citep{Hickson82}, with a diameter of 3.8\arcmin\ (365~kpc), on the same scale as the radio ring. 
\lr{We consider first whether tidal tails could be responsible for the large ring.}
The tidal tails visible in the PanSTARRS image span \lr{a smaller distance of} $\sim73$\arcsec~(115~kpc), \lr{and the lack of distortion in the galaxies suggest that the tidal interactions are still at an early stage.  Tidal tails have been observed, however,  to span large distances, e.g. 200~kpc for the HI tails of the Antennae \citep{2001AJ....122.2969H}, although synchrotron emission has only been detected out to 20~kpc \citep{2017MNRAS.464.1003B}.  Whether or not tidal effects could result in star formation and subsequent cosmic-ray production, along with magnetic field amplification on the scale of the ring, is an open question.}  \editout{ however 
the radio ring appears unlikely to be due to the tidal tails, since it is on much larger scales.  In addition, the galaxies themselves show little distortion, so the interactions must be at a very early stage. Finally, the radio luminosity of the ring is orders of magnitude higher than what would be expected from star formation even from the galaxies, let alone the very sparse tidal structures.}

Galaxy groups\lr{, themselves,} have been found to contain both extended thermal gas and relativistic plasmas, as well as large scale \textup{H\,\textsc{\lowercase{i}}} and cold gas \citep[e.g.,][]{2018A&A...618A.126O}. In some cases, thermal gas which is too diffuse to be detected in X-rays may exist on scales all the way up to 700~kpc \citep{2011ApJ...738..145F}, based on the distortion of radio galaxies. \citet{2019A&A...622A..23N} found diffuse emission in 17 of 20 compact groups, and \citet{2013MNRAS.435..149N} detected diffuse radio emission extending over about 75~kpc in Stephans Quintet. \lr{Group-related emission} has not yet been found on the physical scale or with the luminosity, or structure, shown here. 
The elliptical ring of emission here is brightest on its outer edges, similar to what is seen from shocks on much larger, Mpc,  scales  generated when clusters of galaxies collide (peripheral radio relics). As an alternative to a shock origin, fly-bys of other group galaxies could light up the remnants of previous AGN activity \citep[e.g.,][]{2018MNRAS.473.5248O}.  Determining whether shocks, AGN emission, \lr{tidal tails,} or some other mechanism is responsible for this curious ring-structure in J0416$-$1137 will depend on deeper X-ray observations and radio spectral information.

%


\subsection{The \lr{interaction of ICM} magnetic filaments and radio tails }\label{sec:rg-a194}

\begin{figure}
   \centering
   \includegraphics[width=0.9\columnwidth]{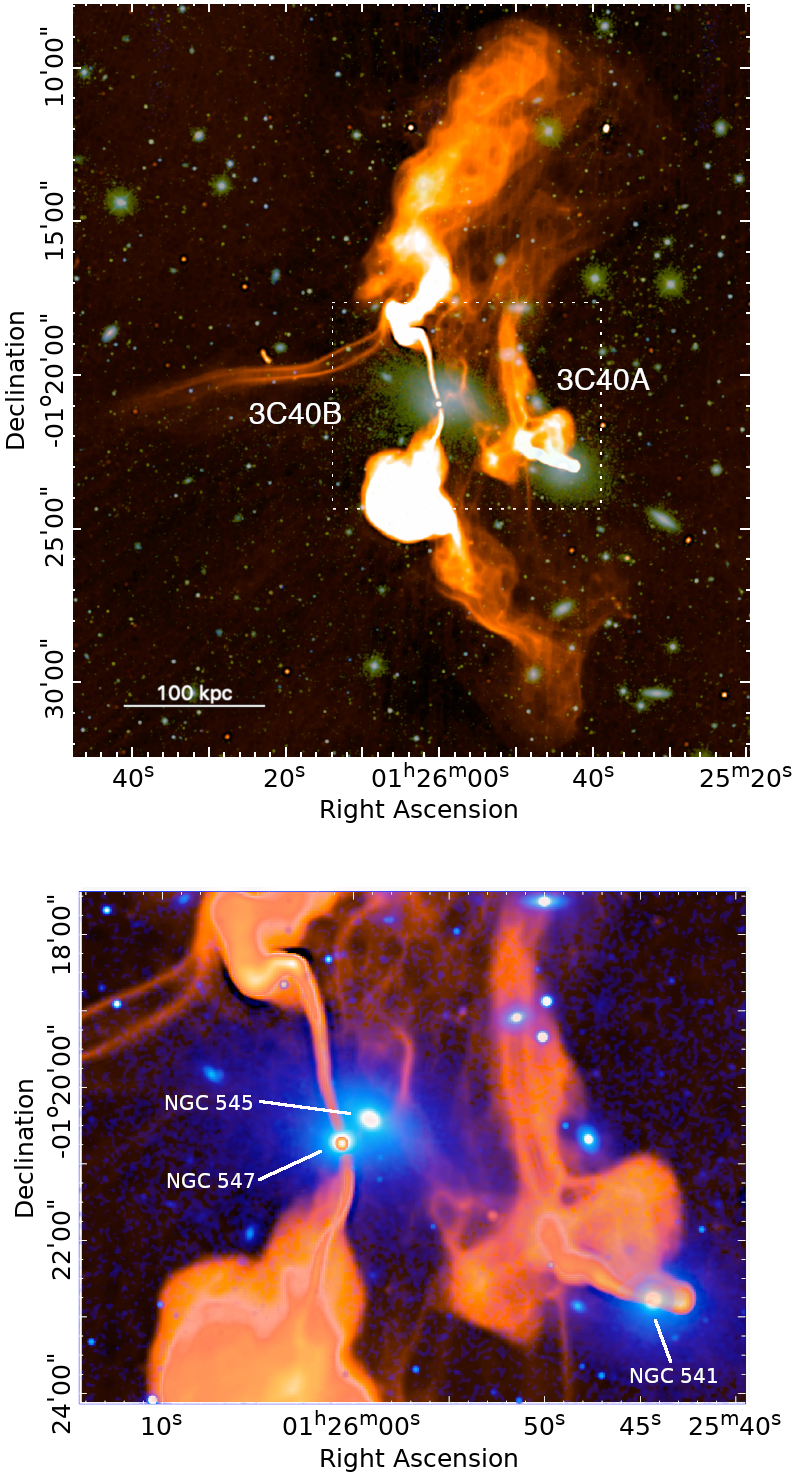}
   \caption{Two radio galaxies in Abell~194, 3C40A and 3C40B, with unusual filamentary appendages. \textbf{Top:} Full-resolution (7.7\arcsec\,$\times$\,7.5\arcsec) MGCLS Stokes~I intensity image (orange), overlaid on a SDSS $gri$ colour composite of the region \citep{SDSSDR12}. The radio brightness is logarithmic, saturating at 2\,mJy\,beam\per. The physical scale at the cluster redshift of $z=0.018$ is indicated at lower left.  
   \textbf{Bottom:} Zoom-in of the boxed region from the top panel, with the full-resolution MGCLS image in orange on a non-monotonic scale, and the SDSS $r$-band image in blue. The optical galaxies near the radio cores are labelled.}  
 
   \label{fig:rg-A194}%
\end{figure}

\lr{With MeerKAT's exceptional combination of sensitivity and resolution, filamentary synchrotron structures are now being discovered in the neighbourhood of radio galaxies \citep{Ramatsokuetal20,2021ApJ...917...18C}. Here, we present the first example where a direct interaction between a filament and the jet flow from a radio galaxy may been seen.} The full-resolution  MGCLS image of the Abell~194 \lr{$z = 0.018$} cluster field is shown in  Figure~\ref{fig:rg-A194}. The newly revealed filamentary structures associated with the cluster radio galaxies 3C40A and 3C40B at the centre of \object{Abell~194} cannot be explained with any current radio galaxy models. Such very large-scale features are not seen in numerical simulations of radio galaxies, nor were such features predicted. These sources were studied at lower resolution at multiple frequencies by \citet{2008MNRAS.384...87S} who first noted unusual extensions around the radio galaxies. 

\begin{figure}
   \centering
   \includegraphics[width=0.95\columnwidth]{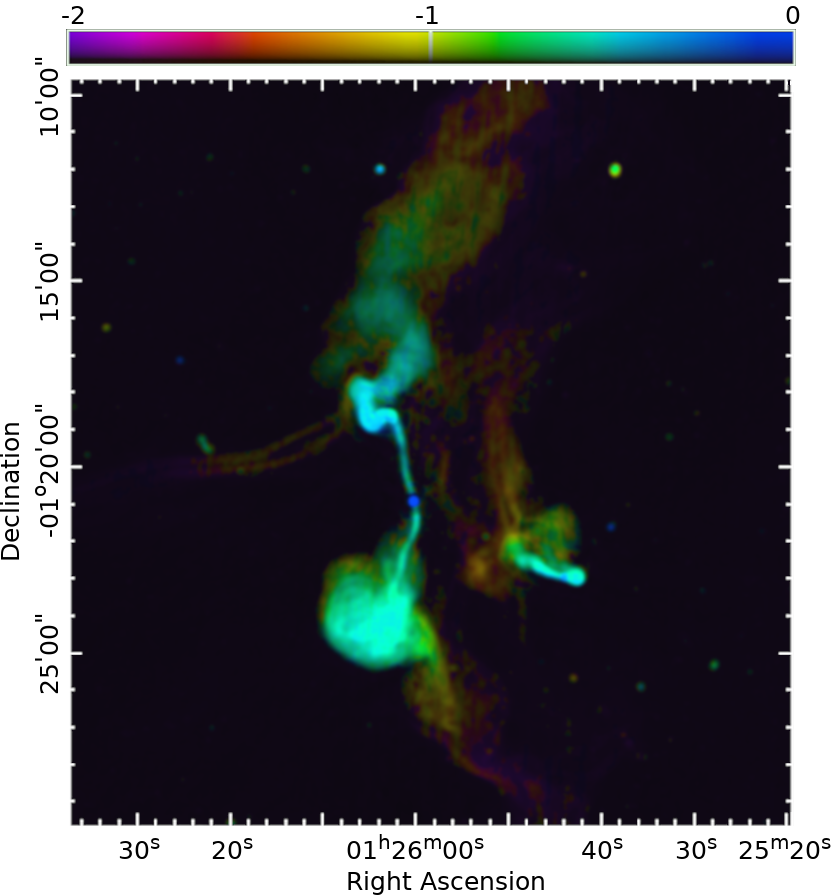}
   \caption{Log intensity-coded MGCLS spectral index map of the 3C40A and 3C40B radio sources in Abell~194, at full resolution (7.7\arcsec\,$\times$\,7.5\arcsec). Colours indicate the spectral index and the brightness indicates the Stokes~I intensity. The colour bar indicates the spectral indices, with the upper (lower) part of the colour bar corresponding to the brighter (fainter) regions. The brightness is on a linear scale, and saturates at 8~mJy\,beam\per. 
   }
   \label{fig:A194alpha}%
\end{figure}

The most spectacular filaments are the parallel curved pair, \lr{each $\sim$100\arcsec~(37~kpc) wide} extending 8.9\arcmin~($\sim$\,200\,kpc)  to the east from the northern lobe of NGC~547 (3C40B). The spectral index map in Figure~\ref{fig:A194alpha} shows that the filamentary structures have steep spectra, similar to the faintest portions of the radio galaxy lobes. \editout{, corresponding to $\sim$\,200\,kpc at its redshift of 0.018.} Where they appear to emerge from the northern jet, the filaments curve due SE, counter to the northern jet flow.  
\editout{Before the filaments merge, the filaments have widths of $\sim$15\arcsec~ (5.5~kpc); to the east, they blend with each other into a bundle of filaments with an overall width of $\sim$100\arcsec~(37~kpc). The high fractional polarisations and tangential field alignment (not shown) suggest that these are magnetic structures.  These properties, along with detailed spectral studies, will be discussed in a followup paper (Rudnick et al., in prep.).}

The second radio galaxy, 3C40A, is associated with NGC~541. At the higher MGCLS resolution, we can clearly see \lr{that the 165\,kpc plumes observed by \citet{2008MNRAS.384...87S} are dominated by}  twin $\sim$\,15\,\arcsec\ (5.5\,kpc) wide filamentary structures.\editout{, each with a width of , similar to the eastern filaments. These sit on top of a faint $\sim$\,40'' envelope.} \editout{To our knowledge, there are no simulations or any other observations of a tailed radio galaxy where the tails re-form into thin streams after dramatically broadening. } In addition, there is a rich network of filaments connecting to and between these two bright radio galaxies, some at lower brightness than visible in the images here. \lr{The origin of these structures is unknown.}

\begin{figure*}
   \centering
   \includegraphics[width=0.95\textwidth]{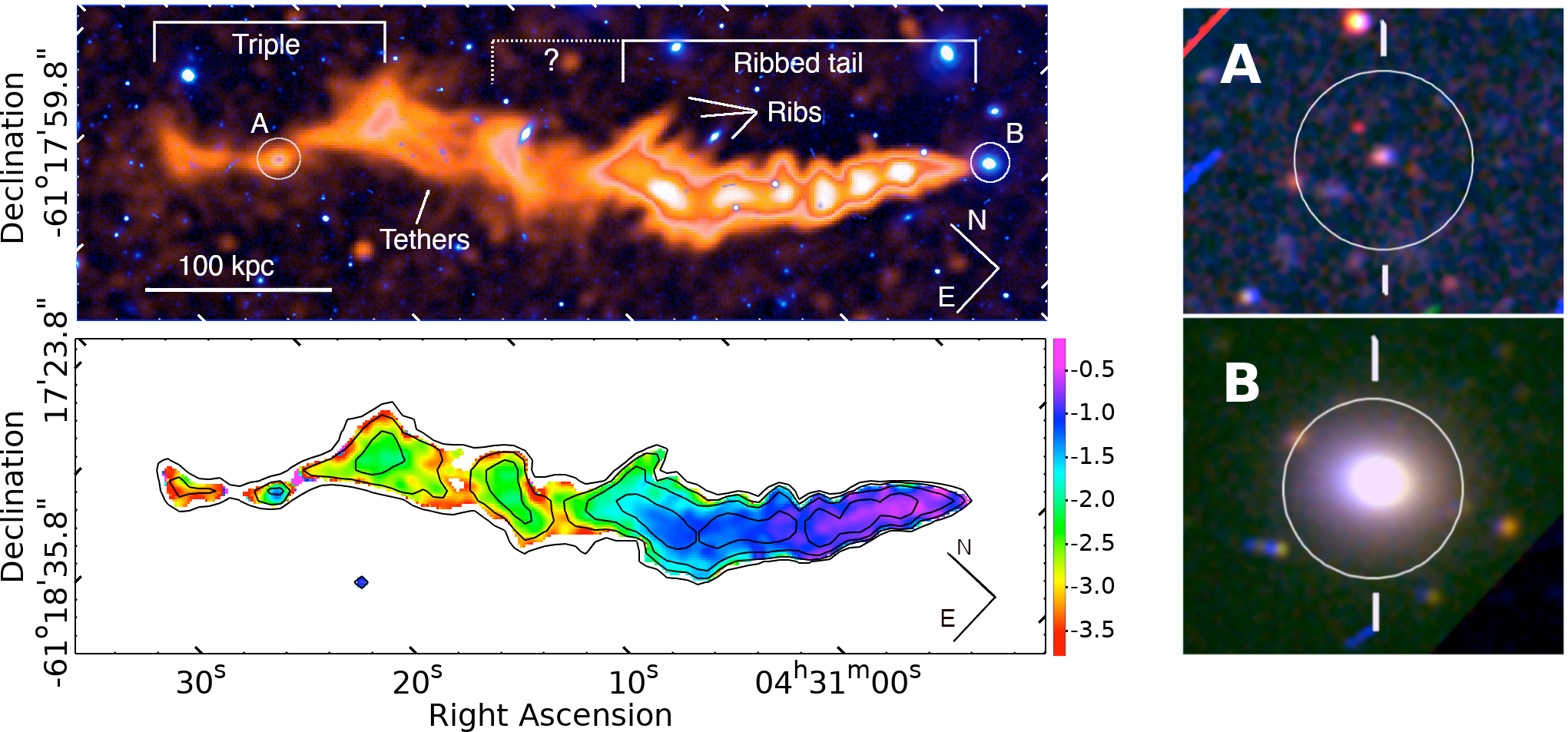}
   \caption{Tailed radio source, \emph{T3266}, NW of centre of MCXC~J0431.4$-$6126 (Abell~3266), which may be an extremely unusual composite of two independent sources. \textbf{Left:} The \textit{top panel} shows the full resolution (7.1\arcsec\ $\times$ 6.7\arcsec) MGCLS Stokes~I intensity of the source (orange) overlaying a DECam $g$-band image (blue). The radio brightness is on a non-monotonic scale, with the faintest regions of the tethers at $\sim$0.15~mJy\,beam\per, and the brightest regions at 2.2~mJy\,beam\per. Circles, labelled A and B, indicate the positions of the two likely host galaxies (see right panels). The physical scale at the cluster redshift is shown at lower left. The \textit{bottom panel} shows the in-band spectral index map, with a strong steepening with distance from host B.  
   \textbf{Right}: DECam \emph{gri} composite images of the two proposed hosts labelled A and B in the top left panel. Host A is at $\rm R.A. = \rm04^h 31^m 18.48^s$, $\rm Dec. = -61^\circ19^\prime18.54$\arcsec. Host B is at $\rm R.A. = \rm04^h30^m45.56^s$, $\rm Dec. = -61^\circ23^\prime35.8$\arcsec. 
   }
   \label{fig:rg-jetrings}%
\end{figure*}
Extensive networks of filamentary structure associated with a tailed radio galaxy are also seen in the LOFAR maps of Abell~1314 (Figure~25 in \citealt{2021A&A...651A.115V}). \citet{Ramatsokuetal20} showed another spectacular example from MeerKAT observations of ESO~137$-$006, with ``collimated synchrotron'' threads connecting the two radio lobes.  The origin of all these features is still unknown.  One speculative possibility mentioned by \citet{Parrishetal12}, \citet{BirkinshawWorrall15}, and \citet{Donnertetal18} is that there are pre-existing magnetic flux tubes in the ICM. Such features would become visible only when there was a sufficient population of relativistic electrons.  \lr{Further spectral and polarisation work, and detailed comparisons with X-ray emission from the ICM will be needed to explain these novel phenomena.}


\subsection{A tail with ribs and tethers}

\lr{At a distance of 350\,kpc from the X-ray peak of Abell~3266 we identify a tailed radio galaxy (hereinafter \emph{T3266}) with features not yet seen elsewhere, and whose physical origins are unclear.} The structure appears in earlier maps by \cite{2016MNRAS.456.1259B} \kk{where it is shown to be elongated parallel to the cluster X-ray emission, positioned between the identified optical sub-clusters}.\editout{ and it lies along the border between the central optical sub-clusters and two of the other sub-clusters.} The Abell~3266 galaxy cluster (listed in the MGCLS catalogue as J0431.4$-$6126) is likely in the midst of a complicated cluster merger, with an elongated X-ray distribution \citep{2002ApJ...577..701H}, two velocity-separated sub-clusters of optical galaxies in its core, and six subclusters on its peripheries \citep{Dehghanetal17}. The cluster core has a mean redshift of 0.0594\,$\pm$\,0.0005 and a velocity dispersion of 1460\,$\pm$\,100\,\kms. 

The full-resolution MGCLS image of \emph{T3266}, overlaid on the $g$-band DECam image, is shown in Figure~\ref{fig:rg-jetrings}. \emph{T3266} has a faint, 65\,$\mu$Jy\,beam\per\ radio core (not visible in this image) associated with \object{WISEA~J043045.39$-$612335.6} (galaxy denoted by B in Figure~\ref{fig:rg-jetrings}), at a redshift of 0.0626. The radio tail extends to the NE (from right to left in the figure), and the entire structure has a projected length of $\sim$\,6.7\arcmin\ ($\sim$\,480\,kpc at the redshift of the presumed host). 
The region of the tail closest to galaxy B (the ``ribbed tail'' in Figure~\ref{fig:rg-jetrings}) shows distinct quasi-periodic changes in brightness and width. The tail is $\sim$\,20\arcsec\ (23\,kpc) wide at its half-intensity in this region, but its transverse profile is flat-topped and shows no signs of bifurcation. No other such features are seen in the MGCLS Abell~3266 field of view, and they are unlikely to be instrumental in origin. The four bright regions nearest the head of \emph{T3266} are separated by $\sim$\,19\arcsec\ (22\,kpc), essentially the same as the jet width.  This suggests that some instability in the flow may be regulating this behaviour. Some faint extensions transverse to the jet are also seen.

Further back along the ribbed tail, three ``ribs'' are clearly seen to extend perpendicular to the tail axis, with a full extent of up to 75\arcsec\ (88\,kpc). \editout{ They are slightly closer together, 17~kpc, than the bright patches at the head.  Given that they extend transversely beyond the confines of the bright tail, w}We speculate that they reflect some type of interaction with the external medium, \lr{or are related to the transverse structures seen in early numerical simulations of intermittently restarting jets \citep{ClarkeBurns91}.} These unusual bright patches and ``ribs'', with their periodicity and transverse extent, stand as a challenge for the current generation of magnetohydrodynamical numerical simulations of jets in a complex ICM.

It is also possible that the full structure of \emph{T3266} may be a serendipitous projection of two separate sources, with the last 2\,\arcmin\ of the tail (left-most in Figure~\ref{fig:rg-jetrings}) potentially a separate triple source. The possible host of this source, indicated as galaxy~A in Figure~\ref{fig:rg-jetrings}, is a faint optical/IR galaxy (DES~J043118.45$-$611917.9 or \object{WISEA~J043118.50$-$611918.2}) with a photometric redshift of 0.78 \citep{2019ApJS..242....8Z}. If that identification is correct, the triple is a background giant radio galaxy, 900\,kpc in length. The region between the ribbed tail and the triple, indicated by the dashed line in Figure~\ref{fig:rg-jetrings}, and the three or more filamentary ``tethers'', might belong to either source. They are each $\sim$\,50\arcsec\ in projection, corresponding to 375\,kpc and 58\,kpc at the two redshifts.

It is not yet clear whether there are one or two individual radio galaxies in \emph{T3266}, or even how to decisively answer that question.  \lr{Either way, the ``tethers'' represent a new physical phenomenon, perhaps related to the ``collimated synchrotron threads'' of \cite{Ramatsokuetal20} mentioned earlier.} \editout{If all the structures belong to Abell~3266, and the connections are physical, then they represent a new astrophysical phenomenon. The ``tethers'' could be related to  the ``collimated synchrotron threads'' mentioned in Section~ref sec:rg-a194 appearing to link the two lobes of ESO~137$-$006 citep Ramatsokuetal20.} Unfortunately, the spectral index behaviour, seen in the bottom left panel of Figure~\ref{fig:rg-jetrings}, does not provide a clear signature. In the region of the ribbed tail, \emph{T3266} shows a monotonic spectral steepening with spectral indices ranging from $-0.75$ at the head to $-1.4$ in the region of the ribs. The spectra \lr{of the ``tethers''} appear very steep ($\alpha < -2$, though the spectral index is uncertain due to the low brightness of the region), and the triple source has spectral indices similar to the end of the tail, which is not expected if these are simply seen in projection.\editout{in the region between the ribbed tail and the triple are preliminary and very steep, ranging from about $-2$ in the brighter regions to $< -3$ in the fainter ones. All of these spectral features are  typical of tailed radio galaxies. The triple, if it were a separate source seen in projection, would be expected to show a break in spectral behaviour; instead, it has steep spectral indices that are very similar to the end of the tail.}

\subsection{Dying radio galaxies}

One of the strengths of the MGCLS is in the detection of radio galaxies in their ``dying phase'', i.e., after the powering jets have been turned off. Studying such galaxies is important for understanding radio galaxy physics, the duty cycle of AGN activity, interactions with the surrounding environment, and for the usefulness of radio galaxies as cosmological probes.  However, one needs a combination of high resolution (to ensure, within observational limits, that there are no significant hot spot regions or jets, and to identify the host galaxy), as well as good surface brightness sensitivity to detect the fading, dying lobes. Here we highlight \lr{two} radio galaxies which might be in a dying phase of their lives, as examples of what is visible in the MGCLS data. 

\begin{figure}
   \centering
   \includegraphics[width=0.95\columnwidth]{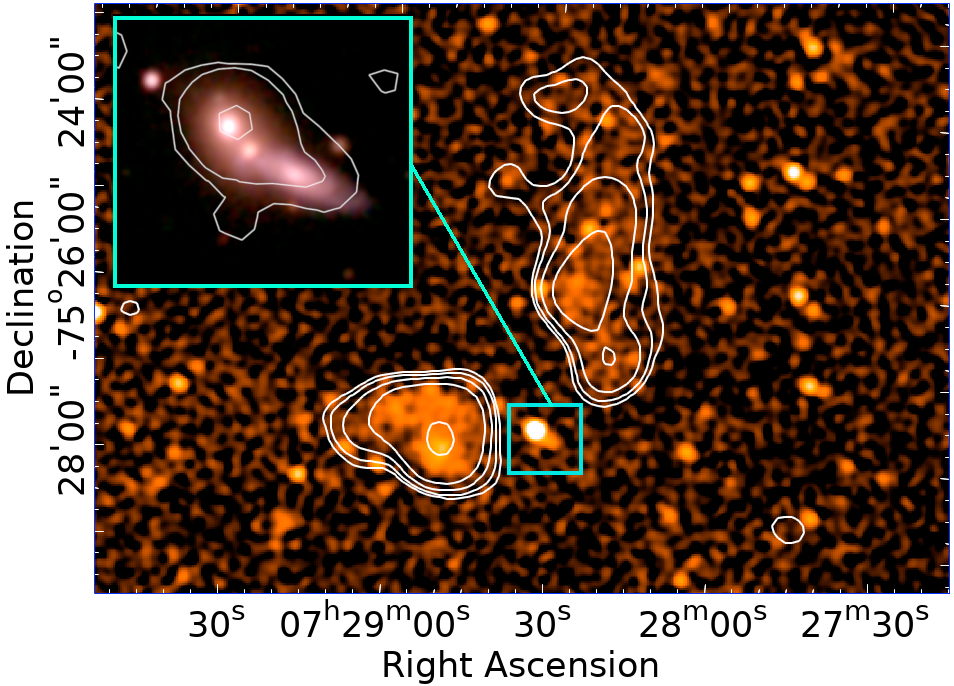}
   \caption{\lr{Dying radio galaxy example, associated with WISEA~J072832.45$-$752740.0, without detectable jets or hotspots in its diffuse lobes. The full resolution (7.5\arcsec) MGCLS Stokes~I intensity image is shown in orange, overlaid with contours showing the filtered  25\arcsec-resolution intensity. It is outside of the primary beam-corrected field of view of MCXC~J0738.1$-$7506. The brightness (\emph{non-primary beam corrected}) is on a logarithmic scale saturating at 0.1\,mJy\,beam\per, and contours are shown at 18, 24, 34, 54, 64\,$\mu$Jy/(25\arcsec\ beam). The inset shows a zoom in on a DECaLS image of the optical host, with full resolution radio contours, showing recent nuclear activity.} }
   \label{fig:dying}%
\end{figure}
\begin{figure*}
    \centering
\includegraphics[width=0.85\textwidth]{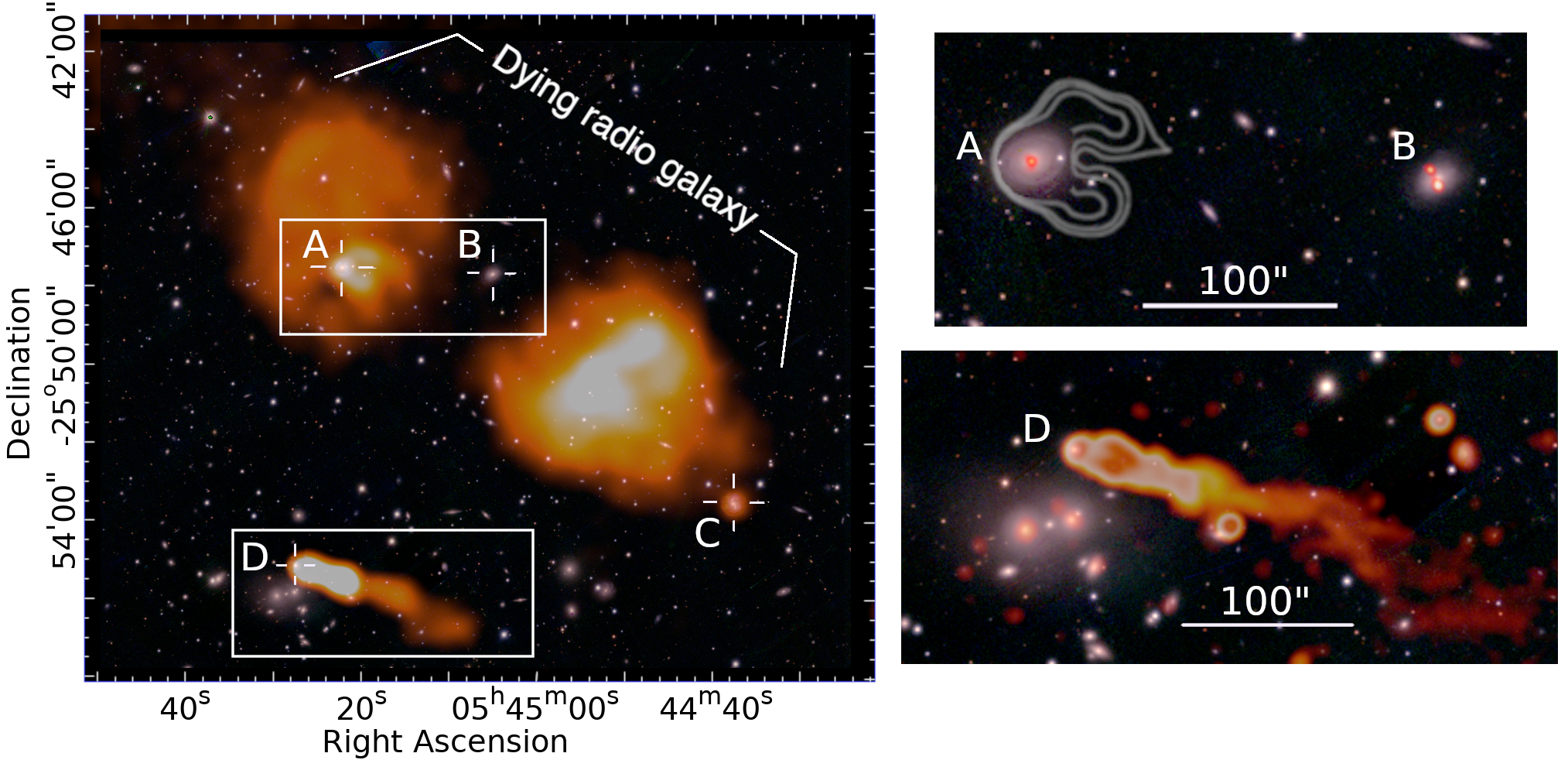}
    \caption{\textbf{Left:} Diffuse emission filtered 25\arcsec-resolution MGCLS Stokes~I intensity image of Abell~548B (orange) overlaid on the false-colour $gri$-composite PanSTARRS image. Previously misidentified structures (see Section~\ref{sec:rg-a548b} for details) are revealed to be a diffuse/dying radio galaxy to the north, and a tailed radio galaxy to the south. The brightness scale is
    logarithmic, saturating at 1\,mJy\,beam\per.  Radio sources A--D have clear optical counterparts, with B the likely host of the diffuse lobes and A a spiral galaxy embedded in the eastern lobe. \textbf{Right, Top}: Zoom in on the boxed region around sources A and B. White contours, showing the 15\arcsec-resolution MGCLS Stokes~I intensity at levels of (0.3, 0.35, 0.5)\,mJy\,beam\per\ and edited for clarity, indicate tailed emission associated with the spiral galaxy A. Red compact structures are from VLASS \citep{2020PASP..132c5001L} at 3~GHz and a resolution of 2.9\arcsec\,$\times$\,1.8\arcsec\ (p.a. 50$^\circ$). The single VLASS component associated with A has a peak flux of 30\,mJy\,beam\per. There is a small double VLASS source associated with B, with a peak flux of 3.3\,mJy\,beam\per, indicating possible recent radio activity. \textbf{Right, Bottom}: Zoom in at full resolution (7.4\arcsec$\times$7.4\arcsec) on the boxed region around the tailed source D. The brightness scale is non-monotonic, and the peak brightness in the frame is 7\,mJy\,beam\per. 100\arcsec\ corresponds to 84\,kpc at the cluster redshift of 0.042.}
    \label{fig:A548}
\end{figure*}

\subsubsection{WISEA J072832.45$-$752740.0} The first example is shown in Figure~\ref{fig:dying}, and is associated with \object{WISEA~J072832.45$-$752740.0}. At $z\,=\,0.0138$ \citep{2009MNRAS.399..683J}, this radio source is found serendipitously in the field of the $z=0.111$ cluster MCXC~J0738.1$-$7506. It has a total extent of $\sim$100~kpc, and the appearance of a wide-angle tail. Its luminosity of $\sim 10^{21}$\;W\,Hz\per\ is orders of magnitude below those of typical extended radio galaxies, although compact AGN emission at such low luminosities is more common \citep{2018MNRAS.479..807L, 2009AJ....137.4450M}. The radio core, shown overlaid with a DECaLS image in the inset of Figure~\ref{fig:dying}, shows extended emission from likely interacting galaxies in a common envelope.
We calculate minimum energy magnetic field strengths for each lobe, as 0.8\,$\mu$G and 2\,$\mu$G for the west and east lobes, respectively, assuming a spectral index of $-1$ and a proton/electron ratio of unity.\footnote{The source is at the edge of where we can make reliable primary beam corrections, and the SNR is low enough that the fluxes could be uncertain by up to $\sim$50\%. This leads to an uncertainty of $\sim$12\% in the derived fields, much less than the uncertainties from the other assumptions.} These imply radiative lifetimes of $\sim$70\,Myr and $\sim$87\,Myr respectively against a combination of synchrotron and Inverse Compton cooling. If the spectra were as steep as $-2.5$, the field strengths would increase by approximately a factor of 5, although the lifetimes would be similar, but now dominated by synchrotron losses.  When the magnetic fields of dying radio galaxies drop below $\mu$G levels, the radio galaxies become very faint and the lifetimes become very short due to Inverse Compton losses \citep{Rudnick04}. The oldest, faintest sources will thus be rare and likely only be found in sensitive large area surveys.

\subsubsection{Abell 548B} \label{sec:rg-a548b}
In our second example, the MGCLS provides a fresh look at Abell~548B\footnote{Abell~548B is the western component of what was originally classified as Abell~548, as described in \cite{1988AJ.....95..985D}.}. The MGCLS images suggest that we are dealing with a large dying radio galaxy $\sim$\,14.2\arcmin\ (650~kpc) in extent. The left panel of Figure~\ref{fig:A548} shows the diffuse emission from this source at a resolution of $25$\arcsec, overlaid on a false colour $gri$-composite PanSTARRS image. The radio galaxy (B) is associated with a 6dFGS source, g0545049$-$254740, at a redshift of 0.036, almost 2000 km\,s\per\ from the central cluster velocity. The radio core is itself double, as seen in the top right panel where red contours are from the VLASS 3\,GHz map at 2.5\arcsec\ resolution. This structure is similar to the classes of Compact- and Medium-Symmetric Objects (CSOs and MSOs, \citealt{2002NewAR..46..263C} and \citealt{2006MNRAS.368.1411A}), which are thought to be young objects. This similarity may be an indication of restarting activity for the large scale radio galaxy, but the double structure is misaligned with respect to the large lobes, so the connection is unclear.

Embedded in the eastern lobe is a compact radio source (A) associated with the spiral galaxy 6dFGS~g0545221$-$254730 at $z\,=\,0.038$. It appears to have radio structures similar to those of a wide-angle-tail, swept towards the west by $\sim$80~kpc; it is not clear whether there is a physical connection to the dying radio galaxy \kk{as there is no obvious interaction with the} diffuse lobe. At the western extremity, another spiral galaxy (C), likely unrelated, is seen in the radio. It is 6dFGS~g0544374$-$255335 at $z\,=\,0.039$. Finally, to the south, nearer the cluster centre, is a 220~kpc long narrow-angle tail, associated with 6dFGS~g0545275$-$255510 (D) at $z\,=\,0.042$.  With their lower resolution and lower sensitivity observations, 
\citet{Ferettietal06} suggested that the radio galaxy lobes in the north and the diffuse emission to the south were instead merger-related relic structures outside of the bright X-ray region of the cluster. Now that MGCLS has elucidated the full structure of these sources, we recognise them as a combination of diffuse-lobed and tailed radio galaxies, as described above.  We expect a closer examination of the MGCLS to unveil more, and fainter, examples of dying radio galaxies.

\subsection{Bulk gas motions far outside clusters}
\begin{figure}
   \centering
   \includegraphics[width=\columnwidth]{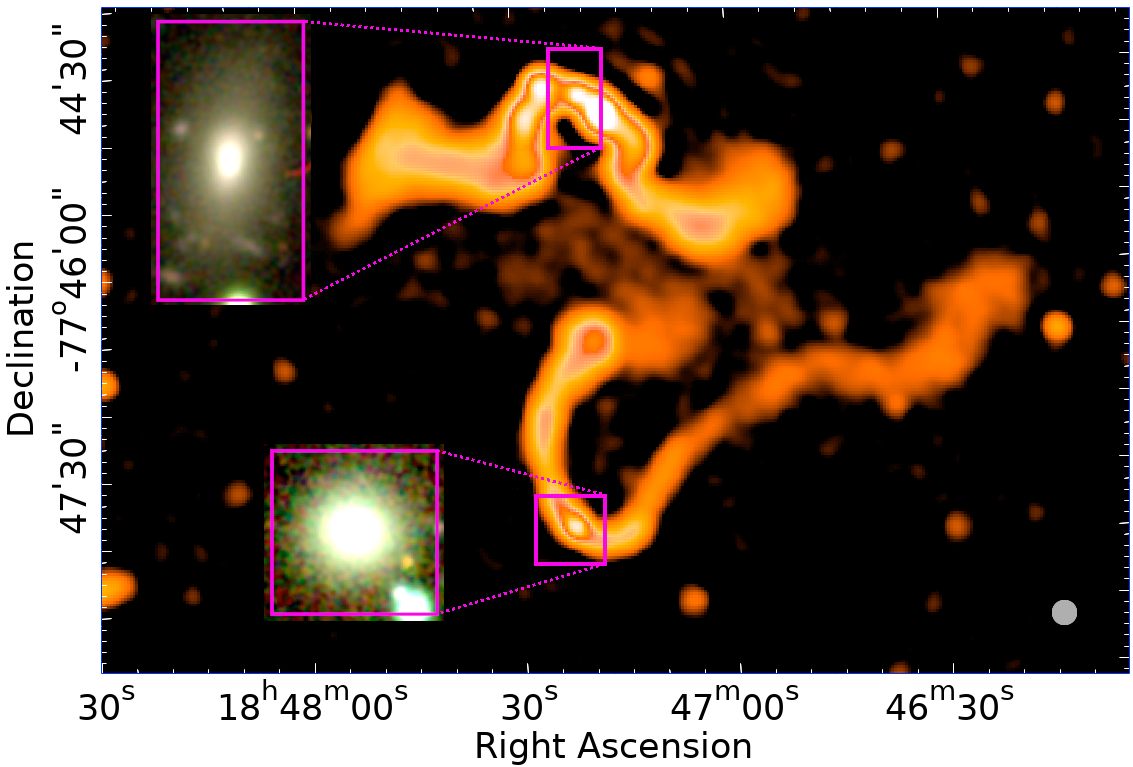}
   \caption{A pair of radio galaxies with multiple bends found serendipitously in the MCXC~J1840.6$-$7709 field, illustrating the likely effects of large-scale motions in the local external medium. The full resolution (8.1\arcsec\,$\times$\,7.6\arcsec) MGCLS Stokes~I intensity image is shown in orange, with a non-monotonic brightness scale and a peak flux of 13~mJy\,beam\per. A false colour $zir$-composite DECam image of each host is shown in the insets. The MGCLS synthesised beam is shown at lower right.} 
   \label{fig:rg-turb}%
\end{figure}

\lr{The pair of radio galaxies shown in Figure~\ref{fig:rg-turb} provide an unusual, and perhaps unique, case of complex radio galaxy / medium interactions far beyond the cluster environment. Early on,} tailed radio galaxies provided evidence of the relative motion of their host galaxies through the ICM \citep{Miley80}.  More recently, modest excess radio galaxy bending has also shown the influence of such motions in local overdensities at $>$\,5~Mpc from the nearest cluster \citep{Garonetal19}. 

The two radio galaxies shown in Figure~\ref{fig:rg-turb} were found $>$\,42\arcmin\ from the centre of the $z\,=\,0.0194$ cluster MCXC~J1840.6$-$7709 (ESO~45$-$11).  The northern source is associated with \object{WISEA~J184720.77$-$774444.2}, at $z\,=\,0.138$, and is therefore not associated with the cluster. The southern source is associated with \object{WISEA~J184722.38$-$774756.1} which has no available redshift; for the purposes of this discussion, we make the plausible assumption that the two radio sources belong to the same system.  There are no other catalogued clusters in the vicinity, and no X-ray emission is visible in the ROSAT All-Sky Survey.

\lr{Multiple-bend sources such as these are important for understanding the dynamics of the diffuse, lower density thermal plasmas both in- and out-side of rich clusters.}
Simple relative motions though an external medium would produce a C-shaped structure, but the bends in these sources require other external forces.  The scale sizes are on the order of several hundred kpc.  In order to see bends such as these, two factors are important.  First, the irregularities in the external medium flows must be on scales comparable to the size of the radio galaxy;  irregularities on much smaller scales would cause small-scale structural variations that are indistinguishable from other instabilities in the jet flows, while flows on much larger scales would simply appear as relative motion, producing C-shapes.  Second, the momentum flux in the external medium must be comparable to those in the jets, or they would not be perturbed.  In clusters of galaxies, where similarly distorted sources are seen, thermal particle densities are of order $10^{-3}$--$10^{-2}$ cm$^{-3}$, with velocities of order 100--1000\,km\,s\per.  It is hard to see how comparable momentum fluxes could be found in the surrounding medium in such an apparently sparse environment.  A more thorough examination of the environment of these galaxies would certainly be useful, as would modeling that examined the potential role of mutual orbital motions of the pair of galaxies.  \editout{Spectral indices at comparable resolutions would also be useful as indicators of the history of the flows, but are not available from the MGCLS for these sources because they were too far out in the primary beam.}

\editout{Beyond simple bending of radio galaxies, the presence of perturbations such as seen here should be part of any census of radio galaxy structures, in order to provide a fuller picture of the diffuse, lower density thermal plasmas both in- and out-side of rich clusters.}
 

\section{Star-forming Galaxies}\label{sec:galaxies}
Star-forming galaxies are fainter at radio frequencies than the AGN or radio galaxies presented in Section~\ref{sec:individsrcs}. With the depth of the MGCLS data, MeerKAT has imaged thousands of star-forming galaxies which are typically not detected in shallower all-sky surveys. Here we present some of the science possible with star-forming galaxies in the MGCLS. 

\subsection{\kk{Nearby s}pirals with MeerKAT}
Radio observations of spiral galaxies provide a dust-free view of their star-forming regions, and have imaged \textup{H\,\textsc{\lowercase{i}}} beyond the optical emission boundary. Most star-forming galaxies are unresolved in the MGCLS images. However several are resolved spirals, and here we present three examples.

\subsubsection{NGC~0685 and NGC~1566}
\begin{figure}
    \centering
    \includegraphics[width=\columnwidth,clip=True,trim=0 0 0 0]{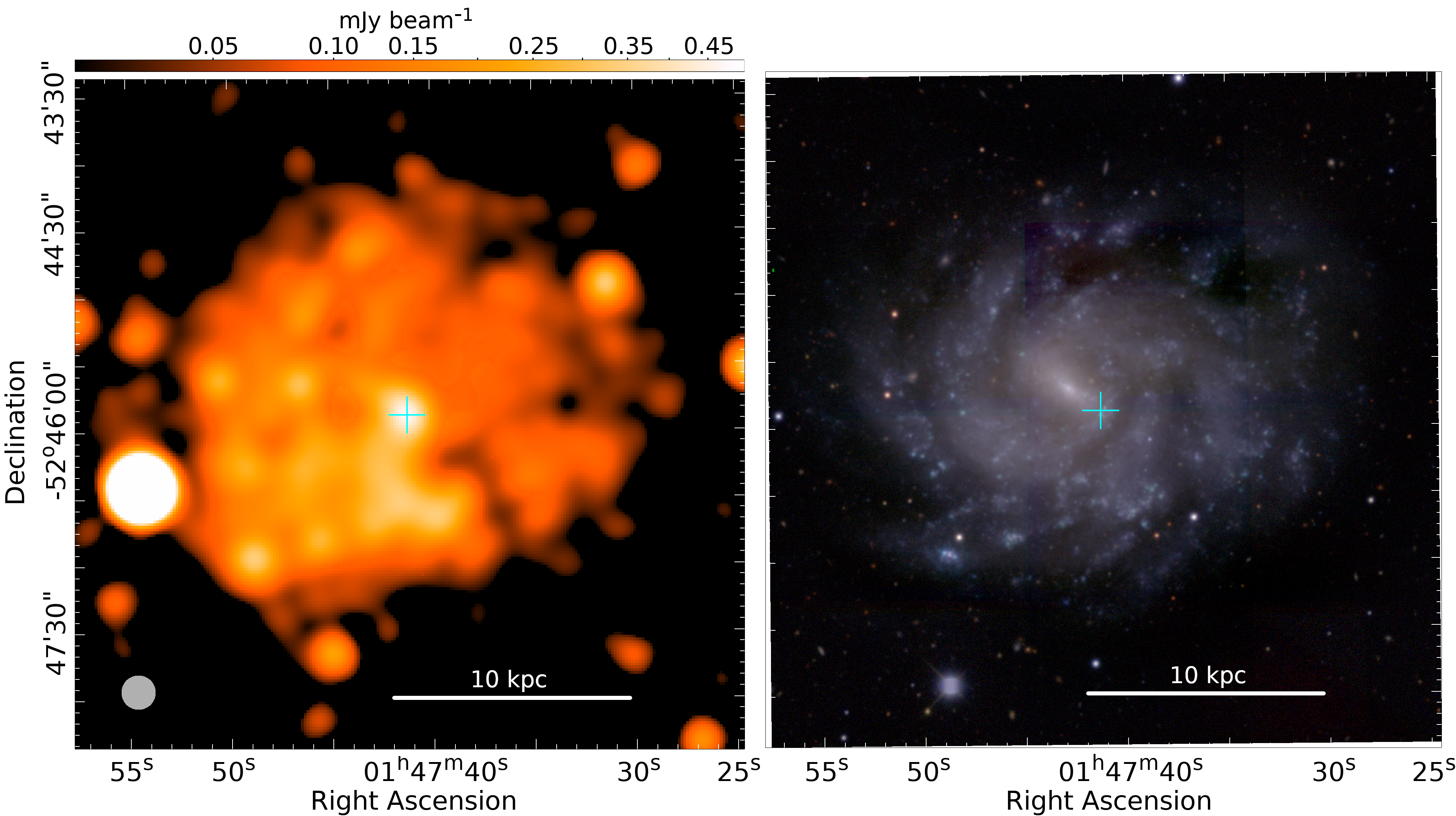}
    \caption{NGC~0685, a face-on barred spiral galaxy at $z=0.0045$. The physical scale is shown at lower right. \textbf{Left:} MGCLS 15\arcsec-resolution Stokes~I intensity image, with the synthesised beam shown by the filled grey ellipse at lower left. \textbf{Right:} DES composite $gri$ image, with star forming regions showing up in blue. The cyan cross indicates the position of the radio peak (470\,$\mu$Jy\,beam\per), aligned with the south-west edge of the galactic bar. }
    \label{fig:spiral-ngc0645}
\end{figure}

The left panel of Figure \ref{fig:spiral-ngc0645} shows the 15{\arcsec}-resolution MGCLS view of the face-on barred spiral galaxy \object{NGC\,0685}, which lies in the field of view and foreground of the MCXC~J0145.0$-$5300 cluster. NGC\,0685 ($z=0.0045$) has a corrected recession velocity of 1297\,\kms\ \citep{2000ApJ...529..786M} and is part of the HIPASS \textup{H\,\textsc{\lowercase{i}}} \citep{2013MNRAS.432.1178H} and GAIA DR2 \citep{2018yCat.1345....0G} catalogues. As expected, radio {peaks} are picked up in the regions of high star formation, indicated by the bluest regions in the composite DES colour image in the right panel of Figure \ref{fig:spiral-ngc0645}. 

\begin{figure}
    \centering
    \includegraphics[width=\columnwidth]{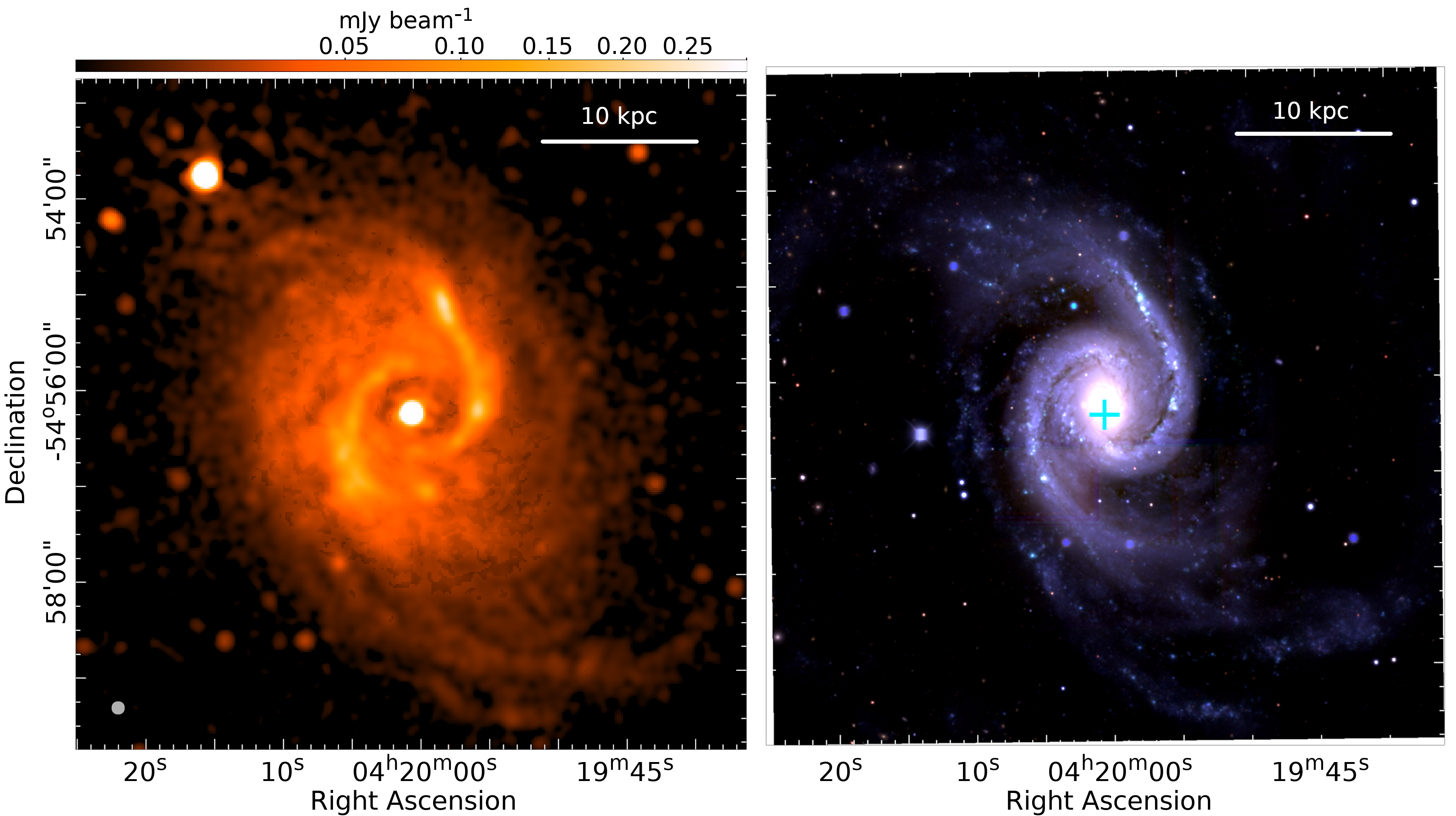}\\
    \caption{NGC\,1566, a face-on Seyfert galaxy at $z=0.0050$. The physical scale is shown at upper right. \textbf{Left:} MGCLS full-resolution (7.2\arcsec\,$\times$\,7.0\arcsec) Stokes~I intensity image, with the synthesised beam shown by the filled grey ellipse at lower left. \textbf{Right:} DES composite $gri$ image, with star forming regions shown in bright blue. The cyan cross indicates the position of the radio peak (31\,mJy\,beam\per), aligned with the galactic core. }
    \label{fig:spiral-ngc1566}
\end{figure}

Another spectacular example of a nearby face-on spiral in the MGCLS is \object{NGC\,1566} ($z = 0.0050$) in the foreground of the MCXC~J0416.7$-$5525 cluster field. This spiral galaxy is part of the IRAS Revised Bright Galaxy Sample \citep{2003AJ....126.1607S,2021arXiv210807891C} and has a Tully-Fisher-derived distance of 5.5\,Mpc \citep{2014MNRAS.444..527S}. The full-resolution MGCLS and $gri$-composite DES images of this source are shown in the left and right panels of Figure \ref{fig:spiral-ngc1566}, respectively. This face-on Seyfert galaxy has high levels of star formation occurring in the innermost regions of the spiral arms.  

\subsubsection{An atypical star formation ring?}

\begin{figure}
    \centering
    \includegraphics[width=\columnwidth,clip=True,trim=0 0 0 0]{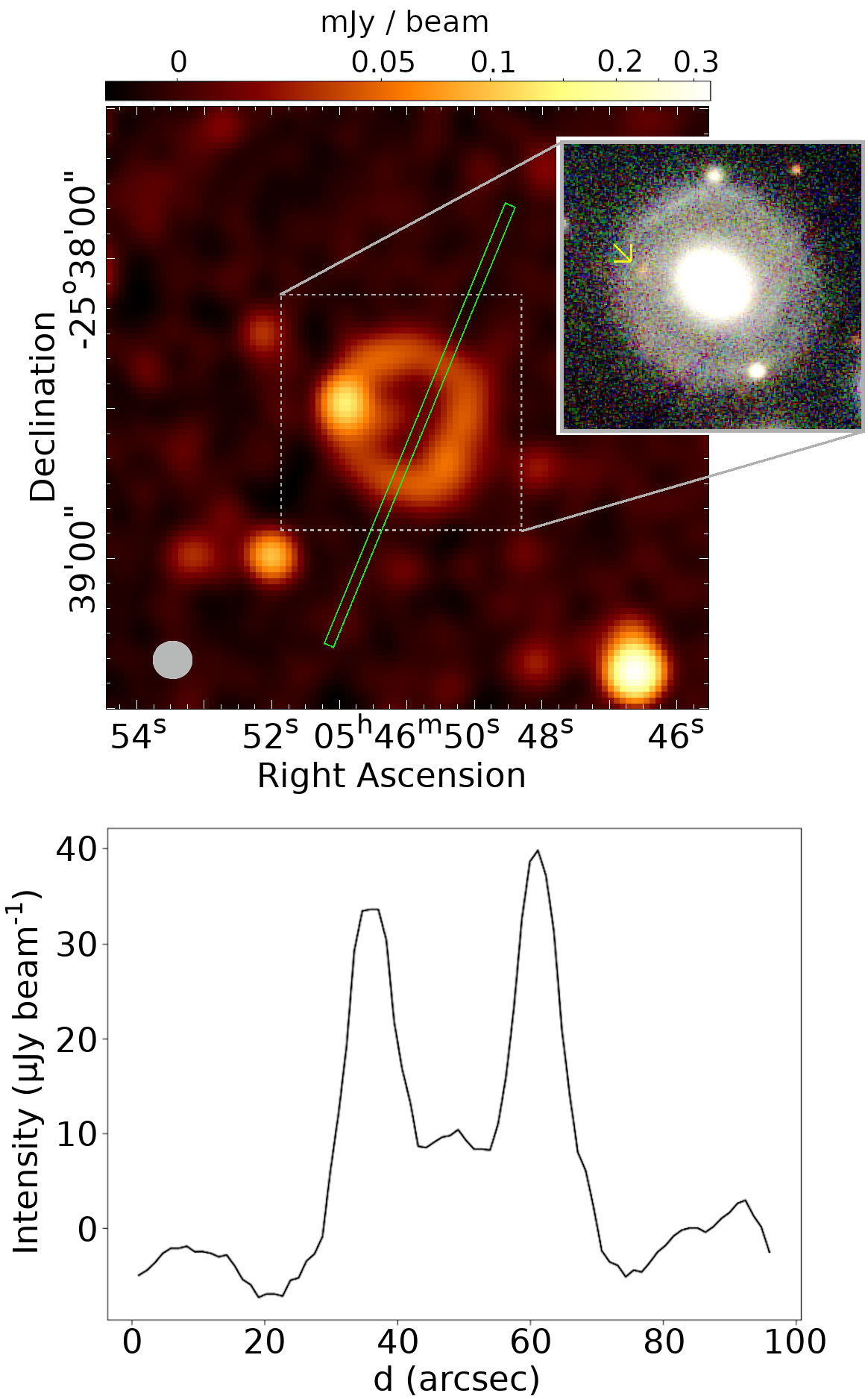}
    \caption{\textbf{Top:} 7.5{\arcsec}-resolution MGCLS Stokes~I intensity image of the ring-like source in the Abell~548 field, centred at $\rm R.A.= 05^h 46^m 50.0^s$, $\rm Dec. = -25^\circ 38^\prime 30.4$\arcsec. The inset shows the PanSTARRS $gri$-composite image. The ring hotspot is coincident with a background galaxy, indicated by the yellow arrow. \textbf{Bottom:} Profile through the centre of the radio ring (green region in the top panel, with the profile starting from the northern end), showing that the structure is filled. }
    \label{fig:spiralring}
\end{figure}

A ring-like radio source, shown in the top panel of Figure \ref{fig:spiralring}, is detected in the central region of the Abell~548 field. The ring, whose centre is at R.A.\,=\,$\rm 05^h46^m50.0^s$, Dec.\,=\,$\rm -25^\circ38^\prime30.4$\arcsec, has a mean brightness of 45\,$\mu$Jy\,beam\per\ in the full resolution (7.5\arcsec) image, with a 130~$\mu$Jy hotspot in the eastern section which has a typical AGN spectral index of $-0.7$. Rings in the radio sky can have several origins, including supernova remnants \citep{2015A&ARv..23....3D}, planetary nebulae \citep{1997MNRAS.284..815B}, star formation \citep{1984A&A...134..207H}, extreme AGN jet bending \citep{Rawes2018ring}, and gravitational galaxy-galaxy lensing \citep{1988Natur.333..537H}. Ring sources with no optical counterpart, dubbed `odd radio circles' \citep[ORCs,][]{Norris2020ORCS} have also been found. They may also be attributed to Lindblad resonances \citep[as in NGC~4736,][]{1976ApL....17..191S} or a past burst of star formation triggered by an interaction, as in the case of M31 \citep{2010ApJ...725..542H}.

The ring in Abell~548 does have an optical counterpart, as seen in the PanSTARRS image of the region shown in the inset of Figure~\ref{fig:spiralring}. The counterpart is a nearby Sab-type galaxy \citep{2006A&A...446...19T}, WISEA~J054650.08$-$253830.8, at $z=0.04653$, which has brighter arc regions on the northern and southern edges of the spiral disc. At this redshift, the ring spans $\sim$\,33\,kpc. The radio hotspot is coincident with a red background galaxy, indicated by the yellow arrow in the PanSTARRS image in Figure~\ref{fig:spiralring}, and therefore unlikely to be linked to the ring itself.

The radio ring emission has a steep spectrum, with the brightest regions (for which we can obtain a reliable fit) having a spectral index of approximately $-1$. Such steep, non-thermal spectra are more typical of Sb and later type galaxies \citep{1988MNRAS.231..465P}. A slice through the centre of the ring, shown in the bottom panel of Figure~\ref{fig:spiralring}, indicates that there is also emission in the interior of the ring, although there is no central peak typically associated with star forming spiral galaxies. If the ring is due to star formation, it may have been triggered by interaction with the companion galaxy ($z = 0.04648$) which lies 75.5\arcsec\ to the northeast, outside of the region shown in Figure~\ref{fig:spiralring}.

\subsection{Population studies in Abell~209}\label{sec:pops}

\kk{Galaxy clusters are populated by two broad classes of galaxies: red elliptical galaxies in which star formation has been quenched, and blue spiral galaxies with on-going star formation \citep[e.g.,][]{Baldry_2004, Taylor_2015, Haines_2017}. The latter types are more often found in the cluster outskirts. This leads to a relation between the fraction of star-forming galaxies and the projected distance from the cluster centre \citep[e.g.,][]{Lewis_2002, Gomez_2003, Haines_2015}.}

Here we present analyses of star formation rates (SFRs) and the radio/far-infrared (FIR) correlation for the Abell~209 cluster field \kk{as an example of the types of population studies possible with the MGCLS. We use} the optically cross-matched MGCLS compact source catalogue (see Section \ref{sec:optxmatch}). 
Abell~209, at a redshift of $z = 0.206$ and with $R_{200} = 2.15$\,Mpc\;\kk{(10.2\arcmin\ on the sky)}, is selected due to the availability of extensive spectroscopic catalogues from the Cluster Lensing And Supernova survey with Hubble \citep[CLASH-VLT,][]{Annunziatella_2016} and the Arizona Cluster Redshift Survey\footnote{\url{https://herschel.as.arizona.edu/acres}} \citep[ACReS, described in][]{Haines_2015} \kk{ for the identification of cluster galaxies.}\editout{; this allows for the classification of detected radio sources as cluster or field galaxies. The $R_{200}$ for Abell~209 corresponds to 10.2' on the sky; g} \kk{G}iven MeerKAT's large primary beam, these studies can therefore be performed out to \editout{ an unprecedented (in the radio regime)}3.5\,$R_{200}$ \kk{in this cluster, unprecedented in the radio regime}. 

\begin{figure}
\includegraphics[width=0.98\columnwidth,clip=True,trim=0 0 40 10]{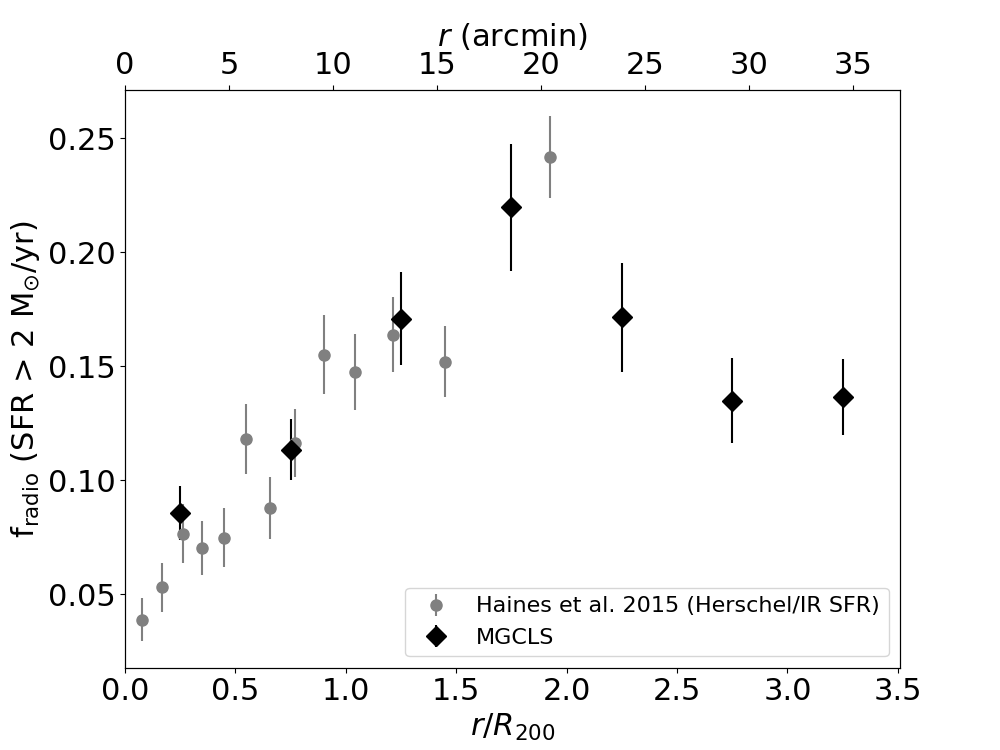}
\caption{The fraction of star-forming cluster members in Abell~209 detected in the radio (black diamonds) as a function of projected radial distance from the cluster centre, in units of $R_{200}$ (bottom) and in angular units (top). We only plot MGCLS sources with SFR greater than the 5$\sigma$ detection limit of SFR$_{5\sigma}\,=$\,2\,M$_{\odot}$\,yr$^{-1}$ (see Appendix~\ref{app:popstudies} for details). Grey circles show the fraction of star-forming galaxies using IR-derived data from \citet{Haines_2015} for 30 clusters. Error bars show $1\,\sigma$ uncertainties. A similar trend is seen out to 2\,$R_{200}$ for both the radio- and IR-derived fractions. 
}
\label{fig:fracsfgs}
\end{figure}

\kk{After assigning cluster membership and excising AGN (see Appendix~\ref{app:popstudies} for details), we obtain a final catalogue of 459 MGCLS-detected star-forming cluster members within the primary beam-corrected field of view (80 within $R_{200}$).}\footnote{\kk{This catalogue forms the basis of a value-added catalogue for Abell~209 being provided with the MGCLS DR1 which includes radio-derived SFRs for each galaxy and the ratio of radio-to-FIR flux densities (where available). See Appendix~\ref{app:popstudies} for details of these additional measurements and Table~\ref{tab:sfrir} for an excerpt of the catalogue.}} Figure~\ref{fig:fracsfgs} shows the fraction of star forming cluster members that are detected by MeerKAT, $f_{\rm radio} = N_{\rm MGCLS} / N_{\rm optIR}$ (with SFR\,$>$\,SFR$_{5\sigma}$), as a function of \kk{angular distance and }projected distance (in units of $R_{200} = 2.15$\,Mpc) from the cluster centre. Here $N_{\rm optIR}$ is the number of cluster member galaxies determined from the optical and IR catalogues after removing AGN contamination (2476 within the field of view). We see that the fraction is lower in the cluster centre ($f_{\rm radio} < 0.1$), and rises to $\sim 0.2$ by $2\,R_{200}$. The fall in the last three radial bins is at least in part due to decreased sensitivity in the MGCLS observations \kk{due to the MeerKAT primary beam}.\editout{farther out in the primary beam, partly due to increased local noise from bright sources at the edge of the MGCLS image.} For comparison, we show results from \citet{Haines_2015} of the fraction of star forming galaxies in a sample of 30 $0.15 < z < 0.30$ clusters, as measured using \textit{Herschel} observations. These probe down to the same SFR limit as the MGCLS data, assume the same \citet{Chabrier_2003} IMF as this work, and are not affected by dust extinction. In the region of overlap, we see a similar trend in the IR-derived results. \citet{Haines_2015} interpreted their result as evidence for relatively slow quenching of star formation in cluster galaxies over a $\approx 2$\,Gyr timescale.

\begin{figure}
\includegraphics[width=0.98\columnwidth,clip=True,trim=0 0 40 60]{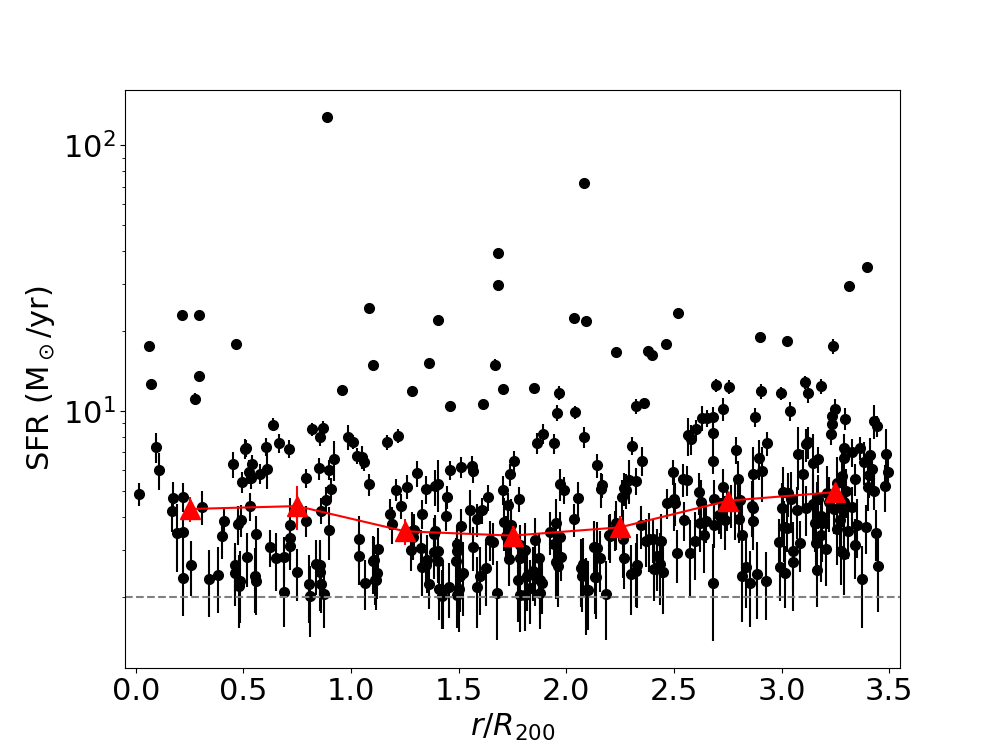}
\caption{SFR versus projected radial distance from the Abell~209 cluster centre, in units of $R_{200}$, \kk{for the 429 star-forming cluster galaxies} in the full primary beam-corrected MGCLS field of view. The red triangles indicate the median SFR plotted in radial bins, and show no evolution with radial distance. The grey dashed line is the SFR at the MGCLS 5$\sigma$ sensitivity limit. All error bars show the 1$\sigma$ uncertainty. }
\label{fig:binned_SFR_colored}
\end{figure}

\subsubsection{Star formation rates}\label{sec:sfrs}

\editout{Galaxy clusters are populated by two broad classes of galaxies: red elliptical galaxies in which star formation has been quenched, and blue spiral galaxies with on-going star formation 
The latter types are more often found in the cluster outskirts. This leads to a relation between the fraction of star-forming galaxies and the projected distance from the cluster centre}

In the last few decades there have been many efforts to measure the \editout{star formation rates (}SFRs\editout{)} for galaxies in a range of environments using various tracers \citep[e.g.,][for H$\alpha$, IR, and radio measurements respectively]{2004A&A...414...23J, Elbaz_2007, Karim2011}. Here we use the MGCLS radio continuum luminosities\editout{of star-forming galaxies in Abell~209}
\editout{A follow-up study (Kesebonye et al., in prep) will expand this to more of the MGCLS clusters to examine star formation as a function of cluster mass and redshift.} to estimate \editout{their }SFRs \kk{for the star-forming galaxies in Abell~209 (details of the SFR determination are provided in Appendix~\ref{app:popstudies})}. The \editout{former}\kk{radio signal} is a combination of thermal free-free (bremsstrahlung) and non-thermal synchrotron \citep[e.g.,][]{Condon1992}. The non-thermal emission typically dominates at frequencies below 5\,GHz.\editout{and the thermal free-free emission contributes only weakly.} The 1.28\,GHz MGCLS luminosities therefore provide us with dust-unbiased measurements over a $\approx$\,1\,deg$^2$ field centred on the cluster, probing \kk{radio-derived SFRs} out to well beyond $R_{200}$ \kk{for the first time}. 

\editout{The top panel of Figure~}
\editout{shows the fraction of star forming cluster members that are detected by MeerKAT, $f_{\rm radio} = N_{\rm MGCLS} / N_{\rm optIR}$ (with SFR\,$>$\,SFR$_{5\sigma}$), as a function of projected distance (in units of $R_{200} = 2.15$\,Mpc) from the cluster centre. Here $N_{\rm optIR}$ is the number of cluster member galaxies determined from the optical and IR catalogues after removing AGN contamination (2476 within the field of view). We see that the fraction is lower in the cluster centre ($f_{\rm radio} < 0.1$), and rises to $\sim 0.2$ by $2\,R_{200}$. The fall in the last three radial bins is at least in part due to decreased sensitivity in the MGCLS observations farther out in the primary beam, partly due to increased local noise from bright sources at the edge of the MGCLS image. For comparison, we show results from }
\editout{of the fraction of star forming galaxies in a sample of 30 $0.15 < z < 0.30$ clusters, as measured using \textit{Herschel} observations. These probe down to the same SFR limit as the MGCLS data, assume the same}
\editout{IMF as this work, and are not affected by dust extinction. In the region of overlap, we see a similar trend in the IR-derived results. }
\editout{interpreted their result as evidence for relatively slow quenching of star formation in cluster galaxies over a $\approx 2$\,Gyr timescale.}

\editout{The bottom panel of }Figure~\ref{fig:binned_SFR_colored} shows the SFR of the MGCLS-detected Abell~209 \kk{star-forming} member galaxies as a function of the projected distance from the cluster centre, under the assumption that the observed radio emission is due to star formation. We find no dependence of the SFR on distance from the cluster centre, based on the median SFR of the member galaxies. This is consistent with the star formation quenching process taking place over an extended time period which is longer than the infall time, i.e., galaxies that are forming stars are not immediately quenched upon encountering the cluster environment. \editnew{A similar result was found for the highest-mass star forming galaxies in nearby clusters from the SDSS \citep{2010MNRAS.404.1231V}, however the lower mass galaxies showed more significant star formation quenching within 0.2$R_{200}$.}

\subsubsection{Radio/FIR ratio}\label{sec:rfir}
\begin{figure}
    \centering
    \includegraphics[width=0.98\columnwidth]{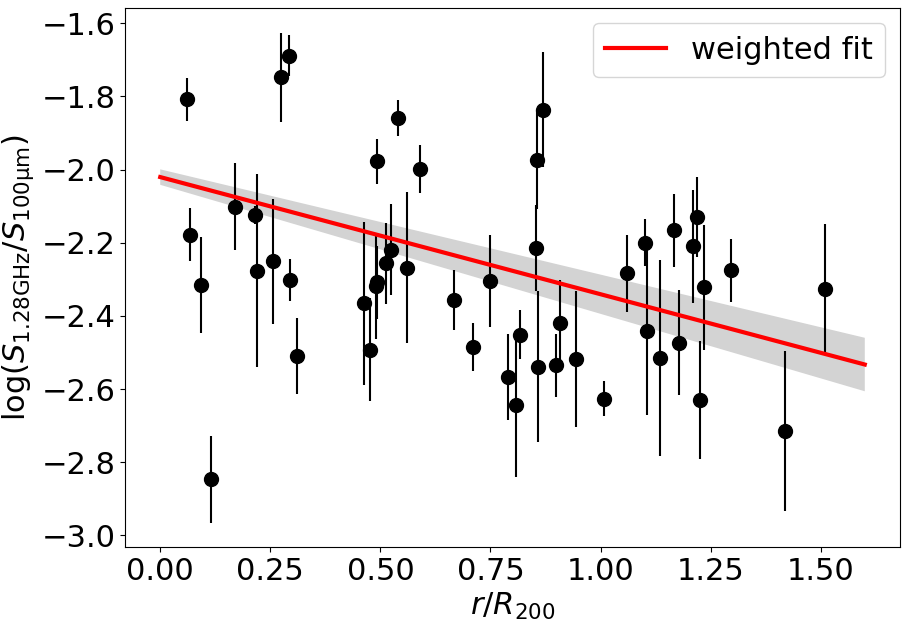}
    \caption{{The log ratio of radio-to-FIR flux densities for 49 star-forming cluster galaxies in Abell~209 as a function of projected radius from the cluster centre in units of $R_{200}$.} Black circles indicate the {star forming} galaxies in the Abell~209 cluster \editnew{which have MGCLS radio and \textit{Herschel} FIR flux densities}. All error bars are 1$\sigma$ uncertainties. The red line is the weighted least-squares fit given in Eqn. \ref{eqn:irfit}, \editnew{with the grey region showing the 1$\sigma$ uncertainty of the fit}.
    }
    \label{fig:qIR}
\end{figure}
The 1.4\,GHz luminosities of star-forming galaxies are tightly correlated with their \editout{far-infrared (}FIR\editout{)} luminosities \citep{Condon1992}, {with} both indicating the rate of star formation. Looking at nearby galaxy clusters ($z < 0.025$), a small but statistically significant enhancement of the radio/FIR ratio has been found for cluster galaxies as compared to the field galaxies \citep{Reddy2004, Murphy2009}. \citet{Gavazzi1991} suggest this enhancement could be due to the ram pressure from the interaction with the ICM, which compresses the star-forming gas and amplifies the embedded magnetic fields. \kk{Using cross-matched \textit{Herschel} 100\,$\mu$m flux densities for 49 Abell~209 star-forming galaxies (see Appendix~\ref{app:popstudies}), w}\editout{W}e examine the \editout{correlation between the ratio of MGCLS versus FIR flux densities, $S_{\rm 1.28\,GHz}/S_{\rm 100\,\mu m}$, and the projected distance from the cluster centre (as a proxy for ICM density). }\kk{radio/FIR correlation looking for evidence of evolution with distance from the cluster centre.}

Figure~\ref{fig:qIR} shows the ratio of radio-to-FIR flux density \kk{for Abell~209} as a function of projected distance from the cluster centre in units of $R_{200}$. We perform a non-linear weighted\footnote{Weights are the 1$\sigma$ uncertainties in the log ratio.} least-squares fit to a straight line using the log ratio vs $r/R_{200}$. We obtain a best-fit relation of 
\begin{equation}
\label{eqn:irfit}
{\rm log}\left(\frac{S_{\rm 1.28\,GHz}}{S_{\rm 100\,\mu m}}\right)=(-0.32\pm0.03) \frac{r}{R_{200}} - 2.02\pm0.02.
\end{equation}
We see a statistically significant trend for the radio/FIR ratio to increase with decreasing projected distance from the cluster centre. Our results are in line with findings from \citet{Murphy2009}, who argued that this primarily arises from ram pressure stripping. The amount of ram pressure is expected to be proportional to the ICM density, but should also be dependent on cluster richness and mass. The MGCLS provides a range of clusters with which to study this further, and potentially disentangle the various effects, as well as a large enough sample of galaxies in individual clusters to look for dependencies on galaxy properties.

\section{HI Science Highlights}\label{sec:hiscience}
\begin{figure*}
    \centering
    \includegraphics[width=0.98\textwidth]{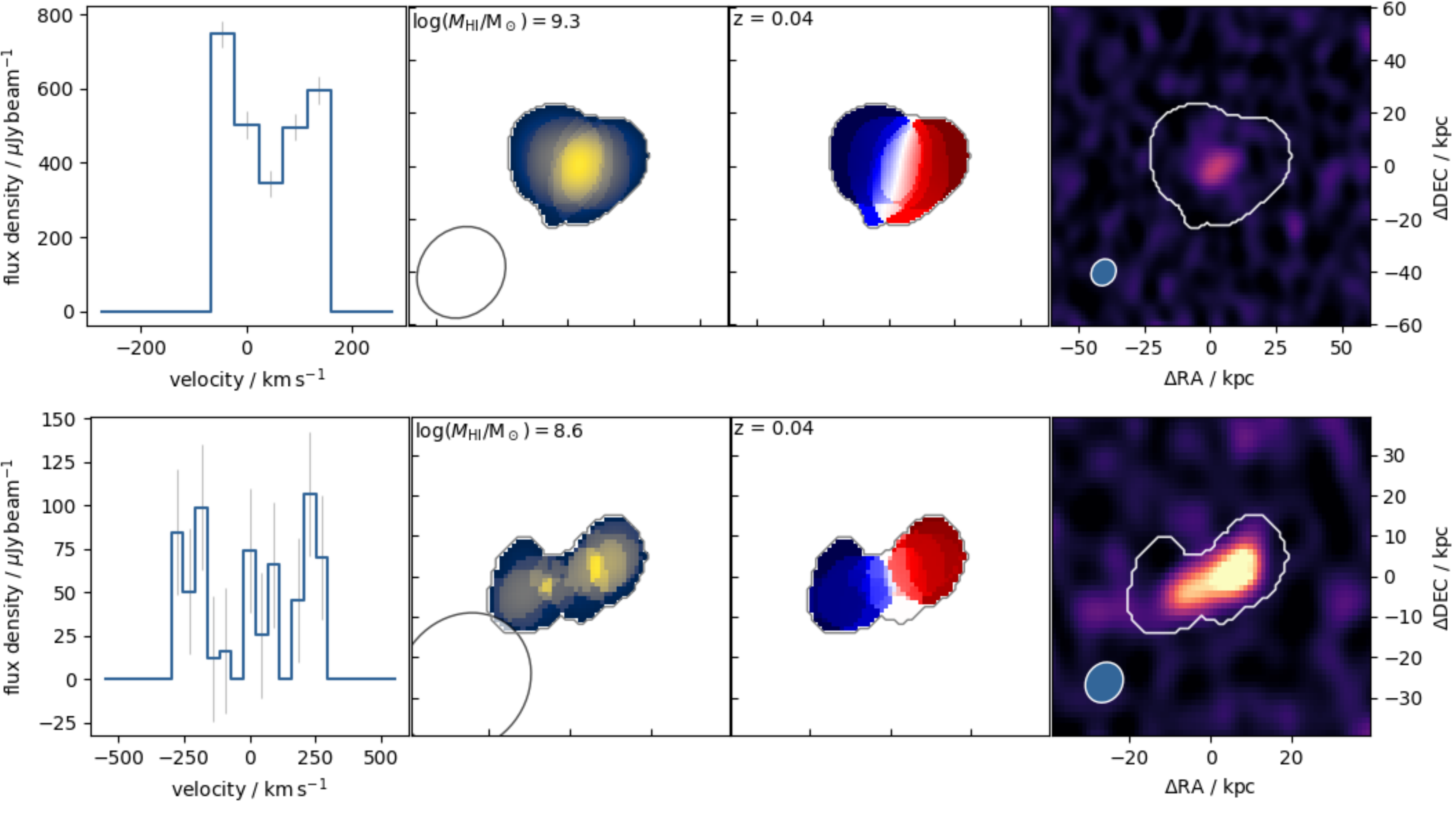}
    \caption{
    Two demonstration examples of complementary \textup{H\,\textsc{\lowercase{i}}} and radio continuum science in the MGCLS from the Abell~3365 field. Both rows show \textup{H\,\textsc{\lowercase{i}}} disks at $z\sim0.04$ with radio continuum counterparts. 
    {\bf Left:} Integrated \textup{H\,\textsc{\lowercase{i}}} spectrum with the per-channel standard deviation indicated by grey vertical bars. 
    {\bf Centre left:} Total intensity \textup{H\,\textsc{\lowercase{i}}} map with the synthesised beam indicated at lower left. The contour outlines the lowest level emission deemed real by the {\sc SoFiA} source finding software (see Section~\ref{sec:hi-srcfinding} for details). The colour scale (from yellow to blue) shows the \textup{H\,\textsc{\lowercase{i}}} flux density from the left panel.
    {\bf Centre right:} \textup{H\,\textsc{\lowercase{i}}} velocity map, with the colour scale (blue to red) set to the velocities for which there is non-zero flux density in the left panel. The contour is the same as in the centre left panel. 
    {\bf Right:} Radio continuum intensity map, with the continuum synthesised beam indicated at lower left. The colour range is between $-20$ and 200 $\mu$Jy\,beam\per, and the contour is the same as the centre left panel. 
    }
    \label{fig:hi-example-AGN-SF}
\end{figure*}

\kk{In conjunction with the continuum science discussed in previous sections, o}ne of the great strengths of the MGCLS dataset is in its usefulness for mapping large volumes in \textup{H\,\textsc{\lowercase{i}}} \kk{(see Section~\ref{sec:uc-hi})}. 
The sensitivity of the observations make the MGCLS a rich resource for galaxy evolution studies employing the \textup{H\,\textsc{\lowercase{i}}} line, especially in cases where there is ancillary data at other wavelengths available. Commensal \textup{H\,\textsc{\lowercase{i}}} science is possible not only within the clusters themselves, but also in the foreground and background as a result of the large \textup{H\,\textsc{\lowercase{i}}} volume probed by the long-track MGCLS observations. \lr{The \textup{H\,\textsc{\lowercase{i}}} data products, themselves, are not included in the current DR1 data release.}

\kk{\textup{H\,\textsc{\lowercase{i}}} MGCLS studies} can make effective use of two low-RFI windows, between 1300--1420\,MHz ($0 < z < 0.09$) and 
between 960--1190\,MHz ($0.19 < z < 0.48$). While the first enables \editout{us to effectively sample}\kk{sampling of} the \textup{H\,\textsc{\lowercase{i}}} mass function at lower redshifts, the second provides a glance into the most massive systems at higher redshift through statistical means and potentially by strong lensing of the clusters themselves. For cases where radio continuum and \textup{H\,\textsc{\lowercase{i}}} sources are spatially resolved,  
\editout{we can study the} \textup{H\,\textsc{\lowercase{i}}} outflows (e.g., in starburst or AGN systems) \kk{can be studied}, and \kk{comparisons can be made between the} star formation and \textup{H\,\textsc{\lowercase{i}}} mass properties on both an individual and statistical basis. To demonstrate this, Figure~\ref{fig:hi-example-AGN-SF} shows two examples of \textup{H\,\textsc{\lowercase{i}}} disks in the MGCLS data, with very different radio continuum properties. 

\lr{To} illustrate the \textup{H\,\textsc{\lowercase{i}}} science possible with this survey \editout{by examining }\kk{we present early results from an examination of} four representative galaxy clusters\kk{, selected primarily to demonstrate data quality and potential science. The clusters, Abell~194, Abell~4038, Abell~3562, and Abell~3365, lie in the redshift range $0.01 < z < 0.1$ and }\lr{were selected from a heterogeneous catalogue of clusters detected in the 0.1--2.4\,keV X-ray band \citep{MCXC}. }\lr{They} \kk{cover a reasonable range of X-ray luminosity ($L_X\,\approx\,(0.07$--1.3)$\,\times\,10^{44}$~erg\,s\per) and cluster virial mass ($M_v \approx 0.4$--2.4$\,\times\,10^{14}\,{\rm M}_\odot$)}. 
\editout{Commensal HI science is possible not only within these clusters, but also in the foreground and background as a result of the large HI volume probed by the long-track MGCLS observations. }

\editout{The HI science presented here includes studying the distribution of HI masses of cluster galaxies, and examining the HI morphology of galaxies as a function of their environment. This is not an exhaustive list of the HI science possible with this legacy dataset, and further exploitation is ongoing. }

\subsection{\kk{\textup{H\,\textsc{\lowercase{i}}} data processing}}
\kk{In the following two sections we describe the methodology followed to extract \textup{H\,\textsc{\lowercase{i}}} cubes for the four cluster datasets from the MGCLS visibilities, and the procedure used to carry out \textup{H\,\textsc{\lowercase{i}}} source finding.}

\subsubsection{\textup{H\,\textsc{\lowercase{i}}} cubes}

\lr{W}e created \textup{H\,\textsc{\lowercase{i}}} data cubes within the frequency interval of 1305--1430\,MHz ($z \lesssim 0.088$). The data reduction was conducted with the \textsc{CARACal}\footnote{\url{https://caracal.readthedocs.io/en/latest/}} \citep{Jozsa2020}, and {\textsc{oxkat}}\footnote{\url{https://github.com/IanHeywood/oxkat}} \citep{Heywood2020} pipelines. The former makes use of \textsc{stimela}\footnote{\url{https://github.com/ratt-ru/Stimela}} which is a radio interferometry scripting framework based on container technologies and Python \citep{Offringa2010}. Within this framework, various open-source radio interferometry software packages were used to perform all necessary procedures from cross calibration to imaging.

The final \textup{H\,\textsc{\lowercase{i}}} cubes of the four clusters cover a field of view of 1.0\,deg$^{2}$ each, have a median RMS noise level of $\sim$\,0.1\,mJy\,beam\per\ per 44.1\,\kms\ channel, and a median FWHM Gaussian restoring beam of $\sim18\arcsec$ (imaged with natural weighting). This noise level results in a typical $5\,\sigma$ \textup{H\,\textsc{\lowercase{i}}} column density sensitivity of $9 \times 10^{19}$\,cm$^{-2}$ (0.72\,M$_\odot$\,pc$^{-2}$) over a line width of $44.1$\,\kms.

\subsubsection{\textup{H\,\textsc{\lowercase{i}}} source finding} \label{sec:hi-srcfinding}
We used the \textup{H\,\textsc{\lowercase{i}}} Source Finding Application \citep[\textsc{SoFiA},][]{Serra2015} to search for line emission from each of the cubes. For this purpose, various tests were conducted with \textsc{SoFiA} noise threshold filters, smoothing kernels, and reliability parameters to ensure optimal source finding which reduced the number of false positives, and ensured that the low \textup{H\,\textsc{\lowercase{i}}} surface brightness is also properly detected. We used the smooth and clip (S$+$C) method \citep{2012MNRAS.422.1835S}, with a noise threshold of 4 times the RMS noise and spatial smoothing kernels corresponding to 1.0, 1.5, and 2.0 times the synthesised beam. Given the coarse 44.1\,\kms\ velocity resolution of the MGCLS data, no smoothing was done in velocity. The catalogue of detections was compiled retaining only the positive voxels with an integrated SNR $>3.5$ and reliability parameter $>0.99$. This high \textsc{SoFiA} reliability was chosen to limit the rate of potential false detections. All detections were additionally validated by eye.

\subsection{\kk{\textup{H\,\textsc{\lowercase{i}}} science examples}}

Here we highlight a select few of the results from this initial set of cluster analysis. As previously emphasised, these include \textup{H\,\textsc{\lowercase{i}}} detections within the clusters themselves, as well in their foregrounds and backgrounds.

\subsubsection{The distribution of \textup{H\,\textsc{\lowercase{i}}} masses }
\begin{figure}
 \centering
 \includegraphics[width=\columnwidth,clip=True,trim=0 0 0 0]{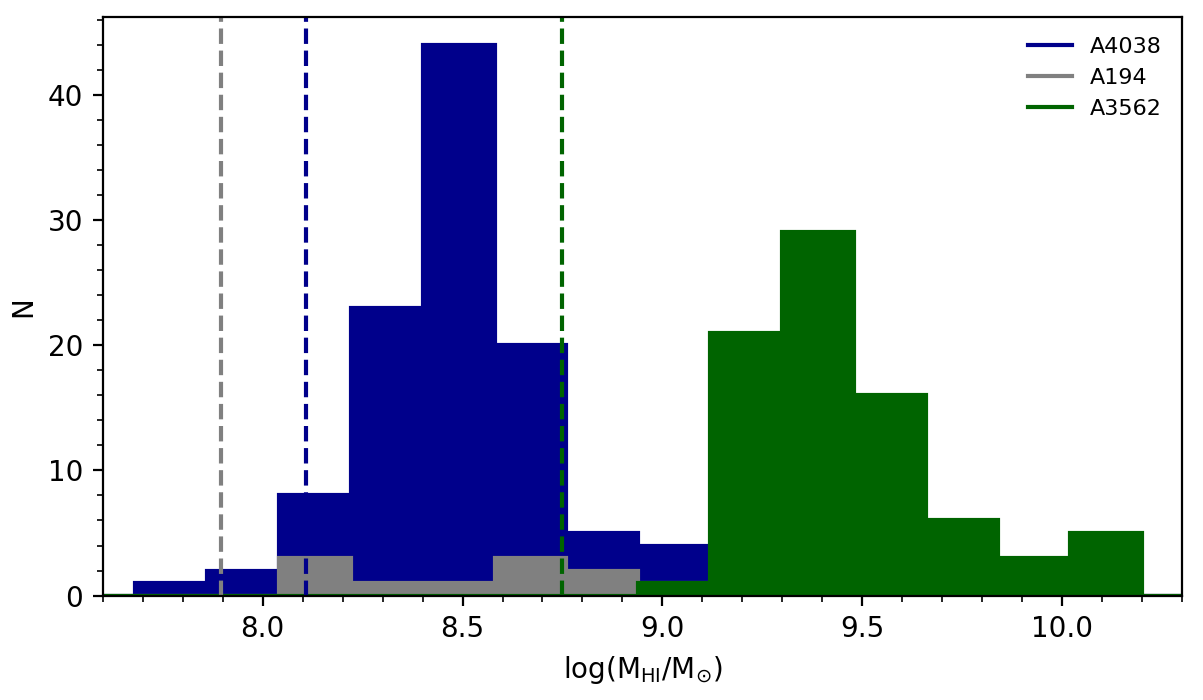}
    \caption{\kk{The H\textsc{i} mass distribution of detected galaxies in Abell~194, Abell~3562, and Abell~4038. Dashed vertical lines indicate the H\textsc{i} mass detection limits at the distances of each cluster, assuming a galaxy line width of 150 km\,s$^{-1}$ detected above 4$\sigma$.}
    }
    \label{fig:hi-MHIcumul}
 \end{figure}

Figure\,\ref{fig:hi-MHIcumul} shows the \textup{H\,\textsc{\lowercase{i}}} mass distribution of all galaxies detected in \textup{H\,\textsc{\lowercase{i}}} in clusters Abell~194 ($z = 0.018$), Abell~4038 ($z = 0.028$), and Abell~3562 ($z = 0.049$). The \textup{H\,\textsc{\lowercase{i}}} detection cluster membership was defined within the velocity dispersion of the clusters and the entire 1\,deg$^2$ field of view and may therefore contain interlopers. Abell~3365 ($z = 0.093$) is not included here as the frequencies corresponding to its redshift were heavily contaminated by RFI -- a demonstration of the upper redshift limit of the $0 < z \lesssim 0.09$ RFI-free window. 

The mass distributions in these three clusters point towards different \textup{H\,\textsc{\lowercase{i}}} populations in different clusters. The \textup{H\,\textsc{\lowercase{i}}} (redshift-dependent) mass detection limit lies below the low-mass drop in the detection rate for Abell~3562, and the lack of low-mass detections in this system is therefore not sensitivity related. By comparison, Abell~4038 shows a deficiency in high-mass \textup{H\,\textsc{\lowercase{i}}} detections. These two systems are of comparable mass ($2.4\,\times\,10^{14}\,{\rm M}_\odot$ vs $2.0\,\times\,10^{14}\,{\rm M}_\odot$), however Abell~3562 is part of the rich Shapley supercluster \citep{1991MNRAS.248..101R} and this environment may impact the \textup{H\,\textsc{\lowercase{i}}} mass in the system. Abell~194 is almost an order of magnitude less massive ($4.0\,\times\,10^{13}\,{\rm M}_\odot$) which may account for the comparatively low number of \textup{H\,\textsc{\lowercase{i}}} galaxies detected in the system.

 \subsubsection{Cluster \textup{H\,\textsc{\lowercase{i}}}: Abell~194}

\begin{figure}
    \centering
 \includegraphics[width=0.85\columnwidth]{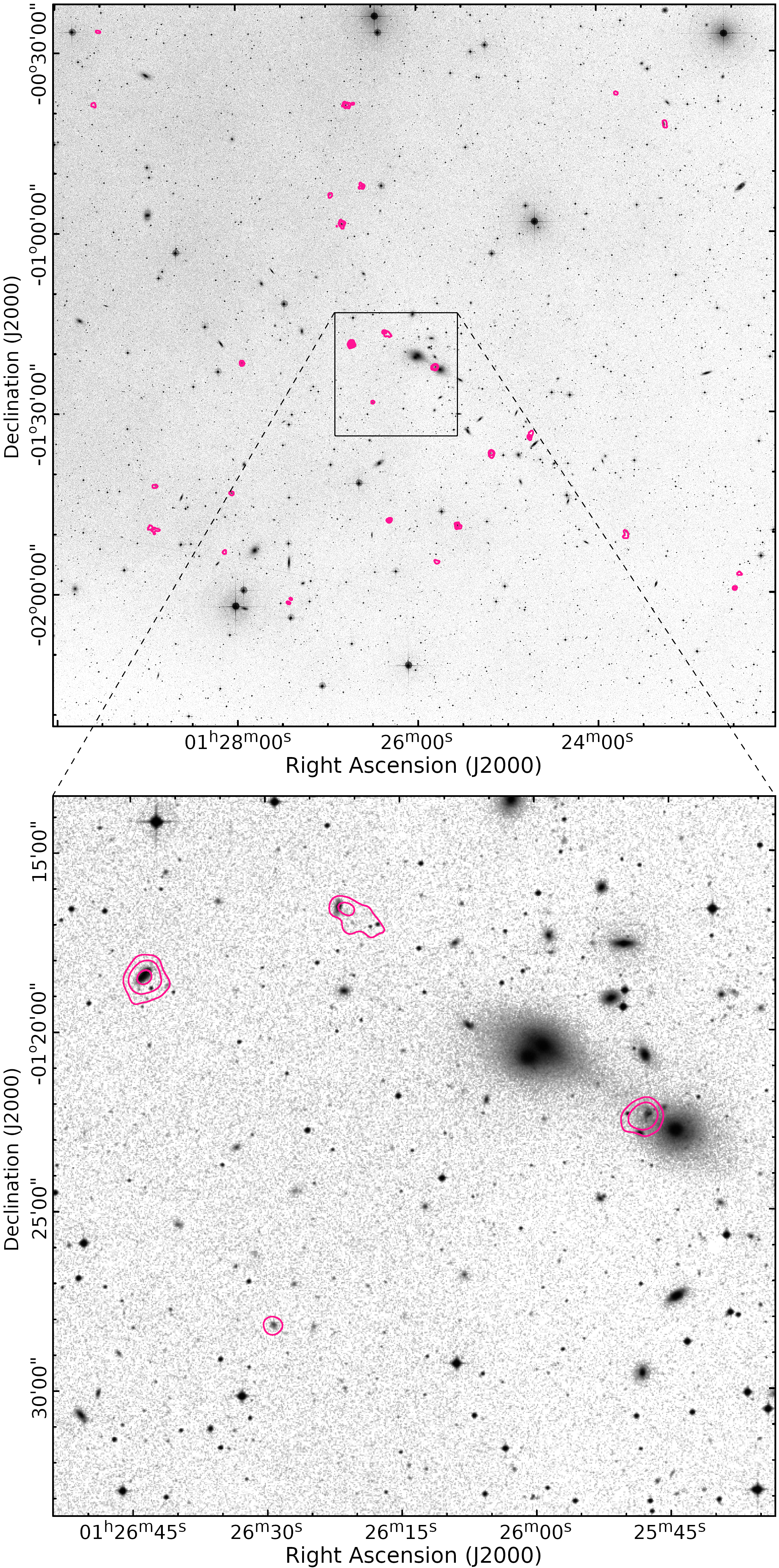}
    \caption{\textup{H\,\textsc{\lowercase{i}}} detections in the Abell~194 field, with contours showing the \textup{H\,\textsc{\lowercase{i}}} column-density ($\rm0.25, 1, 4~M_\odot\,{\rm pc}^{-2}$ levels),
    overlaid on DSS1 optical images. The \textup{H\,\textsc{\lowercase{i}}} resolution is 19\arcsec\,$\times$\,14\arcsec, p.a.\,147$^\circ$. \textbf{Top:} The full cluster region
    showing more than 25 \textup{H\,\textsc{\lowercase{i}}} detections. \textbf{Bottom:} A zoom-in of the boxed region from the top panel, showing the richness of resolved
    structures. Both bound \textup{H\,\textsc{\lowercase{i}}} and that in the process of being stripped from galaxies are evident. The compact \textup{H\,\textsc{\lowercase{i}}} source at
    R.A.\,$\rm=\,01^h25^m47^s$, Dec.\,$=-01^{\circ}22^{\prime}18$\arcsec\ is Minkowski's object, shown also in Figure~\ref{fig:HI}D.} 
    \label{fig:hi-A194}
\end{figure}

In Figure~\ref{fig:hi-A194}, we show the \textup{H\,\textsc{\lowercase{i}}} total intensity map of Abell~194 overlaid on an optical image. This shows the extent, richness, and some of the presumably environment-driven morphological transformation processes that influence the growth and evolution of these cluster members. By studying the full sample of appropriate clusters in the MGCLS sample, we will be able to investigate these effects on a statistical basis with a relatively uniform set of observations, as well as find rare extreme examples, given the large sample size.

\subsubsection{Foreground \textup{H\,\textsc{\lowercase{i}}}: Discovery of a new \textup{H\,\textsc{\lowercase{i}}} group}
\begin{figure}
    \centering
    \includegraphics[width=0.9\columnwidth]{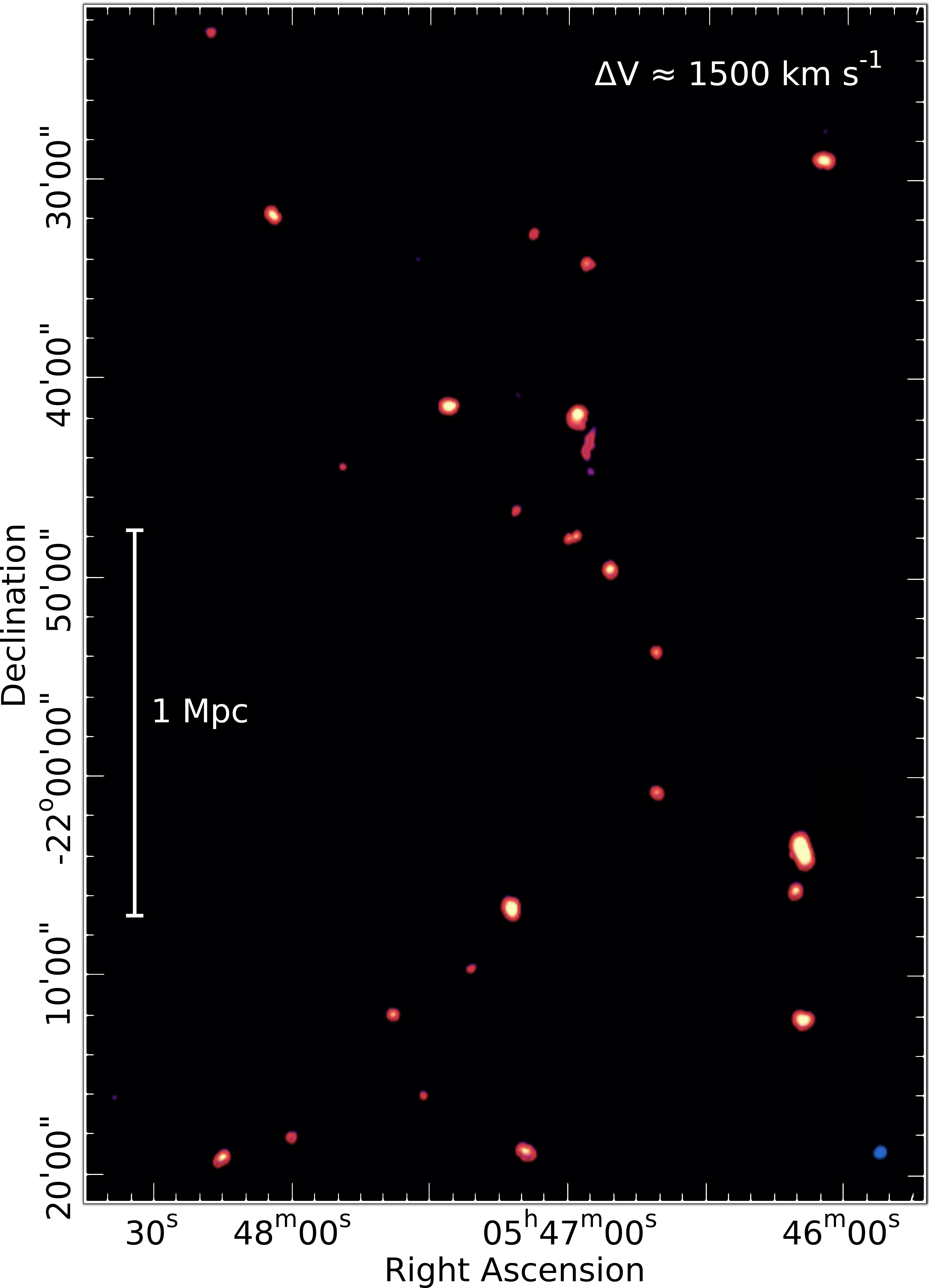}
    \caption{{The \textup{H\,\textsc{\lowercase{i}}} moment-0 map of} the newly-discovered \textup{H\,\textsc{\lowercase{i}}} group at $z = 0.040$, which is in the foreground of the Abell~3365 cluster at $z = 0.093$. The colour scale has a linear stretch and is clipped outside of the range of 0 to 150\,Jy\,beam$^{-1}$\,Hz. A conservative estimate is that this group has a total of 26 members within a radius of $\sim2$ Mpc and a velocity range of $\Delta V \sim 1500$\,km\,s$^{-1}$, with a mass range $8 \lesssim \log(M_{\rm HI}/{\rm M_\odot}) \lesssim 10$. The tapered beam is shown at lower right and has dimension of 30\arcsec\,$\times$\,26\arcsec. The physical scale at the group redshift is indicated. }
    \label{fig:hi-group_a3365}
\end{figure}

The \textup{H\,\textsc{\lowercase{i}}} imaging of Abell~3365 was compromised by severe RFI at the frequency corresponding to $z = 0.0926$. However, this dataset showed the value of our strategy of imaging the full $0 < z \lesssim 0.1$ range through the serendipitous discovery of a massive \textup{H\,\textsc{\lowercase{i}}} group in the foreground of Abell~3365 at $z = 0.040$. This group, with a dynamical mass of $M_{\rm{dyn}} \sim 10^{13}$~M$_\odot$, has at least 26 members, some of which have disturbed and asymmetric \textup{H\,\textsc{\lowercase{i}}} morphologies as seen towards the centre of the group's \textup{H\,\textsc{\lowercase{i}}} moment-0 map shown in Figure~\ref{fig:hi-group_a3365}. 


\section{Summary and Conclusions}\label{sec:conclusion}
We have presented a technical description and initial science results from the MeerKAT Galaxy Cluster Legacy Survey (MGCLS) first data release (DR1), which consists of 1.28\,GHz observations of 115 radio- and X-ray-selected galaxy clusters. The full list of MGCLS targets and associated legacy product status is available at the end of this paper in Table \ref{tab:sample}. We note that the clusters were not selected to be a homogeneous or complete sample, so care must be taken in any statistical investigation of our cluster sample as a whole.

The DR1 products available are comprised of visibilities, full field of view basic cubes without primary beam correction, and primary beam-corrected full-resolution (7--8{\arcsec}) and convolved (15{\arcsec}) Stokes-I intensity and spectral index cubes, and full-resolution primary beam-corrected 12-frequency spectral cubes. For 40\% of the clusters, full Stokes cubes suitable for Faraday rotation analysis, and limited purpose single-plane Q and U maps, are also provided. The data products are available via the DR1 webpage\footnote{\mgclsdoi} and the MGCLS website\footnote{\url{http://mgcls.sarao.ac.za}}. The website will be updated when additional data products become available.

To facilitate community usage of the MGCLS DR1 products we produced a compact source catalogue with more than 626,000 radio sources over the 115 cluster fields (Section~\ref{subsec:compactcat}), as well as optical/IR DECaLS cross-match catalogues for the compact sources in the Abell~209 and Abell~S295 fields (Section~\ref{sec:optxmatch}). We have also provided a list of 59 multi-component extended radio sources in these two cluster fields (Section~\ref{sec:srcs-ext}). Finally, in Table~\ref{tab:diffuse} we have provided a catalogue of diffuse cluster radio emission, containing 99 distinct sources detected in 62 of the MGCLS fields, of which 56 are new.

We also presented some early science results using the DR1 data, with some significant science findings while highlighting the potential for future community study.  In particular, we reported:
\begin{itemize}
    \item The lowest luminosity radio relic candidate detected to date (Section~\ref{sec:de-relics}), by exploiting the excellent surface brightness sensitivity of the MGCLS;
    \item Diffuse structures in several clusters that do not fall cleanly into the typical classes of mini-halo, halo, or relic, indicating the need for new dynamical, particle acceleration, or field amplification processes in the intracluster medium (Section~\ref{sec:clusterde} and Table~\ref{tab:diffuse});
    \item Radio galaxy structures that cannot be explained using current models (Section~\ref{sec:individsrcs}), including trident-like structures, jets that stay well-collimated far past their bending radius, and filamentary features connecting, at least in projection, with otherwise normal radio galaxy structures;
    \item The detection of 459 star-forming galaxies out to $\sim$\,3.5\,R$_{200}$ in Abell~209 (Section~\ref{sec:pops}). We find no star formation rate evolution with distance from the cluster centre, and a reduction in the ratio of radio to 100\,$\mu$m flux densities with increased clustercentric distance;
    \item Early results from \textup{H\,\textsc{\lowercase{i}}} studies of four MGCLS clusters (Section~\ref{sec:hiscience}), including \textup{H\,\textsc{\lowercase{i}}} mass distributions in three clusters, and a new foreground \textup{H\,\textsc{\lowercase{i}}} group in the Abell~3365 field.
    
\end{itemize}

The results presented here represent only a small fraction of what can be achieved with the DR1 legacy data set. Follow-up projects by the MGCLS team and the broader community are likely to make significant contributions to many areas of astrophysics.

\begin{acknowledgements}
The MeerKAT telescope is operated by the South African Radio Astronomy Observatory, which is a facility of the National Research Foundation, an agency of the Department of Science and Innovation. \kk{The authors acknowledge the contribution of those who designed and built the MeerKAT instrument.} The authors thank the anonymous reviewer for their helpful comments which improved the manuscript. \\

\indent KK and ML acknowledge funding support from the South African Radio Astronomy Observatory (SARAO) and the National Research Foundation (NRF). WDC acknowledges support from the National Radio Astronomy Observatory, which is a facility of the National Science Foundation operated by Associated Universities, Inc.
Partial support for LR comes from U.S. National Science Foundation grant AST17-14205 to the University of Minnesota. Basic research in radio astronomy at the US Naval Research Laboratory for TEC is supported by 6.1 Base Funding. RPD, SR, and GL's research is funded by the South African Research Chairs Initiative of the DSI/NRF (grant ID: 77948). MR's research is supported by the SARAO HCD programme via the ``New Scientific Frontiers with Precision Radio Interferometry'' research group grant.
VP acknowledges financial support from the South African Research Chairs Initiative of the Department of Science and Innovation and NRF. Opinions expressed and conclusions arrived at are those of the authors and are not necessarily to be attributed to the NRF. 
TV acknowledges the support from the Ministero degli Affari Esteri della Cooperazione Internazionale - Direzione Generale per la Promozione del Sistema Paese Progetto di Grande Rilevanza ZA18GR02. \\

\indent \kk{We thank E. Egami for providing access to ACReS data. }

Some results published here have made use of the \textsc{caracal} pipeline, partially supported by ERC Starting grant number 679629 “FORNAX”, MAECI Grant Number ZA18GR02, DST-NRF Grant Number 113121 as part of the ISARP Joint Research Scheme, and BMBF project 05A17PC2 for D-MeerKAT. 

We acknowledge the use of computing facilities of the Inter-University Institute for Data Intensive Astronomy (IDIA) for part of this work. IDIA is a partnership of the Universities of Cape Town, of the Western Cape, and of Pretoria. 

This research made use of Astropy (\url{http://www.astropy.org}), a community-developed core Python package for Astronomy (Astropy Collaboration et al. 2013, 2018), and routines from the publicly available Astromatch package authored by A.~Ruiz. It also made use of the SAOImageDS9 \citep{2003ASPC..295..489J} software for image preparation, and the NASA/IPAC Extragalactic Database (NED) and NASA/IPAC Infrared Science Archive (IRSA), which are funded by the National Aeronautics and Space Administration and operated by the California Institute of Technology.

\textit{Chandra} X-ray Observatory data shown in this paper were accessed through the Chandra Data Archive (ObsIDs: 13380, 15173, 16283).  \textit{XMM-Newton} data shown in this paper were accessed through the XMM-Newton Science Archive (ObsID: 0720253401). 

The Digitized Sky Surveys were produced at the Space Telescope Science Institute under U.S. Government grant NAG W-2166. The images of these surveys are based on photographic data obtained using the Oschin Schmidt Telescope on Palomar Mountain and the UK Schmidt Telescope. The plates were processed into the present compressed digital form with the permission of these institutions.

DECam data were obtained through the Legacy Surveys imaging of the DESI footprint, which is supported by the Director, Office of Science, Office of High Energy Physics of the U.S. Department of Energy under Contract No. DE-AC02-05CH1123, by the National Energy Research Scientific Computing Center, a DOE Office of Science User Facility under the same contract; and by the U.S. National Science Foundation, Division of Astronomical Sciences under Contract No. AST-0950945 to NOAO.

DECam data were also obtained from  the Astro Data Lab at NSF's National Optical-Infrared Astronomy Research Laboratory. NOIRLab is operated by the Association of Universities for Research in Astronomy (AURA), Inc. under a cooperative agreement with the National Science Foundation.

The Pan-STARRS1 Surveys (PS1) and the PS1 public science archive have been made possible through contributions by the Institute for Astronomy, the University of Hawaii, the Pan-STARRS Project Office, the Max-Planck Society and its participating institutes, the Max Planck Institute for Astronomy, Heidelberg and the Max Planck Institute for Extraterrestrial Physics, Garching, the Johns Hopkins University, Durham University, the University of Edinburgh, the Queen's University Belfast, the Harvard-Smithsonian Center for Astrophysics, the Las Cumbres Observatory Global Telescope Network Incorporated, the National Central University of Taiwan, the Space Telescope Science Institute, the National Aeronautics and Space Administration under Grant No. NNX08AR22G issued through the Planetary Science Division of the NASA Science Mission Directorate, the National Science Foundation Grant No. AST-1238877, the University of Maryland, Eotvos Lorand University (ELTE), the Los Alamos National Laboratory, and the Gordon and Betty Moore Foundation.

The Very Large Array Sky Survey (VLASS) is a project of the National Radio Astronomy Observatory, which is a facility of the U.S. National Science Foundation operated under cooperative agreement by Associated Universities, Inc.
\end{acknowledgements}

\bibliographystyle{aa}
\bibliography{gcls}

\begin{appendix}
\section{Extended Source Catalogues}\label{app:extsrccats}
Here we present the extended, multi-component source catalogues for the Abell~209 (Table~\ref{tab:extsrcs_A209}) and Abell~S295 (Table~\ref{tab:extsrcs_AS295}) cluster fields. See discussion in Section~\ref{sec:srcs-ext}.
\begin{table*}
\centering
\caption{Extended sources in the Abell~209 MGCLS field. Cols: (1--2) Source J2000 R.A. and Dec. --- the position is that of the optical host, or the flux-weighted centroid if no optical host is identified; (3--4) MeerKAT 1.28\,GHz integrated flux density and its uncertainty; (5) MeerKAT 1.28\,GHz peak brightness; (6) Largest angular size; (7) Optical host identifier where known --- number-only IDs are from DECaLS; (8) Optical host redshift.}
\label{tab:extsrcs_A209}
\begin{tabular}{ccrr.rll}
\hline\hline
(1)      & (2)       & \multicolumn{1}{c}{(3)}  & \multicolumn{1}{c}{(4)} & \multicolumn{1}{c}{(5)}  & \multicolumn{1}{c}{(6)}     & (7)           & (8)  \\
R.A.       & Dec.       & \multicolumn{1}{c}{$S_{\rm int}$} & \multicolumn{1}{c}{${\Delta}S_{\rm int}$} & \multicolumn{1}{c}{$I_{\rm peak}$} & \multicolumn{1}{c}{LAS}     & Optical ID    & $z$    \\
(deg)    & (deg)     & \multicolumn{1}{c}{(mJy) }        & \multicolumn{1}{c}{(mJy) }                & \multicolumn{1}{c}{(mJy\,beam\per)}         & \multicolumn{1}{c}{(\arcsec)}&               &      \\
\hline
22.9860 & $-$13.5807 & 6.049  & 0.053 & 2.411  & 70  & -                         & -    \\
23.0179 & $-$13.6651 & 3.549  & 0.031 & 0.530  & 76  & 7190                      & 0.2  \\
22.8985 & $-$13.4712 & 13.067 & 0.063 & 7.516  & 136 & 4968                      & 0.27 \\
23.1898 & $-$13.3511 & 38.210 & 0.044 & 11.053 & 73  & -                         & -    \\
23.1666 & $-$13.1819 & 3.503  & 0.040 & 1.367  & 56  & 6205                      & 0.17 \\
23.2199 & $-$13.1328 & 2.074  & 0.066 & 0.807  & 54  & 8068                      & 0.08 \\
23.3709 & $-$13.5798 & 1.993  & 0.029 & 0.716  & 54  & 1596                      & 0.46 \\
23.5330 & $-$13.5719 & 2.742  & 0.030 & 1.061  & 82  & WISEA J013407.92$-$133418.8 & -     \\
22.9890 & $-$13.7451 & 3.618  & 0.056 & 1.742  & 107 & 4559                      & 0.2  \\
22.5779 & $-$13.8580 & 0.855  & 0.016 & 0.563  & 37  & 591                       & 0.36 \\
22.5694 & $-$13.8677 & 1.383  & 0.024 & 0.271  & 43  & -                         & - \\
22.7784 & $-$13.0524 & 26.687 & 0.125 & 4.139  & 94  & WISEA J013106.83$-$130308.7 & -     \\
22.7566 & $-$13.3371 & 1.705  & 0.023 & 0.441  & 54  & 1353                      & 0.34 \\
22.6263 & $-$13.4913 & 26.158 & 0.029 & 16.544 & 76  & 4158                      & 0.3  \\
23.4957 & $-$13.2933 & 3.019  & 0.027 & 0.767  & 89  & WISEA J013358.96$-$131736.1  & 0.1  \\
22.5028 & $-$13.5075 & 1.689  & 0.033 & 0.430  & 115 &  -                         &  -    \\
23.3231 & $-$13.6783 & 1.387  & 0.032 & 0.356  & 72  &   -                        &  -    \\
23.0368 & $-$13.3279 & 8.527  & 0.026 & 4.004  & 68  & WISEA J013208.83$-$131940.4 & -     \\
22.9934 & $-$14.1088 & 4.678  & 0.027 & 0.286  & 74  & WISEA J013158.40$-$140631.7  & -     \\
22.9494 & $-$13.9215 & 6.061  & 0.019 & 2.928  & 33  & WISEA J013147.85$-$135517.5 & -     \\
23.3539 & $-$13.3721 & 1.959  & 0.019 & 0.838  & 94  &    -                       & -     \\
22.6013 & $-$13.1741 & 1.300  & 0.021 & 0.889  & 44  & WISEA J013024.30$-$131026.7 & -     \\
23.3064 & $-$13.0585 & 79.000 & 0.083 & 18.792 & 468 & WISE J013313.50$-$130330.5  & -     \\
22.5031 & $-$13.4747 & 13.570 & 0.016 & 7.594  & 58  & WISEA J013000.74$-$132828.9 & -     \\
23.5757 & $-$13.9423 & 6.193  & 0.028 & 1.196  & 93  & WISEA J013418.18$-$135632.2 & -     \\
23.5998 & $-$13.5860 & 1.831  & 0.012 & 0.827  & 41  & WISEA J013423.94$-$133509.6 & -     \\
22.4480 & $-$13.2866 & 28.055 & 0.026 & 8.629  & 65  &     -                      &  -    \\
23.5316 & $-$13.8162 & 25.822 & 0.019 & 10.615 & 47  & WISEA J013407.58$-$134858.4  & -     \\
22.9680 & $-$13.6162 & 20.589 & 0.049 & 12.144 & 83  & 164                       & 0.18 \\
23.4633 & $-$13.6041 & 2.668  & 0.018 & 0.857  & 40  &      -                     &  -    \\
22.6359 & $-$13.8971 & 4.942  & 0.013 & 3.103  & 35  & WISEA J013032.61$-$135349.6 & -     \\
22.6052 & $-$14.0115 & 1.900  & 0.029 & 0.327  & 182 & WISEA J013025.31$-$140042.5 & -     \\
22.9408 & $-$13.6772 & 0.497  & 0.013 & 0.127  & 46  & 6812                      & 0.19 \\
\hline
\end{tabular}
\end{table*}

\begin{table*}
\centering
\caption{Extended sources in the Abell~S295 MGCLS field. Cols: (1--2) Source J2000 R.A. and Dec. --- the position is that of the optical host, or the flux-weighted centroid if no optical host is identified; (3--4) MeerKAT 1.28\,GHz integrated flux density and its uncertainty; (5) MeerKAT 1.28\,GHz peak brightness; (6) Largest angular size; (7) Optical host identifier where known --- number-only IDs are from DECaLS; (8) Optical host redshift.}
\label{tab:extsrcs_AS295}
\begin{tabular}{ccrr.rll}
\hline\hline
(1)      & (2)       & \multicolumn{1}{c}{(3)}  & \multicolumn{1}{c}{(4)} & \multicolumn{1}{c}{(5)}  & \multicolumn{1}{c}{(6)}     & (7)           & (8)  \\
R.A.       & Dec.       & \multicolumn{1}{c}{$S_{\rm int}$} & \multicolumn{1}{c}{${\Delta}S_{\rm int}$} & \multicolumn{1}{c}{$I_{\rm peak}$} & \multicolumn{1}{c}{LAS}     & Optical ID    & $z$    \\
(deg)    & (deg)     & \multicolumn{1}{c}{(mJy) }        & \multicolumn{1}{c}{(mJy) }                & \multicolumn{1}{c}{(mJy\,beam\per)}         & \multicolumn{1}{c}{(\arcsec)}&               &      \\
\hline
41.0580 & $-$53.0237 & 8.243   & 0.021 & 3.486  & 71  & -                          & -         \\
41.7923 & $-$52.8463 & 1.945   & 0.068 & 0.865  & 315 & WISEA J024710.13$-$525046.6 & 0.088    \\
41.2710 & $-$52.8090 & 2.298   & 0.035 & 0.314  & 94  & -                          & -         \\
41.4154 & $-$52.9641 & 123.317 & 0.137 & 16.971 & 587 & ESO 154$-$IG 011 NED01      & 0.096704 \\
42.2267 & $-$52.9842 & 3.624   & 0.098 & 0.590  & 330 & 2686                      & 0.46     \\
42.2926 & $-$53.0489 & 3.934   & 0.063 & 0.380  & 257 & -                          & -         \\
41.3352 & $-$53.1849 & 33.371  & 0.057 & 10.234 & 117 & 3827                      & 0.04     \\
41.3259 & $-$53.2520 & 104.090 & 0.038 & 21.642 & 117 & WISEA J024518.34$-$531505.3 & -         \\
41.6932 & $-$53.2965 & 3.580   & 0.041 & 0.826  & 144 & 1325                      & 0.32     \\
40.9512 & $-$53.5577 & 7.315   & 0.048 & 1.761  & 96  & -                          & -         \\
42.0530 & $-$53.4280 & 6.453   & 0.103 & 0.936  & 179 & -                          & -         \\
41.1320 & $-$53.2208 & 56.355  & 0.031 & 31.709 & 49  & -                          & -         \\
41.4107 & $-$53.6220 & 9.437   & 0.035 & 2.679  & 91  & -                          & -         \\
42.0002 & $-$52.9116 & 3.227   & 0.056 & 2.377  & 174 & 3545                      & 0.01     \\
40.5346 & $-$53.1455 & 3.978   & 0.029 & 1.060  & 53  & 3924                      & 0.44     \\
40.6013 & $-$52.7182 & 1.878   & 0.028 & 0.411  & 57  & WISEA J024224.31$-$524305.4 & -         \\
41.2443 & $-$52.6912 & 4.972   & 0.054 & 2.696  & 117 & 3616                      & 0.16     \\
41.6280 & $-$52.4638 & 16.509  & 0.061 & 3.417  & 140 & -                          & -         \\
41.4335 & $-$52.8585 & 2.895   & 0.020 & 0.989  & 39  & WISEA J024543.99$-$525130.3 & -         \\
41.5287 & $-$53.2896 & 1.414   & 0.030 & 0.284  & 79  &  -                         & -         \\
41.5898 & $-$53.3468 & 0.884   & 0.020 & 0.182  & 44  & 2MASS J02462155$-$5320505   & 0.052199 \\
41.9143 & $-$53.5935 & 33.835  & 0.092 & 9.708  & 160 & WISEA J024739.42$-$533536.5 & -         \\
41.3321 & $-$53.0510 & 3.002   & 0.023 & 1.725  & 36  & 1286                      & 0.26     \\
41.3598 & $-$53.0172 & 1.030   & 0.007 & 0.125  & 28  & -                          & -         \\
40.4822 & $-$53.3950 & 79.906  & 0.033 & 1.045  & 180 & 3502                      & 0.51     \\
41.3927 & $-$52.7395 & 75.244  & 0.054 & 17.551 & 80  & WISEA J024534.30$-$524422.1 & -    \\
\hline
\end{tabular}
\end{table*}

\section{\kk{Star-forming Galaxies Catalogue for Abell 209: SFR and Radio/FIR}}\label{app:popstudies}
\kk{To carry out the environment-dependent studies discussed in Section~\ref{sec:pops}, namely galaxy star-formation rates (SFRs) and the radio-far-infrared (FIR) correlation in Abell~209, we needed to identify which of the MGCLS-detected sources belong to the cluster, remove AGN-dominated sources from that group, and calculate the star formation rate and the radio-FIR ratio. Here we discuss the details of these procedures and the creation of the Abell~209 star-forming galaxies catalogue being released as part of the DR1 products. This catalogue includes the SFRs and radio/FIR ratios for the cluster's star-forming galaxies used in the scientific analyses presented in Section~\ref{sec:pops}. An excerpt of this catalogue is shown in Table~\ref{tab:sfrir}.}

\begin{table*}[ht!]
\caption{Excerpt of the SFR and IR catalogue of star-forming cluster galaxies in Abell~209, used in the SFR and radio/FIR studies in this paper (Sections~\ref{sec:sfrs} and \ref{sec:rfir}, respectively). The first column is an assigned source ID for the radio source. Column descriptions for columns (2)--(28) (not shown here) are the same as in the optical cross-match catalogue, described in Table \ref{table:optxmatch}. The additional columns are: (29) spectroscopic redshift; (30) catalogue from which the redshift was obtained; (31--32) Star formation rate and uncertainty; (33) Radius from centre in units of $R_{200}$; (34--35) 100\,$\mu$m flux density and uncertainty; (36--37) Log of the ratio between the radio and FIR flux densities, and associated uncertainty. 
The full catalogue is available online. } 
\label{tab:sfrir}      
\centering                         
\begin{tabular}{ccccccccccc} 
\hline                 
(1) & (2)-(28) & (29) & (30) & (31) & (32) & (33) & (34) & (35) & (36) & (37)  \\
Source & ... & $z_{\rm s}$ & $z$ Source & SFR & $\Delta$SFR & relR & $S_{\rm100\,{\mu}m}$  & $\Delta S_{\rm100\,{\mu}m}$ & ${\rm log}\left(\frac{S_{\rm1.28\,GHz}}{S_{\rm100\,{\mu}m}}\right)$ & $\Delta {\rm log}\left(\frac{S_{\rm1.28\,GHz}}{S_{\rm100\,{\mu}m}}\right)$ \\    
ID & & & & ($\rm M_\odot\,yr$\per) & ($\rm M_\odot\,yr$\per) &  & (mJy) & (mJy) & & \\
\hline                        
1649 & ... & -       & none  & 4.925 & 0.932 & 3.030 & -      & -     & -      & - \\
1660 & ... & -       & none  & 1.945 & 0.599 & 1.975 & -      & -     & -      & - \\
1665 & ... & 0.21111 & ACReS & 1.541 & 0.442 & 1.136 & 11.217 & 3.709 & $-2.515$ &	0.268 \\
1687 & ... & 0.1963  & CLASH & 6.434 & 0.512 & 1.059 & 30.443 & 4.890 & $-2.284$ &	0.104 \\
1691 & ... & 0.2071  & CLASH & 4.395 & 0.493 & 1.234 & 22.711 & 6.400 & $-2.322$ &	0.171 \\
1706 & ... & -       & none  & 3.301 & 0.953 & 2.263 & -      & -     & -      & - \\
1728 & ... & 0.21197 & ACReS & 2.654 & 0.500 & 1.227 & 27.888 & 5.172 & $-2.630$ &	0.162 \\
1731 & ... & -       & none  & 1.646 & 0.562 & 1.566 & -      & -     & -      & - \\
1732 & ... & 0.17697 & ACReS & 4.664 & 0.510 & 1.781 & -      & -     & -      & - \\
\hline                                   
\end{tabular}
\end{table*}

\subsection{Cluster galaxy membership}\label{sec:a209membership}

We assigned cluster membership using a combination of spectroscopic, $z_{\rm s}$, and photometric, $z_{\rm p}$, redshifts, with the latter determined using the \texttt{zCluster} photometric redshift code \citep{Hilton_2020} and photometry from DECaLS \citep{Dey_2019}. The CLASH-VLT \citep{Annunziatella_2016} and ACReS 
\citep{Haines_2015} spectroscopic datasets contain secure redshifts for 1256 and 345 galaxies in the Abell~209 field, respectively, with a combined total of 1425 {unique} galaxies. We define spectroscopic cluster member galaxies as those with peculiar radial velocities within $\pm 3\,\sigma_{\rm v}$ of the cluster redshift, where
$\sigma_{\rm v} = 1320$\,km\,s$^{-1}$ is the line of sight velocity dispersion \citep{Annunziatella_2016}. 
Photometry-based cluster membership is defined as those galaxies with $|z_{\rm p} - z_{\rm c}| < 3\,\sigma_{\rm bw}(1+z_{\rm c})$, where $z_c$ is the cluster redshift of 0.206 and $\sigma_{\rm bw} = 0.03$ is the scatter in our photometric redshift residuals. This scatter is determined using a biweight scale estimate \citep{Beers_1990}, and the residuals are calculated as $\Delta z/(1+z_{\rm s})$, where $\Delta z = z_{\rm s} - z_{\rm p}$. 

We obtained a final catalogue of 523 MGCLS-detected cluster members within the primary beam-corrected field of view, with 98 members within $R_{200}$. We determine the amount of contamination in the photometric redshift sample using the 91 member galaxies with both spectroscopic and photometric redshifts\footnote{For galaxies which have both a spectroscopic and photometric redshift, inclusion as a cluster member is defined by the spectroscopic redshift.}. We find that 16\% of the members selected using photometric redshifts have spectroscopic redshifts outside the photometric redshift cut range, while 23\% of the spectroscopically-identified cluster galaxies are missed by the photometric redshift selection. 

\subsection{\kk{AGN contamination}}\label{sec:a209agnex}
To separate star-forming galaxies from AGN-dominated galaxies,
we used the `R90' WISE IR-selection criteria by \citet{Assef_2018} \citep[see also][]{Stern_2012,Assef_2013}. These utilise only the WISE W1$-$W2 colour to identify AGN without a threshold that depends on the W2 band magnitude 
\citep[see equation 4 of][which identifies AGNs with 90\% reliability]{Assef_2018}. With this selection method, we classify 64 radio-detected cluster galaxies as AGNs and remove them from the cluster member sample. This leaves a total of 459 MGCLS-detected star-forming cluster members within the primary beam-corrected field of view (80 within $R_{200}$). Table \ref{tab:sfrir} presents an excerpt of the catalogue of star-forming cluster galaxies in Abell~209.

\subsection{\kk{Radio-derived s}tar formation rates}

To estimate the SFR from the radio luminosities we used the \citet{Bell2003} relation \citep[see also][]{Karim2011} scaled to a \citet{Chabrier_2003} IMF:
\begin{equation}
    \rm SFR~(M_{\odot}\,yr^{-1}) = 
        \begin{cases}
            3.18\times10^{-22}L &, ~ L>L_{\rm c}\\
            
            \dfrac{3.18\times10^{-22}L}{0.1+0.9 (L/L_{\rm c})^{0.3}} &, ~ L\leq L_{\rm c}.
        \end{cases}
\end{equation}
$L = L_{\rm 1.4\,GHz}$ is the radio luminosity in W\,Hz$^{-1}$ derived from the MGCLS 1.28\,GHz total flux density, using a power law scaling and assuming a non-thermal spectral index of $-0.8$ \citep{Condon1992}. 
$L_{\rm c}=6.4\times10^{21}$\,W\,Hz$^{-1}$ is taken to be the typical radio luminosity of an $L_{*}$ galaxy.  \citet{Bell2003} argued that galaxies with low luminosities could have their non-thermal emission significantly suppressed and therefore need to be separated from the population with higher luminosities. However, as only a small percentage (39/459) of the final sample have luminosities lower than $L_{\rm c}$, we 
do not separate according to luminosity. We measure SFRs ranging between {1.2 and 432} $\rm M_{\odot}$\,yr\per, within the range of known values. Under the assumption that the radio emission is due to star formation, the median 5$\sigma$ sensitivity limit corresponds to SFR$_{5\sigma}$\,$> 2$\,M$_{\odot}$\,yr$^{-1}$, with 429/459 star-forming members with SFRs above this limit. 

\subsection{\kk{FIR cross-matching for r}adio/FIR correlation}
\editout{In Section~}
\editout{, we identified 459 compact radio sources as star forming galaxies belonging to the Abell~209 cluster. Cross-matching this list}
\kk{In order to study the radio/FIR correlation in Abell~209, we first need to determine FIR flux densities for the cluster star-forming galaxies. We cross-match the catalogue of 459 star-forming members}
with the \textit{Herschel} 100\,$\mu$m catalogue\footnote{The PACS Point Source Catalogue from the NASA/IPAC Infrared Science Archive: \url{https://irsa.ipac.caltech.edu/cgi-bin/Gator/nph-scan?submit=Select&projshort=HERSCHEL}}\kk{. This} yielded 49 star forming cluster galaxies with both 1.28\,GHz and 100\,$\mu$m flux densities. \kk{We use the log of the ratio of these two quantities to investigate evolution in the radio/FIR relation } \kk{The \textit{Herschel} 100\,$\mu$m flux densities for the 49 galaxies and associated radio/FIR ratio are provided in the Abell~209 star-forming galaxies catalogue shown}\editout{An excerpt of this catalogue is presented} in Table~\ref{tab:sfrir}.

\section{Automated Searches for Interesting Objects} \label{app:astronomaly}

\kk{The definition of an `unusual' radio object is subjective and dependent on the science case of interest. In the search for radio galaxies of interest, several of which are presented in Section~\ref{sec:individsrcs},} \editout{although }the MGCLS dataset was small enough for human inspection\kk{. However,} it took a significant amount of time, with all images inspected by several people \kk{and the final set of radio galaxies of interest determined after cross-checking the various choices}. For future surveys, exhaustive \kk{human} searches for unusual objects may be impossible. We therefore used the MGCLS to examine the efficacy of machine learning techniques to find atypical, interesting objects. 

\textsc{Astronomaly} \citep{ASTRONOMALY} is a machine learning framework designed to automatically detect anomalous (rare or unusual) objects in very large datasets. We ran \textsc{Astronomaly} on 128\,px\,$\times$\,128\,px (160\arcsec\,$\times$\,160\arcsec) cutouts around components from the \textsc{pybdsf} source finding (see Section~\ref{sec:srccats}), using the same framework as \citet{ASTRONOMALY}. To compare our results with those from human inspection, we restricted our search to the cluster fields where we had visually identified at least one extended radio galaxy with complex and/or unusual morphology (approximately 40\% of the full sample). For those fields, we centred cutouts on components that were classified as having two or more Gaussian components by \textsc{pybdsf}, which resulted in 21,449 cutouts. 

\begin{figure*}
    \centering
    \includegraphics[width=0.8\paperwidth]{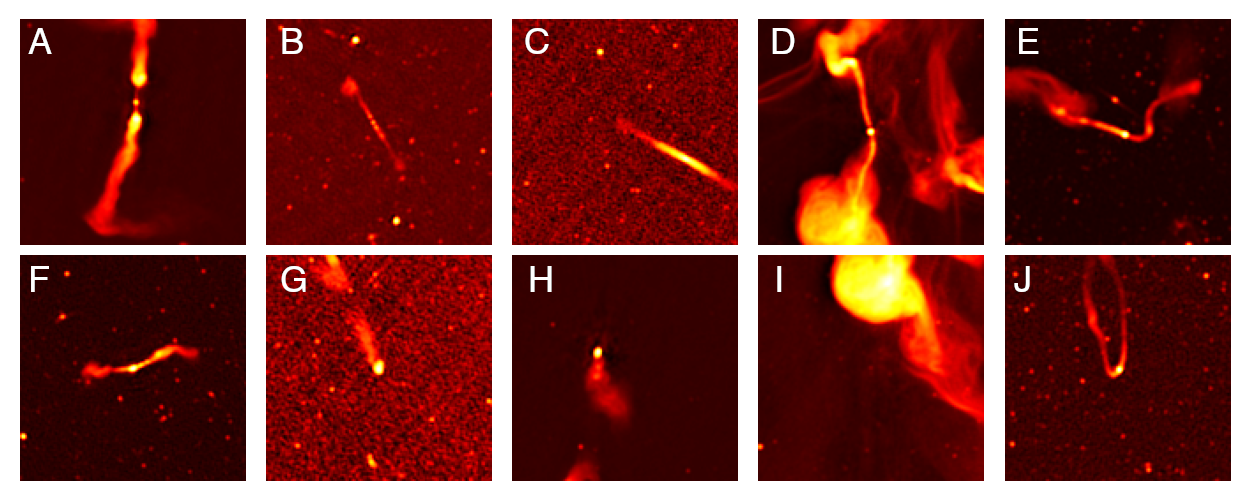}
    \caption{The 10 top-ranked cutouts based on their \textsc{Astronomaly} score. The cutouts are each 160\arcsec on a side, and their respective cluster fields (and cutout centres, in degrees of R.A., Dec.) are as follows:  A: J0216.3$-$4816 (34.190,$-47.836$); B: J1423.7$-$5412 (214.984,$-54.102$); C: Abell~22 (5.055, $-24.925$); D: Abell~194 (21.502, $-1.345$); E: Abell~S295 (41.422, $-52.962$); F: J0607.0$-$4928 (92.267, $-48.900$); G: Pandora (4.279,$-29.934$); H: J0216.3$-$4816 (33.252, $-47.690$); I: Abell~194 (21.522, $-1.446$); J: J0738.1$-$7506 (114.851, $-75.618$). We note that sources \emph{D} (and its extension to the south, identified as a separate source \emph{I}) and source \emph{J} are among the individual sources identified manually (independent of \textsc{Astronomaly}) and selected for discussion in Section~\ref{sec:individsrcs}.}
    \label{fig:astronomaly}
\end{figure*}

We used \textsc{Astronomaly}'s anomaly score to order the data from most to least anomalous, comparing this list with a selected set of 43 sources which were previously identified as `interesting' by one or more team members. Cutouts of the top 10 scoring objects are presented in Figure~\ref{fig:astronomaly}. Of the 43 `interesting' objects, \textsc{Astronomaly} detected 22 in the top 1\% (210 objects) of the rankings. Forty out of 43 sources are found in the first 10\% of the list, while the remaining three could not be distinguished from `normal' sources. We note that the vast majority of the 210 most highly ranked sources\footnote{Available at \url{https://michellelochner.github.io/mgcls.astronomaly}.} are actually visually quite similar to the selected 43. We conclude that \textsc{Astronomaly} can be a useful tool for rapid searches of morphologically interesting objects in large catalogues of radio sources, from which more targeted manual searches can be made efficiently.

\end{appendix}

\clearpage

\end{document}